\documentclass[twocolumn,trackchanges]{aastex701}
\usepackage{subcaption}

\begin{document}

\title{WFST Supernovae in the First Year: II. SN 2024aedt: Systematical Study of a Transitional Type Ia Supernova}

\author{Dezheng Meng}
\affiliation{Department of Astronomy, University of Science and Technology of China, Hefei 230026, China}
\affiliation{School of Astronomy and Space Sciences, University of Science and Technology of China, Hefei, 230026, China}
\email{dezhengmeng@mail.ustc.edu.cn}

\author[0000-0002-9092-0593]{Ji-an Jiang}
\affiliation{Department of Astronomy, University of Science and Technology of China, Hefei 230026, China}
\affiliation{National Astronomical Observatory of Japan, 2-21-1 Osawa, Mitaka, Tokyo 181-8588, Japan}
\textcolor{blue}{\email{jian.jiang@ustc.edu.cn}}

\author{Xu Kong}
\affiliation{Department of Astronomy, University of Science and Technology of China, Hefei 230026, China}
\affiliation{School of Astronomy and Space Sciences, University of Science and Technology of China, Hefei, 230026, China}
\affiliation{Institute of Deep Space Sciences, Deep Space Exploration Laboratory, Hefei 230026, China}
\email{xkong@ustc.edu.cn}

\author{Zelin Xu}
\affiliation{Department of Astronomy, University of Science and Technology of China, Hefei 230026, China}
\affiliation{School of Astronomy and Space Sciences, University of Science and Technology of China, Hefei, 230026, China}
\email{sa21022025@mail.ustc.edu.cn}  

\author[0000-0003-2611-7269]{Keiichi Maeda}
\affiliation{Department of Astronomy, Kyoto University, Kitashirakawa-Oiwake-cho, Sakyo-ku, Kyoto 606-8502, Japan}
\email{keiichi.maeda@kusastro.kyoto-u.ac.jp}

\author{Hanindyo Kuncarayakti}
\affiliation{Tuorla Observatory, Department of Physics and Astronomy, FI-20014 University of Turku, Finland}
\affiliation{Finnish Centre for Astronomy with ESO (FINCA), FI-20014 University of Turku, Finland}
\email{kuncarayakti@gmail.com}

\author{Lluís Galbany}
\affiliation{Institute of Space Sciences (ICE-CSIC), Campus UAB, Carrer de Can Magrans, s/n, E-08193 Barcelona, Spain}
\affiliation{Institut d'Estudis Espacials de Catalunya (IEEC), 08860 Castelldefels (Barcelona), Spain}
\email{lluisgalbany@gmail.com}

\author{Saurabh W. Jha}
\affiliation{Department of Physics and Astronomy, Rutgers, The State University of New Jersey, 136 Frelinghuysen Road, Piscataway, New Jersey 08854, USA}
\email{saurabh@physics.rutgers.edu}

\author{\v{Z}eljko Ivezi\'{c}}
\affiliation{Department of Astronomy, University of Washington, Box 351580, Seattle, Washington 98195-1580, USA}
\email{ivezic@uw.edu}

\author{Peter Yoachim}
\affiliation{Department of Astronomy, University of Washington, Box 351580, Seattle, Washington 98195-1580, USA}
\email{yoachim@uw.edu}

\author{Weiyu Wu}
\affiliation{Department of Astronomy, University of Science and Technology of China, Hefei 230026, China}
\affiliation{School of Astronomy and Space Sciences, University of Science and Technology of China, Hefei, 230026, China}
\email{wwyyy@mail.ustc.edu.cn} 

\author{Zhengyan Liu}
\affiliation{Department of Astronomy, University of Science and Technology of China, Hefei 230026, China}
\email{ustclzy@mail.ustc.edu.cn}

\author{Junhan Zhao}
\affiliation{Department of Astronomy, University of Science and Technology of China, Hefei 230026, China}
\affiliation{School of Astronomy and Space Sciences, University of Science and Technology of China, Hefei, 230026, China}
\email{zjunhan@mail.ustc.edu.cn}

\author{Andrew J. Connolly}
\affiliation{Department of Astronomy, University of Washington, Box 351580, Seattle, Washington 98195-1580, USA}
\email{ajc@astro.washington.edu}

\author{Ziqing Jia}
\affiliation{Department of Astronomy, University of Science and Technology of China, Hefei 230026, China}
\email{zqjia@mail.ustc.edu.cn}

\author{Lei Hu}
\affiliation{McWilliams Center for Cosmology, Department of Physics, Carnegie Mellon University, 5000 Forbes Ave, Pittsburgh, 15213, PA, USA}
\email{leihu@andrew.cmu.edu}

\author{Weiyu Ding}
\affiliation{Department of Astronomy, University of Science and Technology of China, Hefei 230026, China}
\affiliation{School of Astronomy and Space Sciences, University of Science and Technology of China, Hefei, 230026, China}
\email{dingwy@mail.ustc.edu.cn}

\author{Lulu Fan}
\affiliation{Department of Astronomy, University of Science and Technology of China, Hefei 230026, China}
\affiliation{Institute of Deep Space Sciences, Deep Space Exploration Laboratory, Hefei 230026, China}
\email{llfan@ustc.edu.cn}

\author{Feng Li}
\affiliation{State Key Laboratory of Particle Detection and Electronics, University of Science and Technology of China, Hefei 230026, China}
\email{phonelee@ustc.edu.cn}

\author{Ming Liang}
\affiliation{National Optical Astronomy Observatory (NSF’s National Optical-Infrared Astronomy Research Laboratory) 950 N Cherry Ave. Tucson Arizona 85726, USA}
\email{liangming@gmail.com}

\author{Jinlong Tang}
\affiliation{Institute of Optics and Electronics, Chinese Academy of Sciences, Chengdu 610209, China}
\email{ioetang@163.com}

\author{Zhen Wan}
\affiliation{Department of Astronomy, University of Science and Technology of China, Hefei 230026, China}
\email{zhen_wan@ustc.edu.cn}

\author{Hairen Wang}
\affiliation{Purple Mountain Observatory, Chinese Academy of Sciences, Nanjing 210023, China}
\email{hairenwang@pmo.ac.cn}

\author{Jian Wang}
\affiliation{State Key Laboratory of Particle Detection and Electronics, University of Science and Technology of China, Hefei 230026, China}
\affiliation{Institute of Deep Space Sciences, Deep Space Exploration Laboratory, Hefei 230026, China}
\email{wangjian@ustc.edu.cn}

\author{Yongquan Xue}
\affiliation{Department of Astronomy, University of Science and Technology of China, Hefei 230026, China}
\email{xuey@ustc.edu.cn}

\author{Hongfei Zhang}
\affiliation{State Key Laboratory of Particle Detection and Electronics, University of Science and Technology of China, Hefei 230026, China}
\email{nghong@ustc.edu.cn}

\author{Wen Zhao}
\affiliation{Department of Astronomy, University of Science and Technology of China, Hefei 230026, China}
\email{wzhao7@ustc.edu.cn}

\author{Xianzhong Zheng}
\affiliation{Tsung-Dao Lee Institute and Key Laboratory for Particle Physics, Astrophysics and Cosmology, Ministry of Education, Shanghai Jiao Tong University, Shanghai, 201210, China}
\email{xzzheng@sjtu.edu.cn}

\author{Qingfeng Zhu}
\affiliation{Department of Astronomy, University of Science and Technology of China, Hefei 230026, China}
\affiliation{Institute of Deep Space Sciences, Deep Space Exploration Laboratory, Hefei 230026, China}
\email{zhuqf@ustc.edu.cn}

\correspondingauthor{Ji-an Jiang, Xu Kong}
\email{jian.jiang@ustc.edu.cn, xkong@ustc.edu.cn}

\begin{abstract}

We present comprehensive photometric and spectroscopic observations of a transitional type Ia SN 2024aedt, discovered by the 2.5-meter Wide Field Survey Telescope (WFST) within one day of the explosion. Its light curve is characterized by a peak absolute magnitude of $M_B = -18.49 \pm 0.03$ mag and a decline rate of $\Delta m_{15}(B) = 1.53 \pm 0.36$ mag, placing the object on the $\Delta m_{15}(B)$--$M_B$ diagram in the transition region between normal and subluminous SNe Ia. Furthermore, the early-color evolution and host galaxy environment of SN 2024aedt underscore its transitional nature, sharing properties with both normal and 91bg-like SNe Ia. Light-curve modeling with \texttt{MOSFiT} yields a synthesized $^{56}\mathrm{Ni}$ mass of $0.414 \pm 0.042\,M_{\odot}$ and a total ejecta mass of $0.548 \pm 0.108\,M_{\odot}$. A comparison with theoretical models suggests that the evolutionary trend can be broadly explained by both delayed-detonation (DDT) and double-detonation (DDet) scenarios while possible early-excess emissions predicted by DDet cannot be identified given the limited detections soon after the SN explosion. Although the overall spectral evolution of SN 2024aedt is similar to that of other transitional SNe Ia, the spectroscopic comparison reveals diversity in the early-phase blue-end features, which becomes more homogeneous at later phases. The result indicates the importance of early-time observations in understanding the origin of SN Ia diversity.

\end{abstract}



\section{Introduction} \label{sec:intro}

Type Ia supernovae (SNe Ia) are thermonuclear explosions of carbon-oxygen white dwarfs (WDs). Their extreme luminosity places them among the most energetic events observed in the Universe. Because of the homogeneous photometric properties of a majority of SNe Ia \citep{Phillips1993, Phillips1999}, they serve as standardizable cosmological distance indicators \citep{Perlmutter1997, Perlmutter1999, Riess1998}. However, the intrinsic diversity of SNe Ia introduces systematic uncertainties into cosmological measurements by utilizing them as standard candles \citep{2024ApJ...975...86V}.

Based on photometric and spectroscopic observations, SNe Ia are classified into several sub-classes in addition to the homologous, well-defined ``normal" population. SN 1991T-like SNe Ia are characterized by high peak luminosities and broad light curves. Their early-time spectra are dominated by a blue, featureless continuum with strong absorptions of Fe III. While lines of intermediate-mass elements (IMEs) begin to emerge around the time of maximum brightness, they remain weaker than those seen in normal SNe Ia \citep{1992ApJ...384L..15F, Taubenberger2017}. In contrast, SN 1991bg-like SNe Ia are sub-luminous and exhibit rapidly declining light curves. Unlike normal SNe Ia, their early spectra are rich in IMEs. However, they transition to a phase dominated by iron-group elements (IGEs) much earlier than normal SNe Ia. This occurs because the line-forming region recedes more rapidly into the IGE-rich core due to their lower explosion energy. Furthermore, the appearance of a prominent Ti II feature around maximum brightness is a key characteristic of this sub-class \citep{1992AJ....104.1543F, Taubenberger2017}.

Between these two extremes lie several transitional subtypes that bridge the gap between the normal population and the luminous or sub-luminous categories. For instance, SN 1999aa-like SNe initially resemble 91T-like events but evolve to behave like normal SNe Ia at an earlier phase \citep{Krisciunas2000, Garavini2004, Jha2006, Matheson2008}. SN 1986G-like SNe exhibit peak magnitudes and decline rates intermediate between those of normal and 91bg-like SNe Ia \citep{Phillips1987}, and are identified by a strong Si II $\lambda$5972 line and intermediate-strength Ti II lines. These transitional objects play a key role in investigating whether the different subtypes represent distinct classes of objects arising from different progenitor systems or explosion mechanisms. Apart from these, other peculiar subtypes exist, including the 02es-like SNe Ia, which are subluminous but have evolution rates comparable to normal SNe Ia \citep{2012ApJ...751..142G}, exceptionally luminous super-Chandrasekhar SNe Ia (SNe Ia-SC) \citep{AndrewHowell2006, Chakradhari2014}, SNe Iax, which are thought to be partial or failed thermonuclear explosions \citep{Jha2006_02cx}, and SNe Ia-CSM, which show clear features of interaction with circumstellar material \citep{Hamuy2003, Silverman2013}. The diversity of the SN Ia population has been further highlighted by recent studies \citep[e.g.,][]{2024MNRAS.530.5016D, 2025A&A...694A..10D}.

Decades of accumulated observational data have led to the proposal of numerous progenitor scenarios and explosion mechanisms to explain both the common traits and distinct characteristics of SNe Ia. These are broadly divided into two main categories: the single-degenerate (SD) scenario, in which a WD accretes material from a non-degenerate companion \citep{1973ApJ...186.1007W, 1982ApJ...253..798N}; and the double-degenerate (DD) scenario, where the merger of two WDs triggers an explosion \citep{1981NInfo..49....3T, 1984ApJS...54..335I, 1984ApJ...277..355W}.

The mass of the exploding WD can be either near the Chandrasekhar mass($M_{\mathrm{ch}}$) or sub-Chandrasekhar mass (sub-$M_{\mathrm{ch}}$), depending on the thermonuclear triggering mechanism. Among Chandrasekhar-mass models, a leading theory is the delayed detonation (DDT), where an initial subsonic deflagration wave transitions into a supersonic detonation, which can explain a large fraction of normal SNe Ia \citep{1991A&A...245..114K}. Alongside these, sub-$M_{\mathrm{ch}}$ mechanisms were proposed, initially to explain sub-luminous events. The most studied sub-$M_{\mathrm{ch}}$ model is the DDet mechanism, in which the detonation of a surface helium layer on the WD triggers a subsequent, catastrophic detonation of the underlying carbon-oxygen core. If the helium shell is sufficiently thin, this mechanism can explain the main observational properties of normal SNe Ia, not only sub-luminous ones. This scenario may show distinct signatures in early-time observations, e.g., strong Ti II absorption and photometric excess, due to radioactive isotopes and IGE synthesized in the outer layer of the ejecta \citep{2007ApJ...662L..95B, Jiang2017, Jiang2018, Maeda2018}.

In this paper, we present a systematic study of SN 2024aedt, a transitional Type Ia supernova. The evolution of its light curves and spectrum, and the characteristics of its host galaxy are presented. By comparing these features with other transitional supernovae and models, the characterization of this object and its possible origins are discussed. Throughout the work, all magnitudes are reported in the AB System. We adopt a flat $\Lambda$CDM cosmology with $H_0 = 67.66$ km s$^{-1}$ Mpc$^{-1}$, $\Omega_m = 0.30966$ from the Planck 18 results \citep{2020}.

This paper is organized as follows. In Section \ref{sec:obs}, the observational data for our target are detailed. Section \ref{sec:results} describes our analysis of the photometric and spectroscopic data, as well as the properties of the host galaxy. In Section \ref{sec:discussion}, we synthesize the results of the previous sections to discuss the similarities between transitional and normal SNe Ia and explore the possible explosion mechanisms. Finally, a summary of our findings is provided in Section \ref{sec:conclusion}.

\section{Observations and Data Reduction} \label{sec:obs}

The Wide Field Survey Telescope\footnote{https://wfst.ustc.edu.cn} \citep[WFST;][]{Wang2023} is a 2.5-meter optical telescope featuring a 6.5 square-degree field of view and high $u$-band sensitivity. It was jointly built by the University of Science and Technology of China (USTC) and the Purple Mountain Observatory (PMO). This design enables WFST to conduct deep surveys of the northern sky, making it a powerful instrument for exploring the variable universe and discovering extra-galactic transients such as supernovae \citep[SNe;][]{Hu2022} and tidal disruption events (TDEs). WFST started a 6-year WFST transient survey project from Dec 14, 2024.

SN 2024aedt\footnote{https://www.wis-tns.org/object/2024aedt} was officially discovered and publicly reported by the Asteroid Terrestrial-impact Last Alert System \citep[ATLAS;][]{2024TNSTR4922....1T} on UT 2024-12-16 09:00:43 (MJD 60660.376) with $m_o = 18.553 \pm 0.182$. The last non-detection was on UT 2024-12-15 09:53:26 (MJD 60659.412) with a limiting magnitude of $m_o \geq 18.51$.

However, the transient had been independently detected at an earlier epoch by WFST on UT 2024-12-14 16:35:02 (MJD 60658.691, i.e., the first night of the WFST formal survey) through the Deep High-cadence $ugr$-band Survey project (``DH$ugr$;" J. Jiang et al. 2026, in prep), a key project for the formal surveys. The first detection has a $g$-band magnitude $m_g = 19.59 \pm 0.05$ (corrected for Galactic extinction). The last non-detection was on UT 2024-12-08 16:17:45 (MJD 60652.679) with a 5$\sigma$ limiting magnitude of $m_g \geq 22.43$, as measured from one of the individual images used for stacking the reference image. The primary observing strategy involved daily/hourly-cadence 90s exposures in $u$, $g$, and $r$ bands. Standard data processing was implemented through the WFST data pipeline, which is built on a modified version of the Large Synoptic Survey Telescope (LSST) software stack \citep{2010SPIE.7740E..15A,2018PASJ...70S...5B,2019ApJ...873..111I}. More details on WFST data processing can be found in \cite{2025arXiv250115018C} and Z. Xu et al. (2025, in prep). Team members carried out visual inspections of all WFST DH$ugr$ transient candidates after the real-bogus classification. These inspections were based on specific criteria (including host-center offset, redshift, light-curve morphology, source catalog cross-match, etc.).

\begin{figure}[ht]
\centering
\includegraphics[width=0.9\columnwidth]{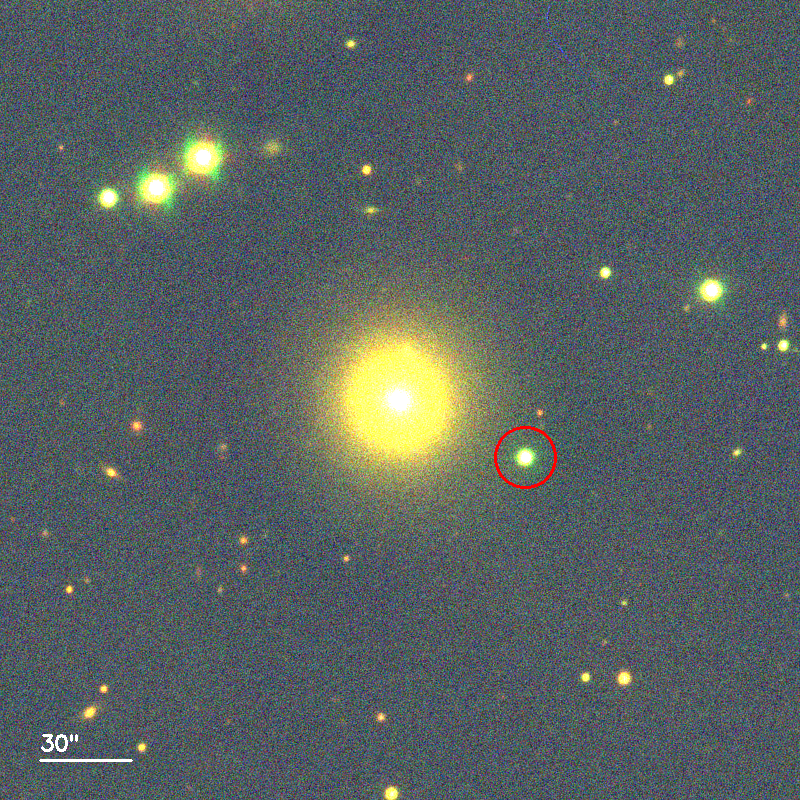}
\caption{A three-color composite image of SN 2024aedt, observed by WFST on 2024-12-25. The image is constructed from $u$ (blue), $g$ (green), and $r$ (red) filter data. The position of the supernova is marked by the red circle. A $30^{\prime\prime}$ scale bar is shown in the bottom-left corner.}
\label{fig:20241225}
\end{figure}

\subsection{Light Curves} \label{subsec:lightcurves}

After the discovery of SN 2024aedt, a multi-band photometric and spectroscopic follow-up campaign was promptly initiated. The SN was monitored by WFST until 2025 March 07, and the field was revisited by WFST in 2025 August.

The transient is located 46.49 arcsec from the center of its host, the elliptical galaxy UGC 1325 \citep{1991rc3..book.....D}, which has a spectroscopic redshift of $z=0.018413 \pm 0.000053$ \citep{2003AJ....126.2268W}, as shown in Figure \ref{fig:20241225}. While the NASA/IPAC Extragalactic Database (NED) provides a redshift-independent distance modulus of $\mu_{\text{TF}} = 34.70 \pm 0.45$ mag based on the Tully-Fisher relation \citep{2007A&A...465...71T}, we utilize the redshift-derived distance $\mu_z = 34.59 \pm 0.12 \text{ mag}$ for our primary analysis given its higher precision. The measurement uncertainty of the redshift contributes only $\sim 0.006$ mag to $\mu$, which is negligible. We incorporate an uncertainty of $300 \text{ km s}^{-1}$ for the peculiar velocity, corresponding to an uncertainty of $\sim 0.12$ mag in $\mu$. This is in good agreement with the Tully-Fisher distance modulus with a discrepancy of only $0.24\sigma$.

\begin{figure*}[ht]
\centering
\includegraphics[width=0.85\textwidth]{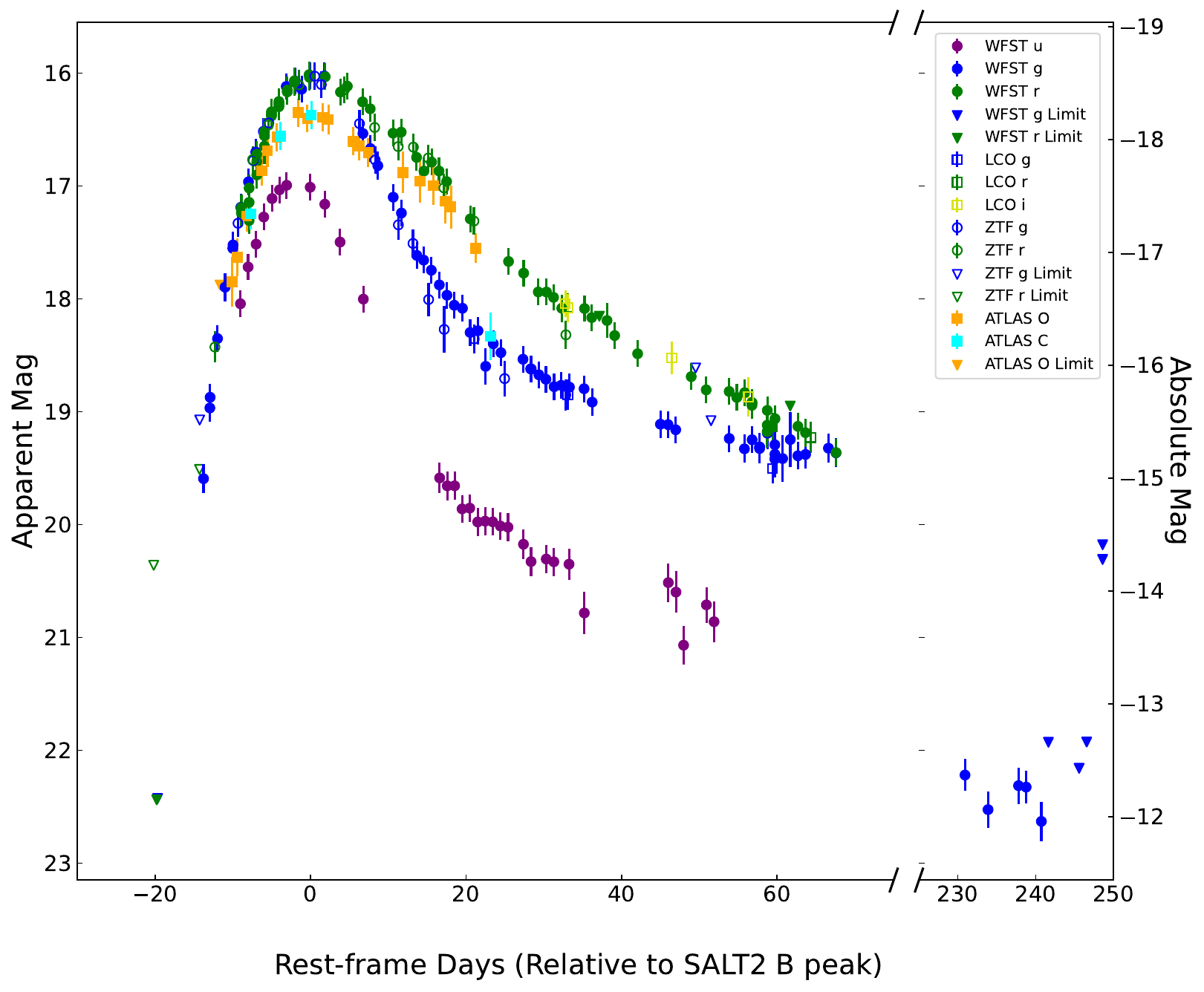}
\caption{Multi-band light curves of SN 2024aedt. The photometry for each band is color-coded as shown in the legend. Triangles denote 5$\sigma$ upper limits.}
\label{fig:lc}
\end{figure*}

Figure \ref{fig:lc} shows the multi-band light curves of SN 2024aedt. Due to the observation schedule for reference image acquisition, no early-time $u$- and $r$-band observations were obtained. By fitting the light curves with a quartic polynomial, the time of maximum brightness for each band is determined to be MJD $60670.71 \pm 0.59$ ($u$-band), MJD $60672.71 \pm 1.02$ ($g$-band), and MJD $60672.85 \pm 0.93$ ($r$-band). Additionally, the $B$-band peak time, as determined by a SALT2 fit \citep{refId0}, is MJD $60672.71 \pm 0.06$ (see Section \ref{subsec:salt2}).

The photometric data were corrected for Milky Way (MW) extinction. The specific extinction values for each band were retrieved from the NASA/IPAC Infrared Science Archive (IRSA) using the \texttt{astroquery} package \citep{2019AJ....157...98G}. This service applies the \cite{1999PASP..111...63F} (F99) extinction law to the \cite{Schlafly2011} dust map, assuming a default $R_V = 3.1$. The mean Galactic reddening of $E(B-V) = 0.0890 \pm 0.0056$ mag, which corresponds to a V-band extinction of $A_V = 0.2824$ mag. Given the large projected distance from its host galaxy, no correction for host galaxy extinction was applied. Given the low redshifts, no $K$-corrections were applied to SN 2024aedt or the comparison sample used in the subsequent analysis.

We also queried the ZTF Forced Photometry Service \citep{2023arXiv230516279M} and the ATLAS Forced Photometry Service\footnote{https://fallingstar-data.com/forcedphot/} \citep{2018PASP..130f4505T, 2020PASP..132h5002S, 2021TNSAN...7....1S} to obtain supplementary data. Following the official guidelines, these data were processed and corrected for MW extinction and redshift using the same procedure described previously.

Late-time photometric follow-up observations were conducted using telescopes from the Las Cumbres Observatory Global Telescope Network \citep[LCOGT;][]{2013PASP..125.1031B}. The data processing pipeline began with background subtraction on individual frames using \texttt{SWarp} \citep{2002ASPC..281..228B}. Subsequently, \texttt{Source Extractor} \citep{1996A&AS..117..393B} was employed to detect sources and perform astrometric calibration against the Pan-STARRS DR1 catalog \citep{https://doi.org/10.48550/arxiv.1612.05560, https://doi.org/10.48550/arxiv.1612.05243}. This allowed for the calculation of the photometric zero point for each frame, which was determined via linear regression after a 2$\sigma$ clipping process and normalized to a reference magnitude of 29. Images in the same band taken on the same night were then co-added. The zero point of the stacked image was recalculated to ensure consistency. Finally, aperture photometry was performed at the SN position using \texttt{Photutils} \citep{larry_bradley_2024_13989456}. An aperture with a diameter of 5 times the full width at half maximum (FWHM) was used, where the FWHM for each stacked image was measured using \texttt{Source Extractor}. Some images suffer from fringe artifacts. The correction procedure is described in Appendix \ref{sec:fringe}.

It should be noted that the photometric procedures vary across the different surveys. While external data from ZTF and ATLAS utilize their respective automated template-subtraction pipelines, the photometry for WFST and LCO was performed using forced photometry on the science images. For WFST, this approach was chosen because the transient was already present in the data used to construct the $r$-band reference images; forced photometry also allowed us to retrieve the flux during the reference-building period. Furthermore, given the substantial spatial offset of the SN from the host galaxy, the background contribution from host galaxy at the transient's location is negligible. We also note that despite the differences in filter transmission profiles across these facilities, the resulting light curves show consistent evolution, with data points from different telescopes agreeing within approximately $1\sigma$ uncertainties.

\subsection{Color Evolution} \label{subsec:colorevolution}

\begin{figure}[ht]
\centering
\includegraphics[width=0.9\columnwidth]{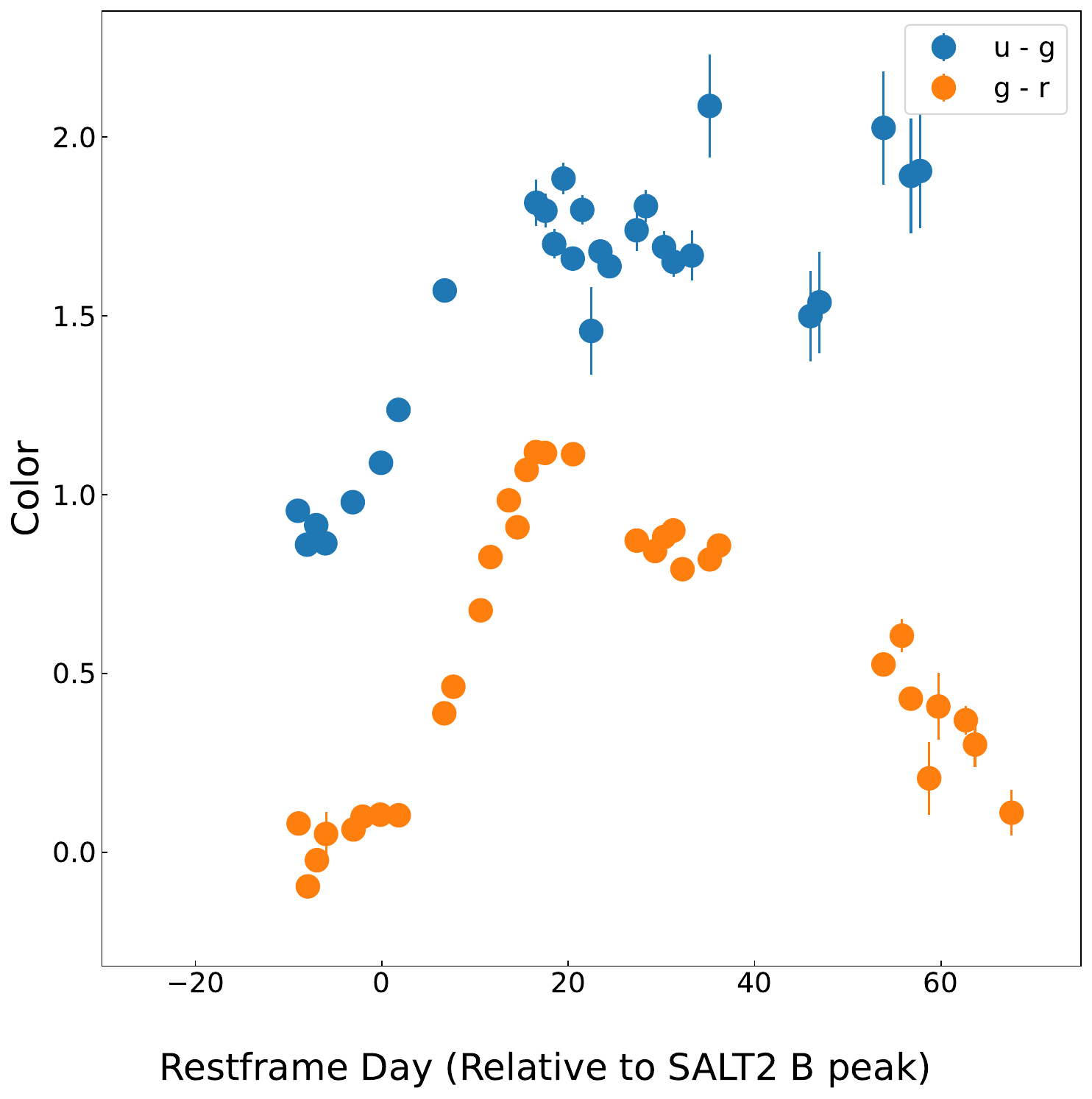}
\caption{The $u-g$ and $g-r$ color evolution of SN 2024aedt, calculated from nightly-binned WFST photometry.}
\label{fig:cl}
\end{figure}

Figure \ref{fig:cl} presents the $u-g$ and $g-r$ color evolution of SN 2024aedt. The color indexes were calculated from nightly-binned WFST photometry, and the phase is measured in days relative to the time of $B$-band maximum. The $u-g$ color begins at approximately 1.0 mag at a phase of $-9$ days, becomes slightly bluer, and then steadily reddens to a peak around $+17$ days. The $g-r$ color starts at a value of about 0.0 mag at $-9$ days and continuously reddens, reaching a peak of approximately 1.12 mag around $+16$ days.

\begin{figure}[ht]
\centering
\includegraphics[width=0.9\columnwidth]{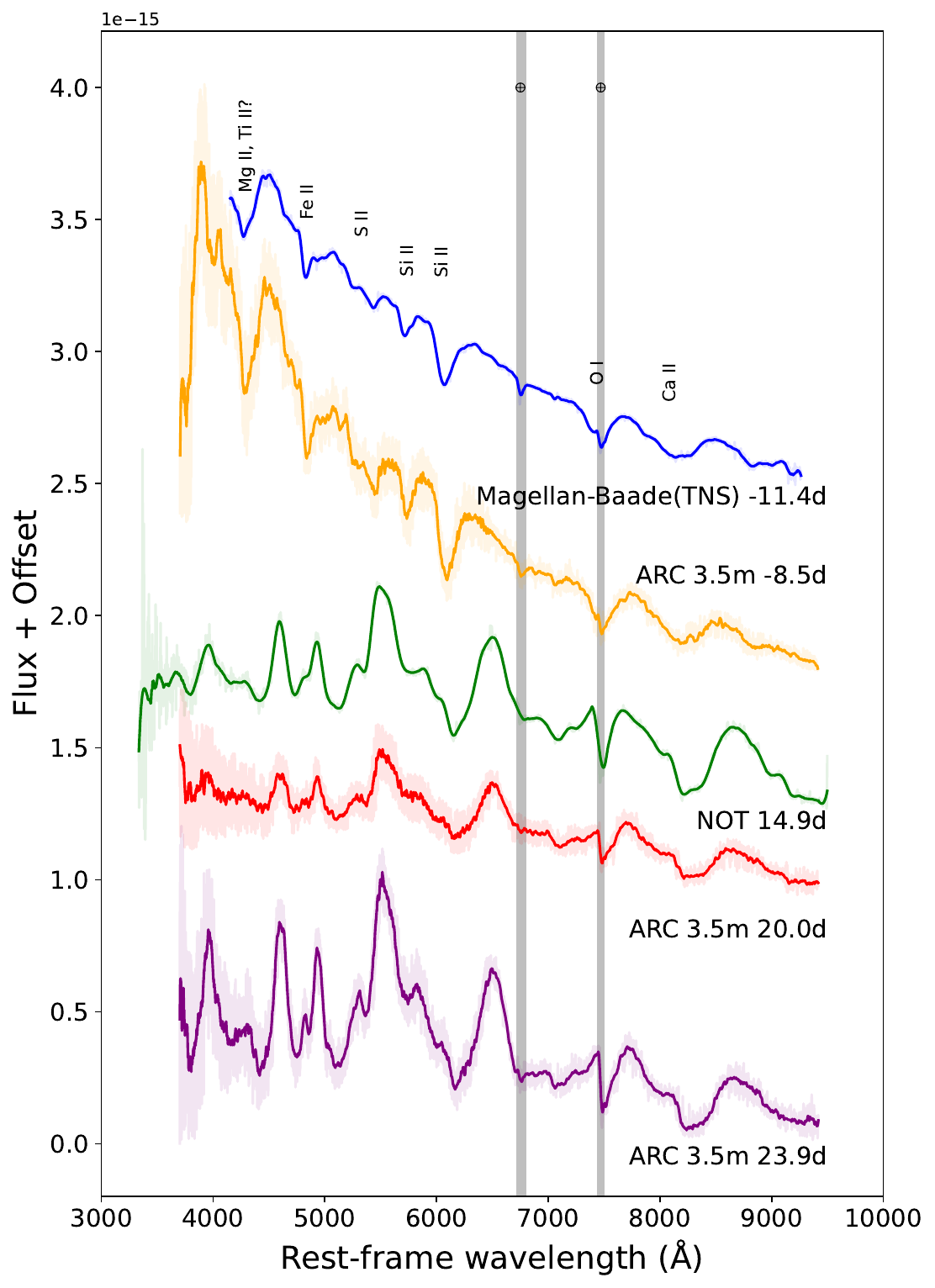}
\caption{Spectra of SN 2024aedt from multiple epochs and instruments. For display purposes, an arbitrary offset has been applied to the flux of each spectrum. The original data are shown as faint lines, while the solid lines are smoothed versions using a Savitzky-Golay filter \citep{Savitzky1964}. Shaded regions mark the locations of telluric absorption. }
\label{fig:spec}
\end{figure}

\subsection{Spectroscopy} \label{subsec:spectroscopy}

We obtain three spectra of SN 2024aedt using the Kitt Peak Ohio State Multi-Object Spectrograph (KOSMOS) on the 3.5-meter Astrophysical Research Consortium (ARC) Telescope at Apache Point Observatory (APO). KOSMOS is a medium dispersion spectrograph (resolving power up to 2600). Our observations, using the center slit, cover the wavelength ranges 3800--6600\,\AA\ (blue) and 5600--9400\,\AA\ (red). The observations were conducted on UT 2024-12-18, 2024-12-20, 2025-01-18, and 2025-01-22. The raw data are reduced using the \texttt{PyKOSMOS} \citep{davenport_2021_5120786} pipeline. The resulting spectra are then corrected for Milky Way (MW) extinction using the model of \cite{2007ApJ...663..320F} via the \texttt{extinction} package and subsequently transformed to the rest frame. Although no discernible features can be extracted from the earliest KOSMOS spectrum, the other three were successful. One spectrum was also obtained on 2025-01-12 using the Alhambra Faint Object Spectrograph and Camera (ALFOSC) on the Nordic Optical Telescope \citep[NOT;][]{2010ASSP...14..211D}, with a setup (Grism \#4, 1.0 arcsec slit) that provided a wavelength range of 3200--9600\,\AA\ and a resolution of $R = 360$.

In addition to our own observations, a publicly available spectrum from the Transient Name Server (TNS) is included in analysis. This spectrum was taken by the POISE group with the Magellan-Baade/IMACS instrument approximately one day before our first KOSMOS observation \citep{2024TNSCR4946....1M}. The full spectroscopic sequence is presented in Figure \ref{fig:spec}.

The pre-maximum spectra show prominent features characteristic of a normal SN Ia before maximum light. These include Si II $\lambda$5972, Si II $\lambda$6355, the S II "W" trough, O I $\lambda$7774, and the Ca II near-infrared (NIR) triplet. The feature around 4820\,\AA\ is identified as Fe II, while the absorption near 4285\,\AA\ is a blend of Mg II and Ti II. In the post-maximum spectra, Ca II and Si II lines remain visible. A blended absorption feature around 5750\,\AA\ is identified as Na I D, and the 4500--5500\,\AA\ region is dominated by Fe II lines. Our final spectrum, taken at a phase of +24.0 days relative to $B$-band maximum, indicates that the supernova had not yet entered the nebular phase.

\section{Results} \label{sec:results}

\subsection{Explosion Time} \label{subsec:exp}

\begin{figure}[ht]
\centering
\includegraphics[width=0.9\columnwidth]{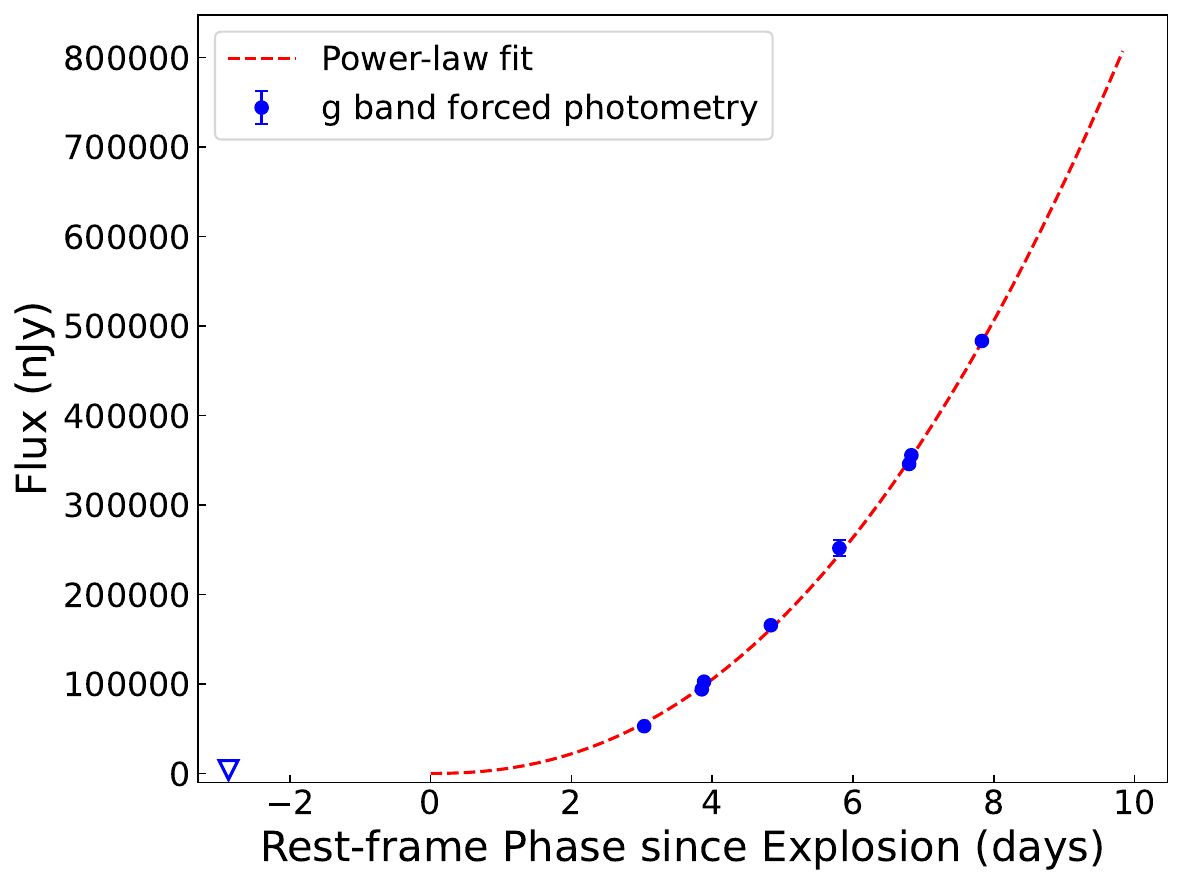}
\caption{Power-law fit to the early $g$-band light curve. The fitting procedure and results are described in Section \ref{subsec:exp}.}
\label{fig:exp}
\end{figure}

To determine the explosion time ($t_0$) and characterize the early light curve, we fit the rising-phase photometry with the following power-law function:
\begin{equation}
f(t) = A(t-t_0)^\alpha
\end{equation}
Following the methodology described in \cite{Olling2015}, only photometric points with a flux below 40\% of the peak were included in the fit. The fit was performed in the rest-frame time, presented in Figure \ref{fig:exp}, yields the following best-fit parameters: an amplitude of $A = 4545.557 \pm 1314.776~\text{nJy}$, a power-law exponent of $\alpha = 2.266 \pm 0.105$, and a time of first light of $t_0 = \text{MJD}~60655.603 \pm 0.263$.

The resulting exponent, $\alpha = 2.266$, is consistent with the predictions of an expanding fireball model for a supernova shortly after explosion \citep{1999AJ....118.2675R}, also consistent with result from \citep{2020ApJ...902...47M} of $2.08^{+0.15}_{-0.13}$ around $1.02\sigma$. While the smooth power-law provides an excellent fit to the detected early-time data, we note that the gap between the last non-detection and the inferred explosion time means that a short-lived early flux excess—potentially arising from multiple possibilities \citep{Kasen2009, Jiang2017, Jiang2018, Jiang2021}—cannot be definitively ruled out.

\subsection{SALT2 and SNooPy Fitting} \label{subsec:salt2}

\begin{figure}[ht]
\centering
\includegraphics[width=0.9\columnwidth]{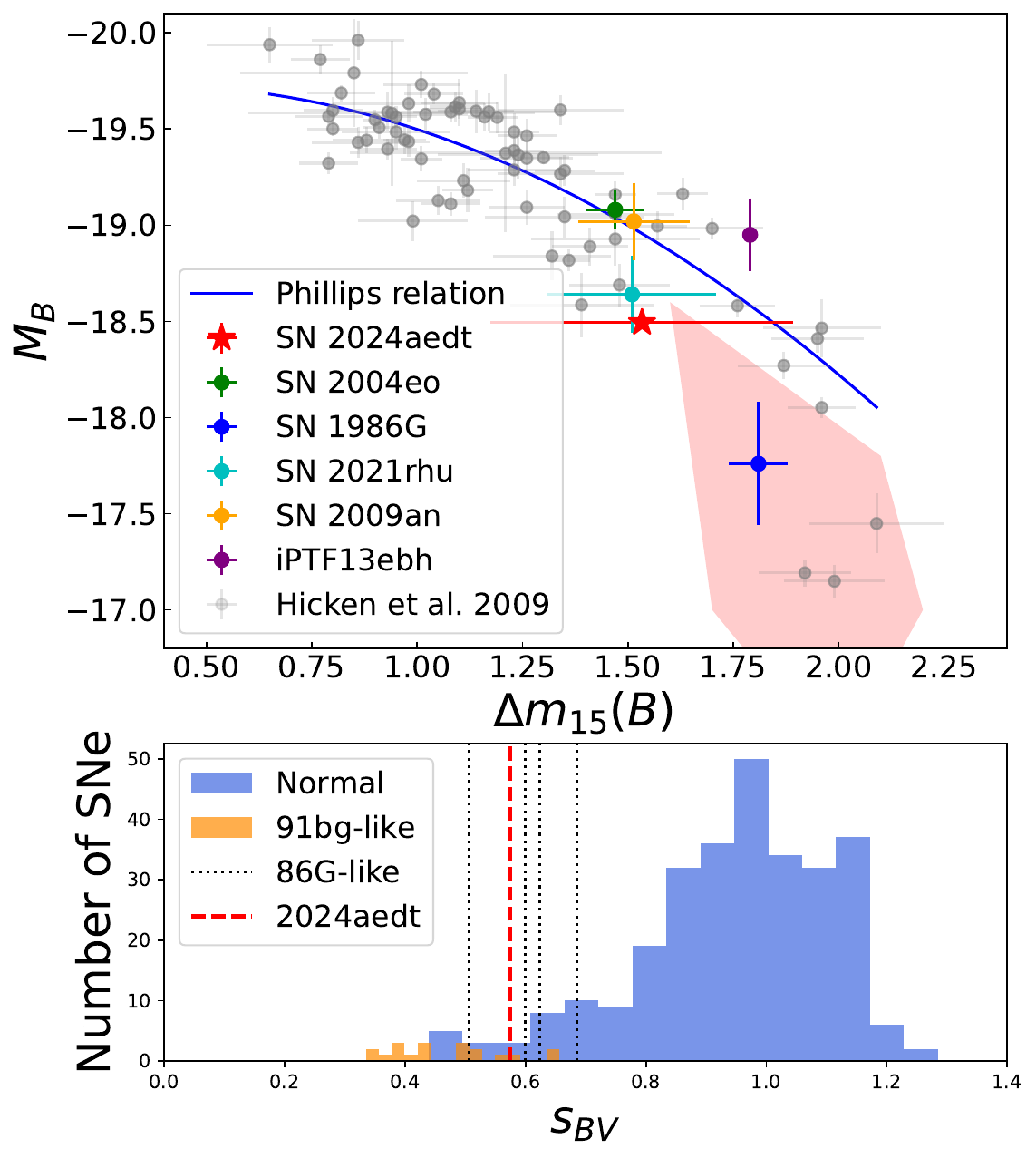}
\caption{\textbf{Upper panel:} The $\Delta m_{15}(B)$--$M_B$ diagram (Phillips relation) for SN 2024aedt and the \citep{Hicken_2009} comparison sample (excluding objects at $z < 0.01$). SN 2024aedt is indicated by the red star; see Section \ref{subsec:salt2} for details. The blue line represents the Phillips relation \citep{Phillips1999}. The `91bg-like' region from \cite{Taubenberger2017} is shown in red.\footnote{It should be noted that this region is not strictly '91bg-like' but also includes part of the parameter space for transitional SNe.} Details of the transitional sample can be found in Section \ref{subsec:compare}.
\textbf{Lower panel:} Histogram of the $s_{BV}$ distribution for SN 2024aedt and the \cite{Uddin2024} sample. SN 2024aedt is indicated by a red dashed line, and 86G-like objects from \cite{Uddin2024} sample are marked by black dotted lines.}
\label{fig:dm15}
\end{figure}

We fit WFST $g$- and $r$-band light curves using the \texttt{SALT2} model \citep{Guy2007} implemented in the \texttt{snfit} (v2.4.2) package. The best fit yields a peak absolute $B$-band magnitude of $M_B = -18.49 \pm 0.03$ mag, occurring at MJD $60672.711 \pm 0.058$. The light-curve shape parameters are $x_1 = -2.253 \pm 0.043$ and $c = 0.001 \pm 0.028$. A decline rate parameter of $\Delta m_{15}(B) = 1.53 \pm 0.36$ mag is calculated using the relation from Section 5 of \cite{Guy2007}. To cross-validate this result, we performed a direct fourth-order polynomial fit to the $g$-band photometry around the peak, yielding $\Delta m_{15}(g) = 1.614 \pm 0.003$~mag and $M_g = -18.53 \pm 0.06$~mag. The error was derived from Monte Carlo simulations. The result of \texttt{SALT2} fitting is plotted on the $\Delta m_{15}(B)$--$M_B$ diagram in the upper panel of Figure \ref{fig:dm15}, along with a comparison sample from \cite{Hicken_2009} (excluding objects at $z < 0.01$) and other transitional events. As shown in the figure, SN 2024aedt lies at the faint, fast-declining end of the distribution, below the canonical Phillips relation.

A comparison with the SNe Ia population in \cite{Taubenberger2017} (see their Figure 1) places SN 2024aedt on the boundary between normal SNe Ia and the subluminous 91bg-like class. It is slightly more luminous and slower-evolving than the prototypical transitional object, SN 1986G \citep{Phillips1987, Taubenberger2008}. Based on these characteristics, SN 2024aedt is classified as a subluminous transitional SN Ia of the 86G-like subclass.

We also fitted the light curves with \texttt{SNooPy} \citep{2011AJ....141...19B} to determine the color stretch parameter, $s_{BV}$ \citep{Burns2014}. Using the WFST $ugr$ data, this fit yields $s_{BV}=0.574$. This value is plotted as a red dotted line, along with the sample from \cite{Uddin2024}, in the lower panel of Figure \ref{fig:dm15}. SN 2024aedt, together with other 86G-like events, falls in the region between the 91bg-like and normal populations.

\subsection{Light Curves Fitting} \label{subsec:mosfit}

We modeled the light curve of SN 2024aedt using \texttt{MOSFiT} \citep[Modular Open-Source Fitter for Transients;][]{2018ApJS..236....6G}. This Python-based package employs Monte Carlo methods to generate ensembles of semi-analytical light curve fits and derive Bayesian posterior distributions for model parameters. For this work, the \texttt{ia} model \citep{1994ApJS...92..527N} is applied to the $ugr$-band photometric data at phases earlier than +40 days.

\begin{figure}[ht]
\centering
\includegraphics[width=0.9\columnwidth]{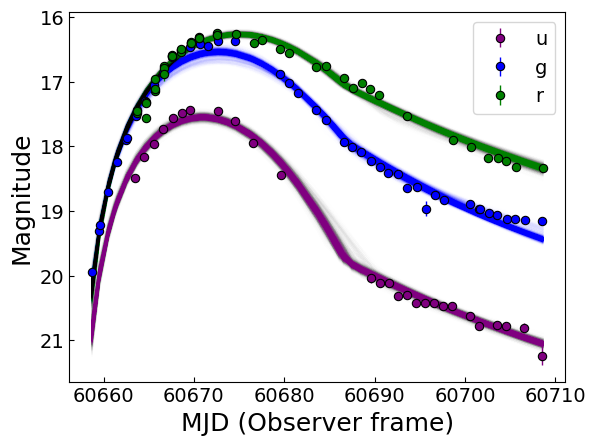}
\caption{The observed $ugr$-band photometry (solid circles) overlaid with a representative sample of model light curve realizations (solid lines) from the \texttt{MOSFiT} posterior distribution.}
\label{fig:mos}
\end{figure}

The resulting model realizations are shown against the data in Figure \ref{fig:mos}, while the parameter posterior distributions are presented in Figure \ref{fig:param}. The model provides a good overall fit, although the observed $g$-band data begins to deviate slightly from the best-fit model after a phase of +35 days. From the posterior distributions, we derive a synthesized $^{56}\mathrm{Ni}$ mass of $0.414 \pm 0.042\,M_{\odot}$ and a total ejecta mass of $0.548 \pm 0.108\,M_{\odot}$. The explosion time given by \texttt{MOSFiT} has a gap of around 1.5 days with the power-law fitting result. This might be because \texttt{MOSFiT} is constrained by the overall light-curve morphology and diffusion physics, whereas the power-law fit only provides the time of "first light".

\begin{figure}[htbp]
\centering
\includegraphics[width=0.8\columnwidth]{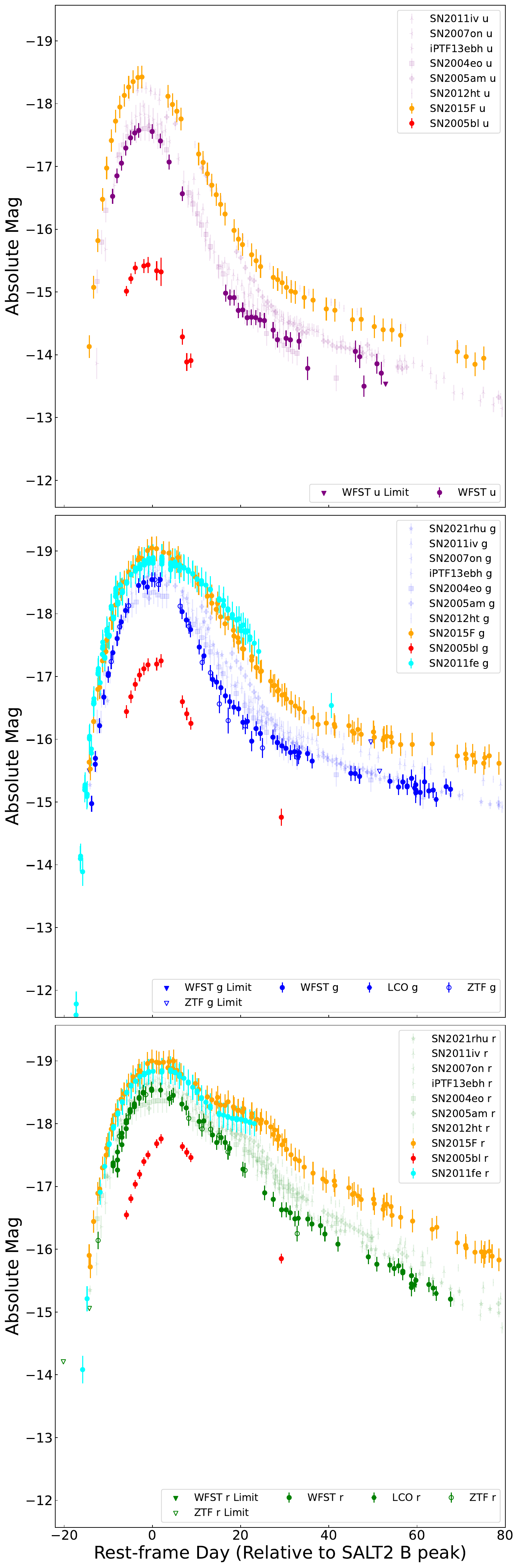}
\caption{Light curve comparison for SN 2024aedt and other SNe Ia. Data are coded by color and opacity as follows: SN 2024aedt (solid purple, blue and green points); a transitional SN Ia sample (semi-transparent points); normal SNe Ia 2011fe (cyan) and 2015F (orange); and a 91bg-like SN 2005bl (red).}
\label{fig:comp_lc}
\end{figure}

\subsection{Light Curves and Spectra Comparison Sample} \label{subsec:compare}

We compiled a sample of transitional SNe Ia from the literature for comparison, including: SN 2021rhu \citep{10.1093/mnras/stad1226}, SN 1986G \citep{Phillips1987}, SN 1998bp \citep{2002PhDT........10J}, SN 2003gs \citep{Krisciunas2009}, SN 2007on \citep{Gall2018}, SN 2011iv \citep{Gall2018}, iPTF13ebh \citep{Hsiao2015}, SN 2003hv \citep{Leloudas2009}, SN 2004eo \citep{Contreras2010, Pastorello2007}, SN 2005am \citep{Contreras2010}, SN 2009an \citep{Sahu2013}, SN 2012ht \citep{Burns2018}, and SN 2015bp \citep{Wyatt2021}.

For broader comparison, examples of normal and 91bg-like SNe Ia are also included. The photometric data for these objects were obtained from the Open Supernova Catalog \citep{2017ApJ...835...64G}. This part of the sample includes the normal SNe 2015F \citep{Burns2018, Graham2017} and 2011fe \citep{Graham2017, 2015MNRAS.446.3895F}, and the 91bg-like SN 2005bl \citep{Contreras2010}.

The data for the comparison sample were processed as follows. Unless otherwise specified, the distance modulus for each supernova was adopted from its primary reference cited above. For the following objects, however, the distance moduli were sourced from different papers: SN 1998bp \citep{2013MNRAS.433.2240G}, SN 2004eo and SN 2005am \citep{2017ApJ...846...58H}, SN 2015F \citep{2017MNRAS.464.4476C}, SN 2005bl \citep{2008ApJ...689..377W}, and SN 2011fe \citep{2016A&A...588A..84D}. For SN 1986G, which suffers from significant host galaxy extinction, we adopted the host extinction values ($A_B$ and $E(B-V)$) directly from \cite{Phillips1987}. For the remaining objects, photometry was corrected for Milky Way extinction using the procedure described previously, with the exception of SN 2004eo, SN 2005am, and SN 2005bl, for which pre-corrected data were used.

\begin{figure}[htbp]
\centering
\includegraphics[width=0.9\columnwidth]{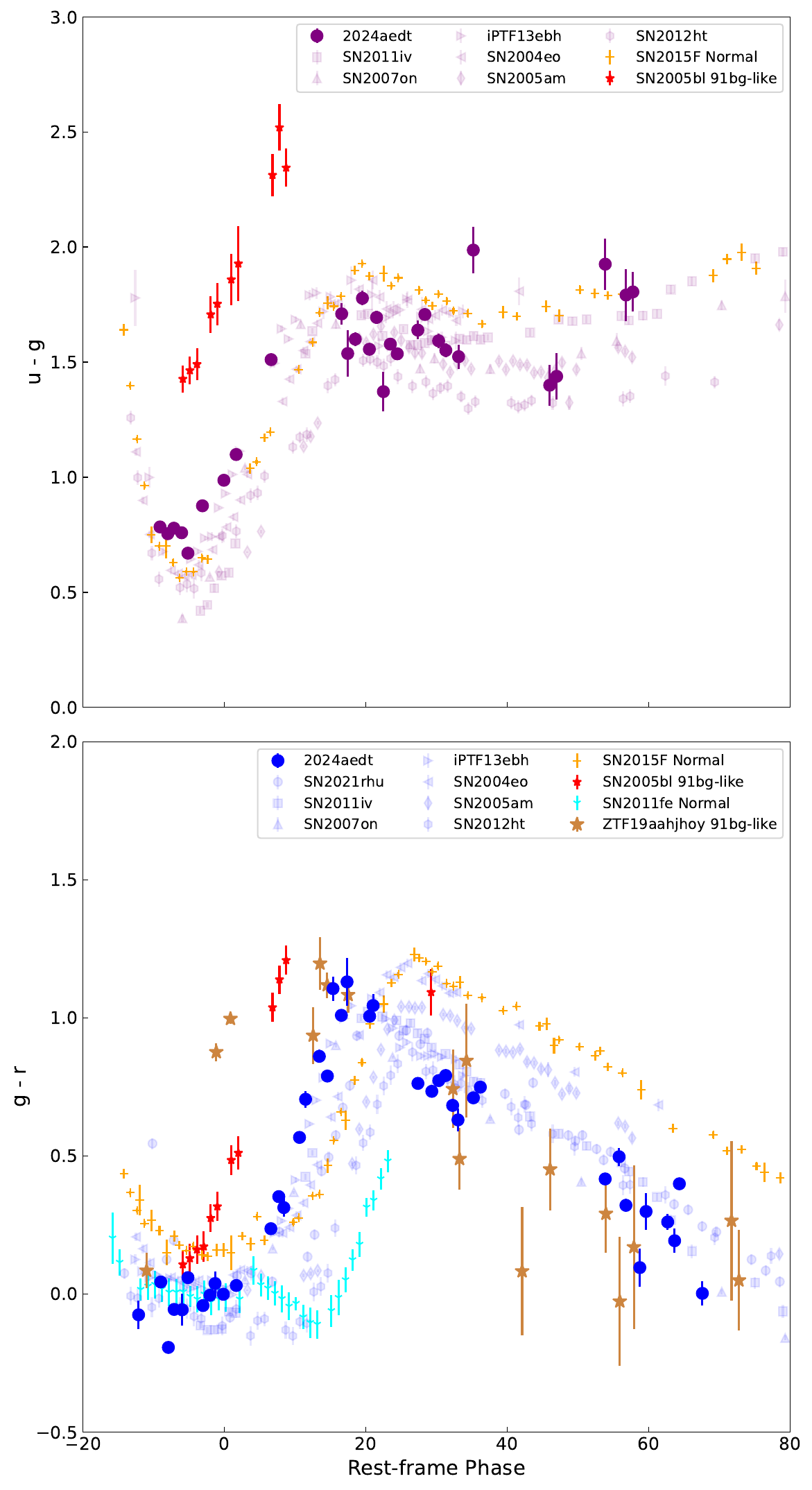}
\caption{Color evolution comparison for SN 2024aedt and other SNe Ia. The data are coded by color and opacity as follows: SN 2024aedt (solid purple and blue points); the transitional SN Ia sample (semi-transparent points); the normal SN 2011fe (cyan); the slightly subluminous normal SN 2015F (orange); and the 91bg-like SN 2005bl (red), ZTF19aahjhoy (peru)}.
\label{fig:comp_cl}
\end{figure}

\subsubsection{Multi-band photometric comprision} \label{subsubsec:comp_photo}

The absolute magnitude light curves of the full comparison sample are presented in Figure \ref{fig:comp_lc}. For clarity, data obtained from various facilities are grouped by their respective filters ($u, g,$ and $r$). The figure shows a clear separation based on supernova subtype: normal SNe Ia populate the luminous, slow-evolving region at the top of the plot, while the 91bg-like SN 2005bl occupies the subluminous, fast-evolving region. The transitional SNe Ia are distributed between these two extremes. The light curve evolution of SN 2024aedt is consistent with this transitional group.

A feature of SN 2024aedt is the weak secondary maximum in the redder bands. In the rest frame, the $r$-band light curve between +15 and +20 days shows a shoulder that is much less prominent than those seen in normal SNe Ia. This is thought to be a consequence of a smaller synthesized mass of IGEs compared to normal events, as well as the IGE distribution and metallicity \citep{Kasen2006, Deckers2025}.

The comparison also highlights the diversity within the transitional class itself, which is most apparent in the bluer bands. The $u$-band, in particular, reveals the largest separation between the 91bg-like SN 2005bl and the transitional events. The transitional SNe themselves exhibit a peak magnitude spread of up to 1 mag in this band, while the dispersion is smaller in the $g$ and $r$ bands. However, we note that since no host galaxy extinction correction was applied (with the exception of the heavily reddened SN 1986G), this observed dispersion—particularly in the bluer filters—may be partially influenced by extrinsic dust reddening rather than intrinsic diversity alone.

Finally, the absolute peak magnitudes for some objects in the comparison sample should be treated with caution, as they may be uncertain. This is due to the significant uncertainties in the distance moduli for some nearby supernovae; values available in public databases like the NASA/IPAC Extragalactic Database (NED) can have a large scatter, which directly impacts the calculated absolute magnitudes. For SN 2024aedt, the impact of a typical peculiar velocity on its redshift-derived distance modulus is estimated to be at most 0.1–0.2 mag. This level of uncertainty is insufficient to alter its classification as a transitional object.

Figure \ref{fig:comp_cl} shows the $u-g$ and $g-r$ color evolution for the full comparison sample. As with the light curves, a clear separation between the subtypes is evident, with the transitional SNe Ia occupying a region between the normal and the 91bg-like populations. The 91bg-like SN 2005bl exhibits the fastest color evolution, cooling much more rapidly than any other object in the sample. Due to the lack of late-time photometric coverage for SN 2005bl, we supplemented our comparison with the 91bg-like object ZTF19aahjhoy (SN 2019arb) from the ZTF DR2 archive \citep{Rigault2025}. Using the $g$-band peak as the reference epoch, the $g-r$ color evolution of SN 2019arb shows the rapid cooling seen in SN 2005bl during the early phases, reaching a red color peak before evolving blueward at a rate consistent with the overall sample.

The normal SNe Ia also show diverse behavior. SN 2011fe, a prototypical normal event, displays a particularly complex $g-r$ evolution: it first rapidly becomes bluer until a phase of approximately $-12$ days, after which its color changes much more slowly for the next 15 days. This is followed by a brief period of reddening, a subsequent 10-day blueing trend, and finally a rapid evolution towards redder colors. Interestingly, the normal SN 2015F, reported as a slightly subluminous SN Ia \citep{2017MNRAS.464.4476C}, follows the general trend of the transitional events, though it remains in a parameter space closer to SN 2011fe.

Within this context, SN 2024aedt is situated on the faster-evolving side of the transitional population. This rapid color evolution corresponds to its relatively subluminous nature as observed in the light curve comparison.

\subsubsection{Spectra for Comparision}  \label{subsec:compspec}

\begin{figure*}[htbp]
    \centering
    \begin{subfigure}{0.49\textwidth}
        \centering
        \includegraphics[width=\textwidth]{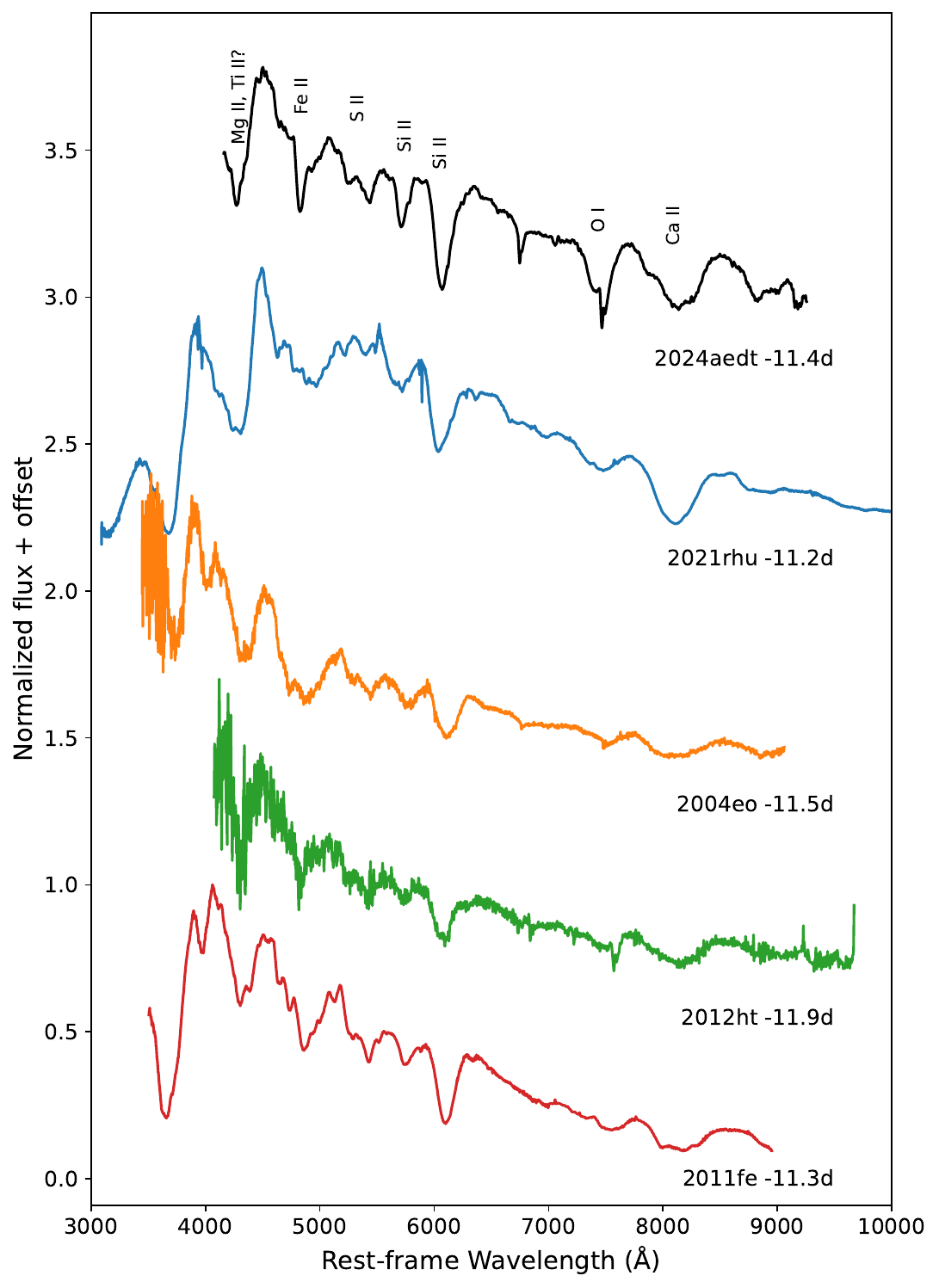}
        \label{fig:spec_0}
    \end{subfigure}
    \hfill
    \begin{subfigure}{0.49\textwidth}
        \centering
        \includegraphics[width=\textwidth]{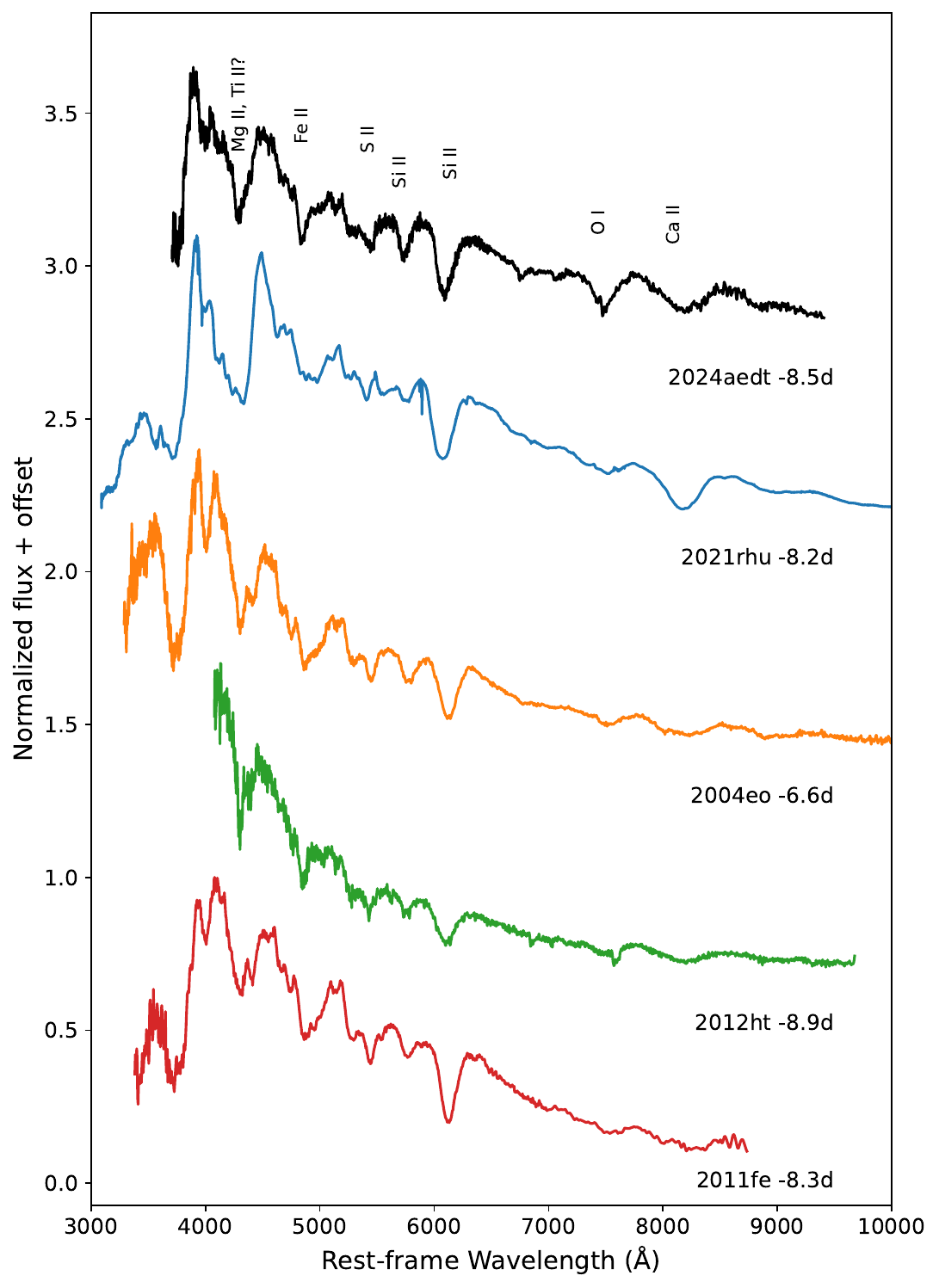}
        \label{fig:spec_1}
    \end{subfigure}
    \caption{Comparison of spectra at similar phases. The spectrum of SN 2024aedt is shown in black, plotted against other SNe Ia from the comparison sample. Each spectrum is normalized and vertically offset for clarity. See Section \ref{subsec:compspec} for details.}
    \label{fig:comp_spec_1}
\end{figure*}

\begin{figure*}[htbp]
    \centering
    \begin{subfigure}{0.49\textwidth}
        \centering
        \includegraphics[width=\textwidth]{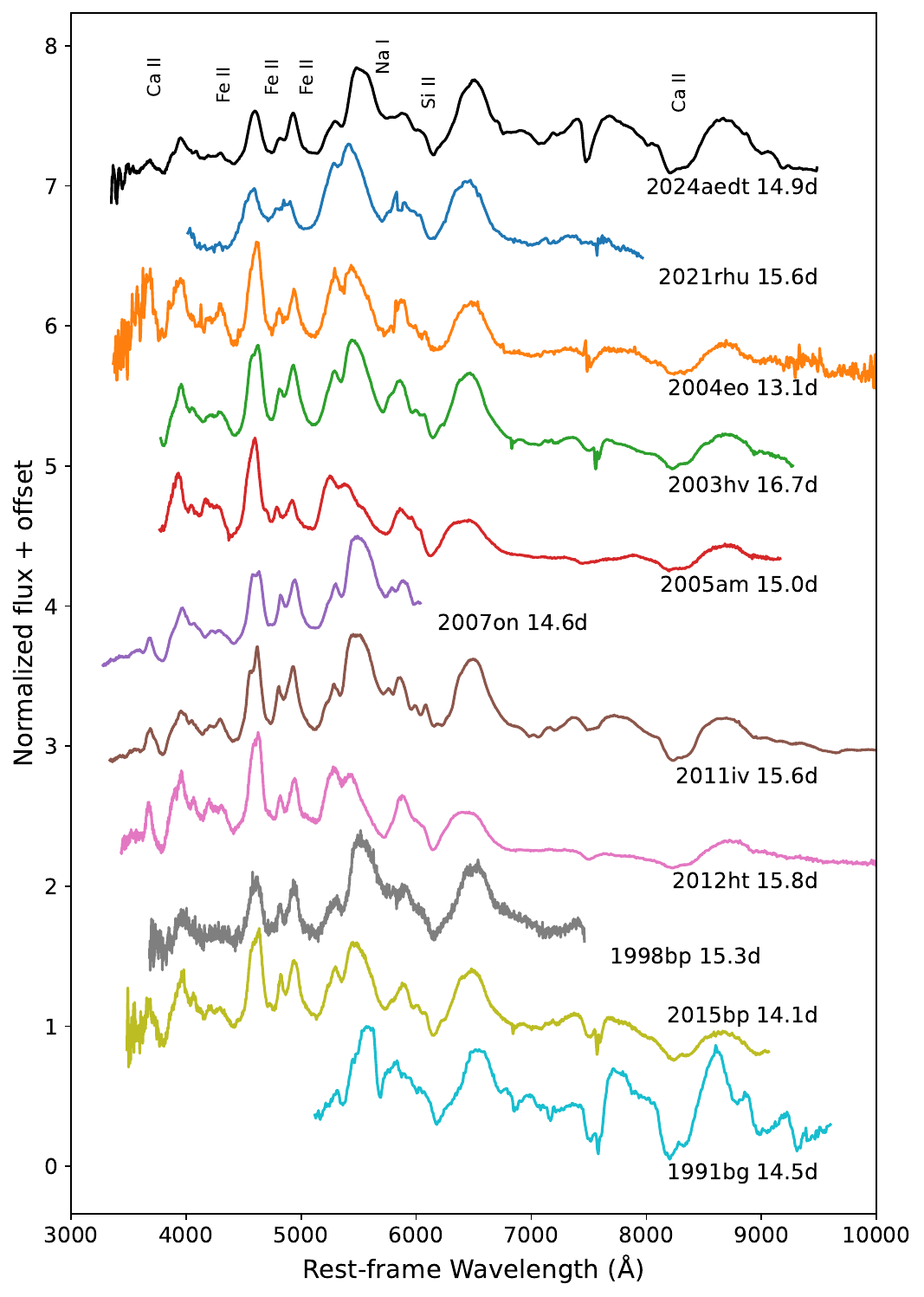}
        \label{fig:spec_2}
    \end{subfigure}
    \hfill
    \begin{subfigure}{0.49\textwidth}
        \centering
        \includegraphics[width=\textwidth]{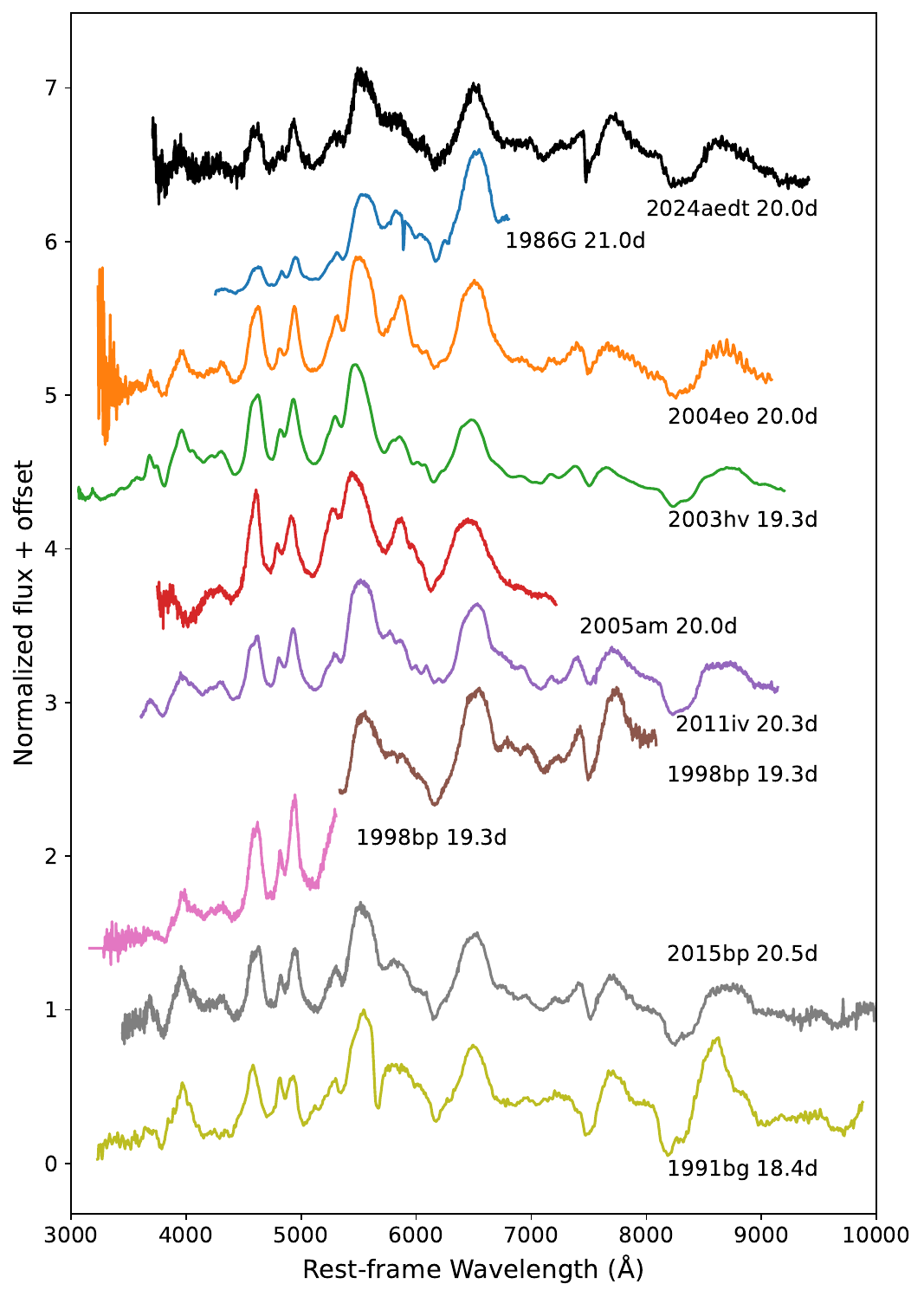}
        \label{fig:spec_3}
    \end{subfigure}
    \caption{Spectral comparison of SN 2024aedt (Continued). The plotting conventions are the same as in the previous figure.}
    \label{fig:comp_spec_2}
\end{figure*}

\begin{figure}[htbp]
    \centering
    \begin{subfigure}{0.45\textwidth}
        \centering
        \includegraphics[width=\textwidth]{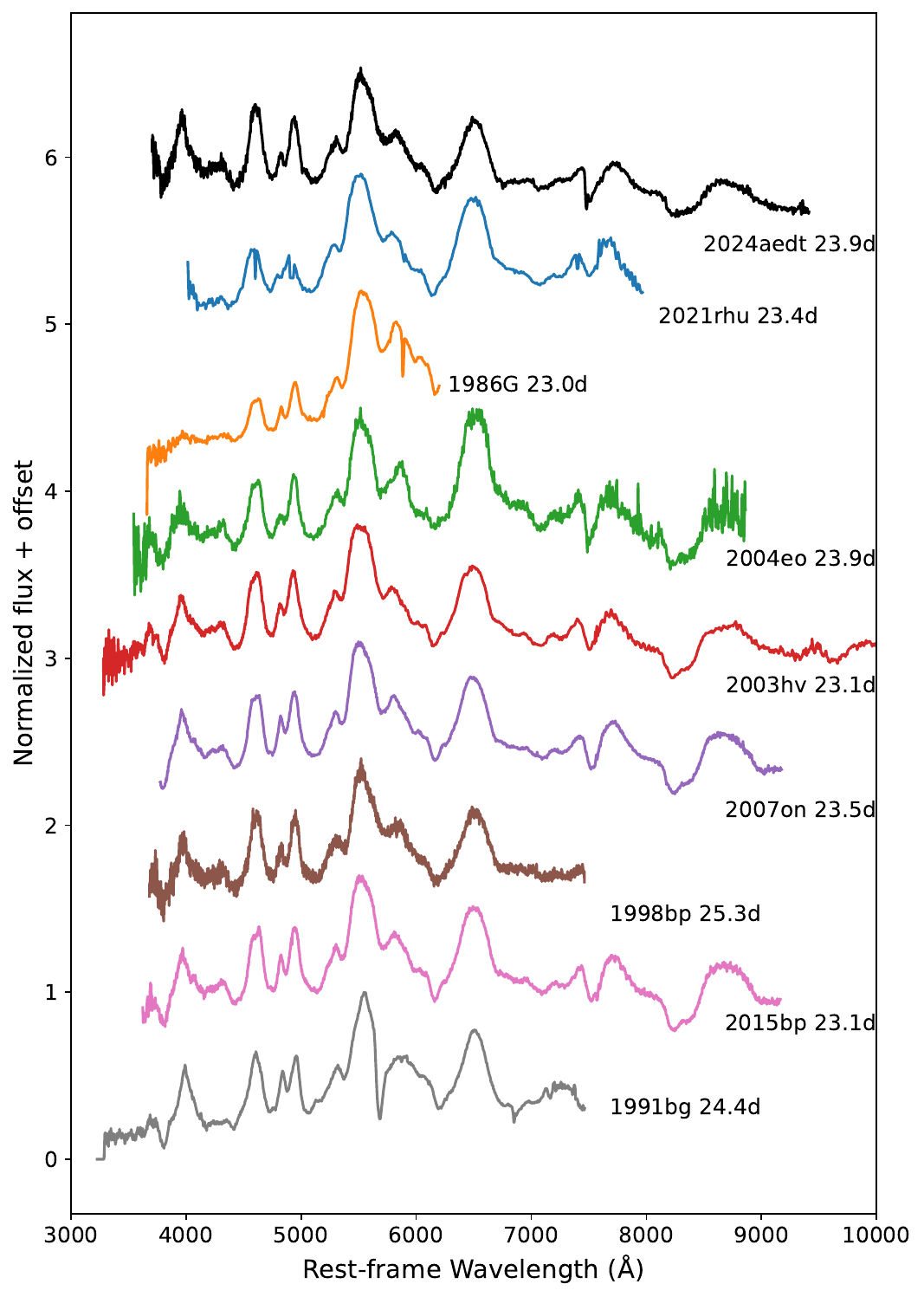}
        \label{fig:spec_4}
    \end{subfigure}
    \hfill

    \caption{Spectral comparison of SN 2024aedt (Continued). The plotting conventions are the same as in the previous figure.}
    \label{fig:comp_spec_3}
    \hfill
\end{figure}

To perform a spectral comparison, we collected all available spectra for our comparison sample from the Weizmann Interactive Supernova Data Repository\footnote{https://www.wiserep.org} \citep[WISeREP;][]{2012PASP..124..668Y}. The redshift correction status of each spectrum was verified by examining the wavelengths of telluric absorption features. After verification, the spectra were corrected for redshift and Milky Way extinction as needed, using the procedure described previously. For each SN 2024aedt observation, the comparison spectrum with the closest phase within a $\pm$2-day window was selected from each source in the sample. The resulting spectral comparisons are presented in Figures \ref{fig:comp_spec_1} to \ref{fig:comp_spec_3}.

In the pre-maximum epochs, the ejecta velocity of SN 2024aedt, measured from the absorption minimum of the Si II $\lambda$6355 line, is similar to that of the transitional SN 2021rhu and SN 2012ht, and slightly higher than that of the transitional SN 2004eo and the normal SN 2011fe. Although SN 2021rhu exhibits a similar photometric evolution to SN 2024aedt (Figure \ref{fig:comp_lc}), there is significant diversity in the features between 4500\,\AA\ and 5000\,\AA. This combination of overall similarity and diversity in the blue-end features persists into the post-maximum spectra. However, as the supernovae evolve, their spectral features become more homogeneous, a trend that is particularly evident in the latest-phase comparison.

This evolutionary trend suggests that early-time spectra are more sensitive probes of the explosion physics than late-time spectra. The initial diversity may be due to differences in the chemical composition or physical state of the outer ejecta, which are in turn influenced by progenitor properties and explosion mechanisms \citep[e.g.,][]{Jiang2017}.


We measured the pseudo-equivalent widths (pEWs) and velocities of the silicon lines (Si~{\sc ii} $\lambda$5972 and $\lambda$6355) in the pre-maximum spectra using a direct interactive procedure. Each spectrum was first corrected for Milky Way extinction and redshift and then smoothed using a Savitzky-Golay filter \citep{Savitzky1964}. A pseudo-continuum was subsequently defined by interpolating between interactively selected local maxima, and the absorption features were then fitted with Gaussian profiles to measure their properties on the raw spectrum. The fitting region and initial guesses for the Gaussian parameters were also selected interactively. To quantify the uncertainties inherent in this interactive process, a Monte Carlo simulation with 10,000 trials is performed. In each trial, the position of every manually selected point (e.g., continuum anchors, fitting boundaries) was randomly perturbed, assuming a characteristic uncertainty of 20\,\AA\ for each selection. The final measurements are taken as the median of the resulting distributions, with the $16^{th}$ and $84^{th}$ percentiles adopted as the lower and upper $1\sigma$ uncertainties, respectively. Then we manually check the results and remove the failed ones.

\begin{figure}[htbp]
\centering
\includegraphics[width=0.9\columnwidth]{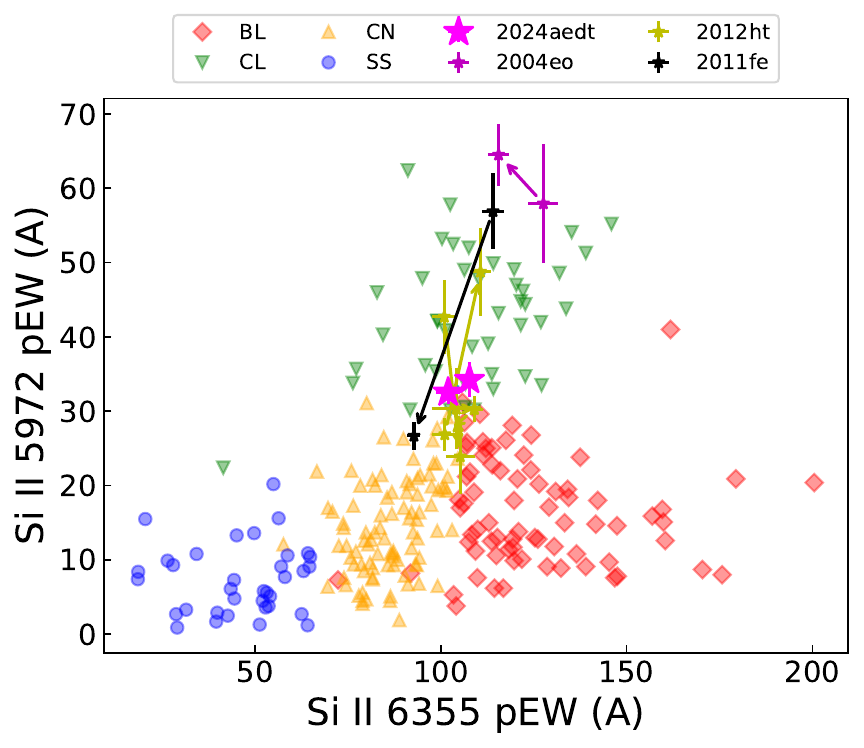}
\caption{Early-time evolutionary tracks on the Branch diagram. The track for SN 2024aedt is shown in magenta, plotted alongside the normal SN 2011fe (black) and two other transitional SNe Ia. See Section \ref{subsec:compspec} for a detailed description.}
\label{fig:branch}
\end{figure}

In Figure \ref{fig:branch}, the evolutionary track of SN 2024aedt is plotted on the Branch diagram, which shows pEW(Si II $\lambda$5972) versus pEW(Si II $\lambda$6355) \citep{2006PASP..118..560B}. The background comparison data are from \cite{Blondin2012}. It is important to note that the Branch diagram is conventionally plotted using measurements taken at the time of maximum brightness. Since our observations do not cover this specific epoch, the end point of our measured track does not represent the supernova's final classification position on this diagram. Nevertheless, the pre-maximum track shows that the transitional SNe—SN 2024aedt, SN 2004eo, and SN 2012ht—are distributed across the ``cool" (CL) region, in good agreement with the results of \cite{Ogawa2023, Burgaz2025}. In contrast, the normal SN 2011fe rapidly evolves from the edge of the CL region towards the "core-normal" (CN) region, where it eventually resides at maximum light \citep{Pereira2013}.

\begin{figure*}[htbp]
    \centering
    \begin{subfigure}{0.49\textwidth}
        \centering
        \includegraphics[width=\textwidth]{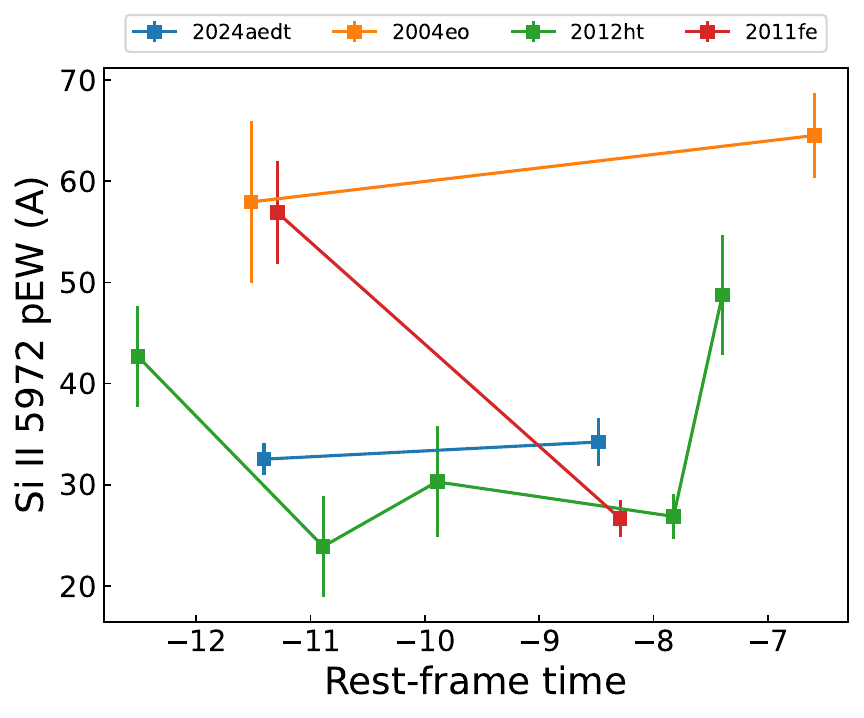}
        \label{fig:pew_5972}
    \end{subfigure}
    \hfill
    \begin{subfigure}{0.49\textwidth}
        \centering
        \includegraphics[width=\textwidth]{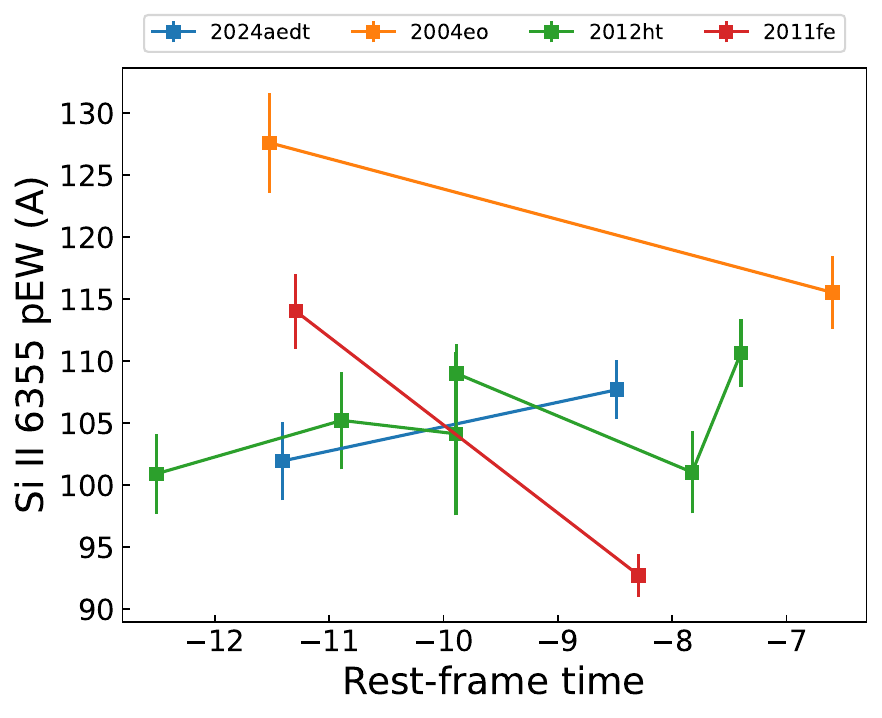}
        \label{fig:pew_6355}
    \end{subfigure}
    \\[0.5cm] 
    \begin{subfigure}{0.49\textwidth}
        \centering
        \includegraphics[width=\textwidth]{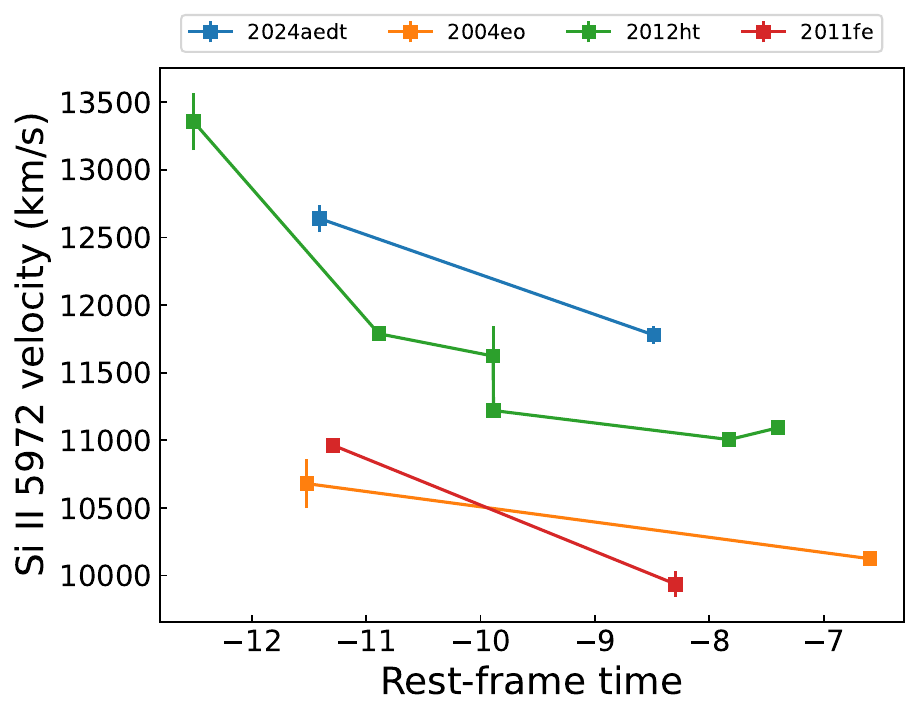}
        \label{fig:v_5972}
    \end{subfigure}
    \hfill
    \begin{subfigure}{0.49\textwidth}
        \centering
        \includegraphics[width=\textwidth]{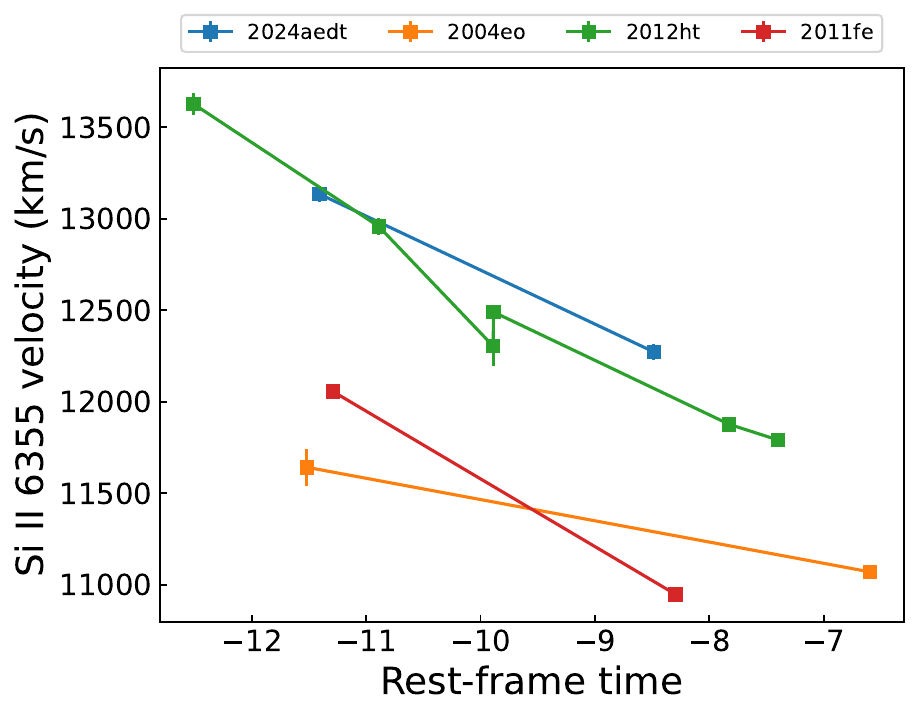}
        \label{fig:v_6355}
    \end{subfigure}
    \caption{Evolution of the properties of the Si II doublet for SN 2024aedt and the comparison sample. The top panels show the pseudo-equivalent width (pEW) evolution for the Si II $\lambda$5972 (left) and Si II $\lambda$6355 (right) lines. The bottom panels show the corresponding velocity evolution measured from each line.}
    \label{fig:lines}
\end{figure*}

The evolution of the pEWs and line velocities are presented in Figure \ref{fig:lines}. The SNe in our sample exhibit diverse pEW evolution. For the normal SN 2011fe, the pEW of Si II $\lambda$6355 declines significantly over approximately four days, consistent with the findings of \cite{Zhao2020}, while the other SNe show no clear trend for this feature. In contrast, for the Si II $\lambda$5972 line, both SN 2024aedt and SN 2004eo show a slight increase in pEW; for the Si II $\lambda$6355 line, however, this increasing trend is only apparent for SN 2024aedt. Compared with the measurements of the CN and CL subtypes from the first phase of the Carnegie Supernova Project (CSP-I) \cite{Folatelli2013}, our results are consistent with the distribution and evolutionary trends reported in their work for the Si lines of both normal and transitional events. Furthermore, it should be noted that the pEW values presented here may have significant uncertainties, which can arise from the subjective definition of the pseudo-continuum and potential light contamination from the host galaxy, especially given our limited number of spectra.

The line velocities, measured from the Si II $\lambda$6355 feature, all show a clear decreasing trend over time. This behavior is in good agreement with the general trend for SNe Ia presented in \cite{Silverman2012} and \cite{Folatelli2013}. The velocity values for SN 2024aedt lie well within the range spanned by the comparison sample.

\subsection{Host Galaxy} \label{subsec:compare}

We obtained multi-band photometry for the host galaxy, UGC 1325, using the \texttt{hostphot} package \citep{Müller-Bravo2022}. The photometric measurements were compiled from several surveys, including SDSS \citep[$ugriz$;][]{Abdurrouf2022}, WISE \citep[W1--W3;][]{2010AJ....140.1868W}, 2MASS \citep[$J, H, K_s$;][]{2006AJ....131.1163S}, and GALEX \citep[NUV;][]{Martin2005}. The complete photometric data set is presented in Table \ref{tab:hostphot}.

\begin{figure}[ht]
\centering
\includegraphics[width=0.9\columnwidth]{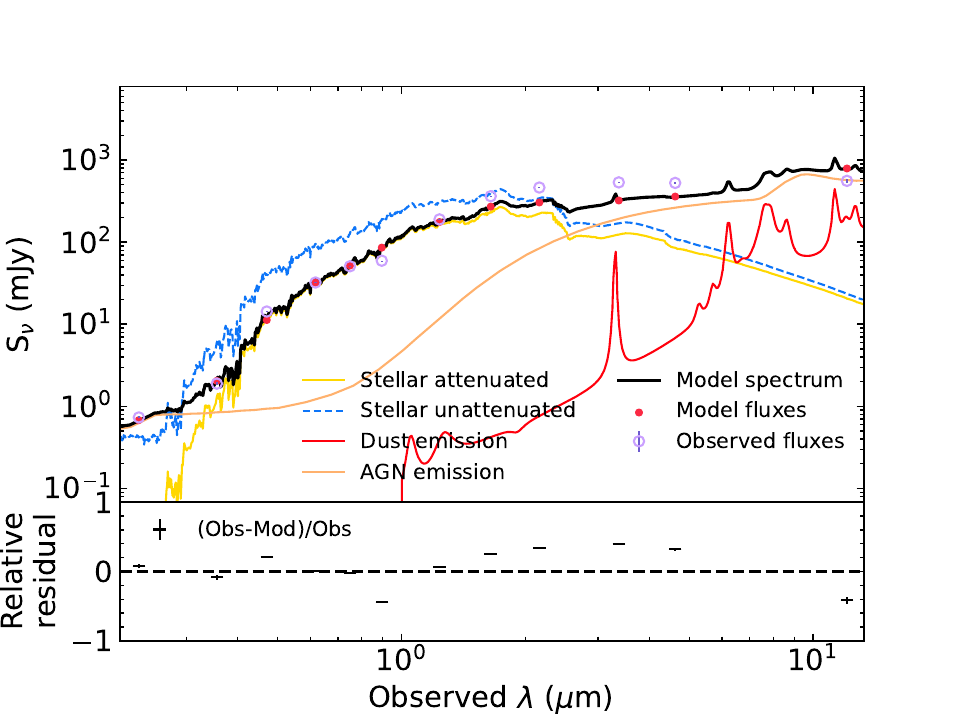}
\caption{The best-fit Spectral Energy Distribution (SED) model for the host galaxy UGC 1325, generated with \texttt{CIGALE}.}
\label{fig:sed}
\end{figure}

The spectral energy distribution (SED) of the host galaxy was then modeled using \texttt{CIGALE} \citep{2019A&A...622A.103B}. The model was constructed using the simple stellar population (SSP) models of \cite{Bruzual2003} with a Chabrier initial mass function \citep[IMF;][]{Chabrier2003}, a modified dust attenuation law from \cite{Charlot2000}, and dust emission templates from \cite{2014ApJ...780..172D}. An initial fit revealed a flux excess around 2\,$\mu$m, which was significantly improved by the inclusion of an AGN component based on the models of \cite{2012MNRAS.420.2756S} and \cite{Stalevski2016}. The final best-fit SED is shown in Figure \ref{fig:sed}. The fitting result indicates that UGC 1325 is a massive, quiescent elliptical galaxy with a stellar mass of $M_{\star} \approx 3.914 \times 10^{11}\,M_{\odot}$. The SFH shows that the galaxy is ancient, having begun forming stars approximately 13\,Gyr ago over a timescale of $\tau_{\mathrm{main}} \approx 500$\,Myr. This resulted in a stellar population with a mass-weighted age of 12\,Gyr and a high metallicity ($Z=0.050$). The galaxy is now fully quenched, with a star formation rate (SFR) averaged over the last 100 Myr of only $0.024 \pm 0.054\,M_{\odot}\,\mathrm{yr}^{-1}$. The galaxy's IR emission is dominated by an obscured Active Galactic Nucleus (AGN), which contributes 70\% of its total IR luminosity.

\begin{table}[h!]
\centering
\begin{tabular}{cc}
\hline
Filter & AB Magnitude \\
\hline
GALEX NUV & $16.737 \pm 0.031$ \\
SDSS $u$ & $15.707 \pm 0.040$ \\
SDSS $g$ & $13.510 \pm 0.010$ \\
SDSS $r$ & $12.627 \pm 0.006$ \\
SDSS $i$ & $12.132 \pm 0.004$ \\
SDSS $z$ & $11.963 \pm 0.005$ \\
2MASS $J$ & $10.704 \pm 0.002$ \\
2MASS $H$ & $9.998 \pm 0.003$ \\
2MASS $Ks$ & $9.734 \pm 0.003$ \\
WISE W1 & $9.577 \pm 0.013$ \\
WISE W2 & $9.593 \pm 0.018$ \\
WISE W3 & $9.530 \pm 0.056$ \\
\hline
\end{tabular}
\caption{Photometry of the host galaxy UGC 1325 in the adopted filter set. The measurements were obtained with the \texttt{hostphot} package.}
\label{tab:hostphot}
\end{table}

\section{Discussion} \label{sec:discussion}

\subsection{SN 2024aedt as a Link Between the Normal and Subluminous Classes} \label{subsec:preferredsubtype}

The \texttt{SALT2} fit yields a peak absolute magnitude of $M_B = -18.49 \pm 0.03$ mag and a decline rate of $\Delta m_{15}(B) = 1.53 \pm 0.36$ mag, placing SN 2024aedt in the transitional region between normal and 91bg-like SNe Ia. This is also evident from the distribution of $s_{bv}$ in the lower panel of Figure \ref{fig:dm15}. The light curves are consistent with other transitional events, lying on the subluminous, fast-evolving end of the distribution. A subtle secondary maximum is observed in $r$ band, the strength of which appears related to the peak magnitude, consistent with the findings of \cite{Deckers2025}. This consistency is also reflected in the color evolution, which shows high similarity to other transitional events. The \texttt{MOSFiT} analysis yields a synthesized $^{56}\mathrm{Ni}$ mass of $0.414\,M_{\odot}$, a value intermediate between those of normal and 91bg-like SNe Ia.

\begin{figure}[ht]
\centering
\includegraphics[width=0.9\columnwidth]{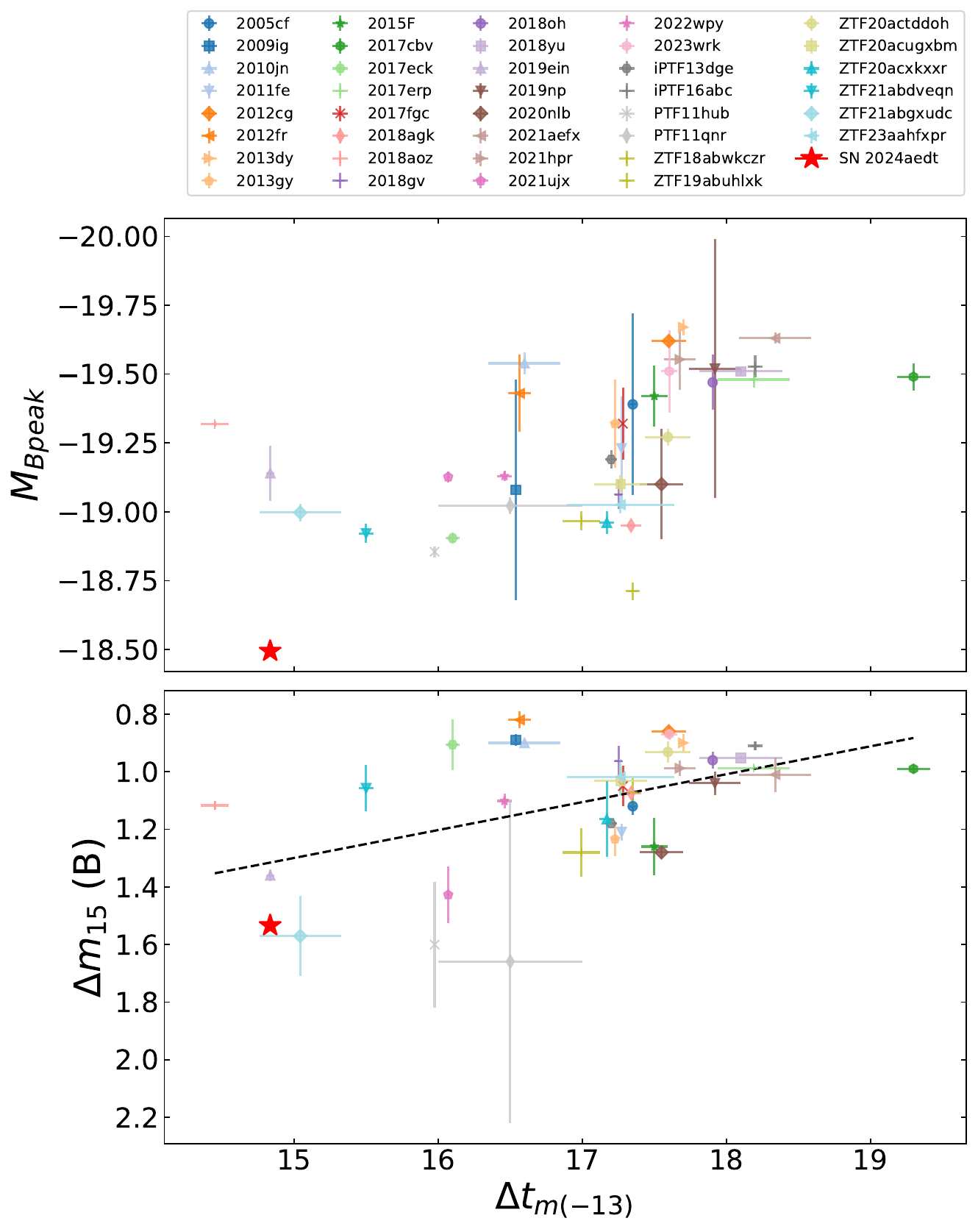}
\caption{Top panel: $M_{B, \text{max}}$ versus $\Delta t_{m(-13)}$. Bottom panel: $\Delta m_{15}(B)$ versus $\Delta t_{m(-13)}$. The position of SN 2024aedt is shown for comparison with the sample from \citep{https://doi.org/10.48550/arxiv.2507.15609}.}
\label{fig:wy}
\end{figure}

Following the method of \cite{https://doi.org/10.48550/arxiv.2507.15609}, we calculated the rest-frame rise time from $-13 \ \text{mag}$ to the peak in the $g$-band to be $\Delta t_{m(-13)} = 14.834$ days. This result places SN 2024aedt in the bottom-left corner of both subplots in their Figure 2 (see Figure \ref{fig:wy}). In the $M_{B, \text{max}}$ versus $\Delta t_{m(-13)}$ plane, SN 2024aedt appears separate from the normal SN Ia population, due to the selection criteria (i.e., the limitation to spectroscopically normal SNe Ia at $z<0.03$ with early-phase detections, which removes most fast-evolving transitional events). In the $\Delta m_{15}(B)$ versus $\Delta t_{m(-13)}$ plane, SN 2024aedt seems follow the linear trend of the normal group and is close to a normal SN Ia SN 2019ein \citep{2020ApJ...893..143K, Xi2022}. This similarity is further supported by its early-time color evolution. The early colors of normal SNe Ia like SN 2011fe and SN 2015F show a trend similar to transitional events until the secondary maximum in SN 2011fe's light curve becomes dominant, whereas the 91bg-like SN 2005bl shows a much more rapid reddening. However, a closer look at the color diagram reveals that an evolutionary trend is still present. If the colors of some sources with faster-evolving light curves were included, this trend would likely be more pronounced, which also reflects the diversity within the transitional group.  Furthermore, host properties of our compiled transitional sample (Table \ref{tab:hostmor}) suggest no significant preference for elliptical hosts, which are typical for 91bg-like SNe Ia \citep{Senzel2025, GonzlezGaitn2011}. This underscores the diversity and highlights the intermediate nature of the class, which shares properties with both the normal and 91bg-like populations. Therefore, the analysis of its photometric properties and derived physical parameters robustly highlights the transitional characteristics of SN 2024aedt. 

\begin{table}[h!]
\centering
\begin{tabular}{ccc}
\hline
SN & Host Type & Source \\
\hline
SN 2011iv & E1 & \cite{Weidner2013} \\
SN 2007on & E1 & \citep{Weidner2013} \\
SN 2012ht & SBm & \citep{2015ApJS..217...27A} \\
SN 2021rhu & S & \citep{2019MNRAS.488..590B} \\
SN 2003hv & SA0$^0$(r)? & \citep{1991rc3..book.....D} \\
SN 2005am & SB(rs)a & \citep{1991rc3..book.....D} \\
SN 1998bp & E & \citep{1991rc3..book.....D} \\
SN 2015bp & SAB0$^0$?(rs) & \citep{1991rc3..book.....D} \\
SN 1986G & 	S0pec & \citep{2013MNRAS.435.2274W} \\
SN 2004eo & SB(s)ab & \citep{1991rc3..book.....D} \\
SN 2003gs & SB0$^+$(rs) & \citep{1991rc3..book.....D} \\
iPTF13ebh & SAB0$^-$(r)? & \citep{1991rc3..book.....D} \\
SN 2009an & SB(s)a & \citep{1991rc3..book.....D} \\

\hline
\end{tabular}
\caption{Host galaxy morphological type of collected transitional sample. Data are find from \texttt{SIMBAD} and \texttt{NED}. }
\label{tab:hostmor}
\end{table}

\subsection{Transitional SN 2024aedt: DDT or DDet Origin?} \label{subsec:preferredmodel}

\begin{figure}[ht]
\centering
\includegraphics[width=0.9\columnwidth]{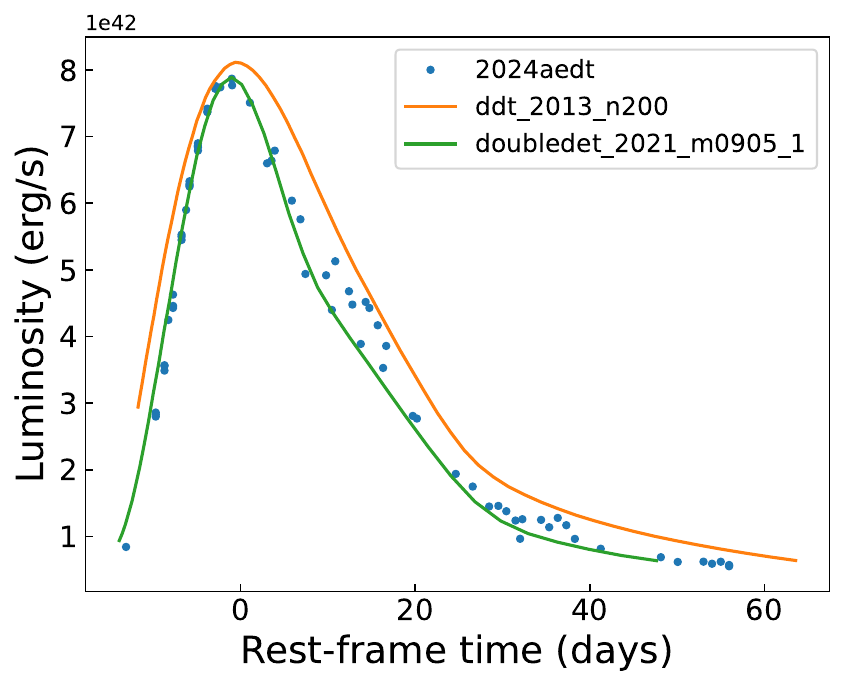}
\caption{Best-fitting model light curves from the HESMA for SN 2024aedt. The observed bolometric light curve (blue dots) is shown alongside the best-fitting DDT (orange) and DDet (green) models.}
\label{fig:model_lc}
\end{figure}

We constructed the pseudo-bolometric light curve using the \texttt{superbol} code \citep{Nicholl_2018}. The fit incorporated our well-sampled $u, g,$ and $r$-band data, as these bands provide the best coverage, corresponding to a wavelength range of approximately 3000--7000\,\AA. \footnote{Due to strong absorption features and the limited passbands used, bolometric parameters derived via blackbody approximation should be treated as qualitative guides.} We compared the result with theoretical models from the Heidelberg Supernova Model Archive \citep[HESMA;][]{https://doi.org/10.48550/arxiv.1706.09879}, considering two main scenarios: DDT and DDet. We use the angle-averaged bolometric light curve data directly as provided in the HESMA. We plotted and inspected the full suite of available models, most of which showed significant discrepancies with SN 2024aedt, only two models show a good match to our data (Figure \ref{fig:model_lc}). The first is a near-$M_{\mathrm{ch}}$ DDT model \citep[\texttt{ddt\_2013\_n200};][]{2013MNRAS.429.1156S, 2013MNRAS.436..333S} from a non-rotating WD with 200 ignition kernels, which produces $0.41\,M_{\odot}$ of $^{56}\mathrm{Ni}$ and a kinetic energy of $1.34 \times 10^{51}$ erg. The second is a sub-$M_{\mathrm{ch}}$ DDet model \citep[\texttt{doubledet\_2021\_m0905\_1};][]{2021A&A...649A.155G} with a $0.899\,M_{\odot}$ core and a $0.053\,M_{\odot}$ helium shell, synthesizing $0.38\,M_{\odot}$ of $^{56}\mathrm{Ni}$. Of the two, the DDet model provides an excellent match to the data, while the DDT model light curve is broader than that of the observation.

\begin{figure}[ht]
\centering
\includegraphics[width=0.9\columnwidth]{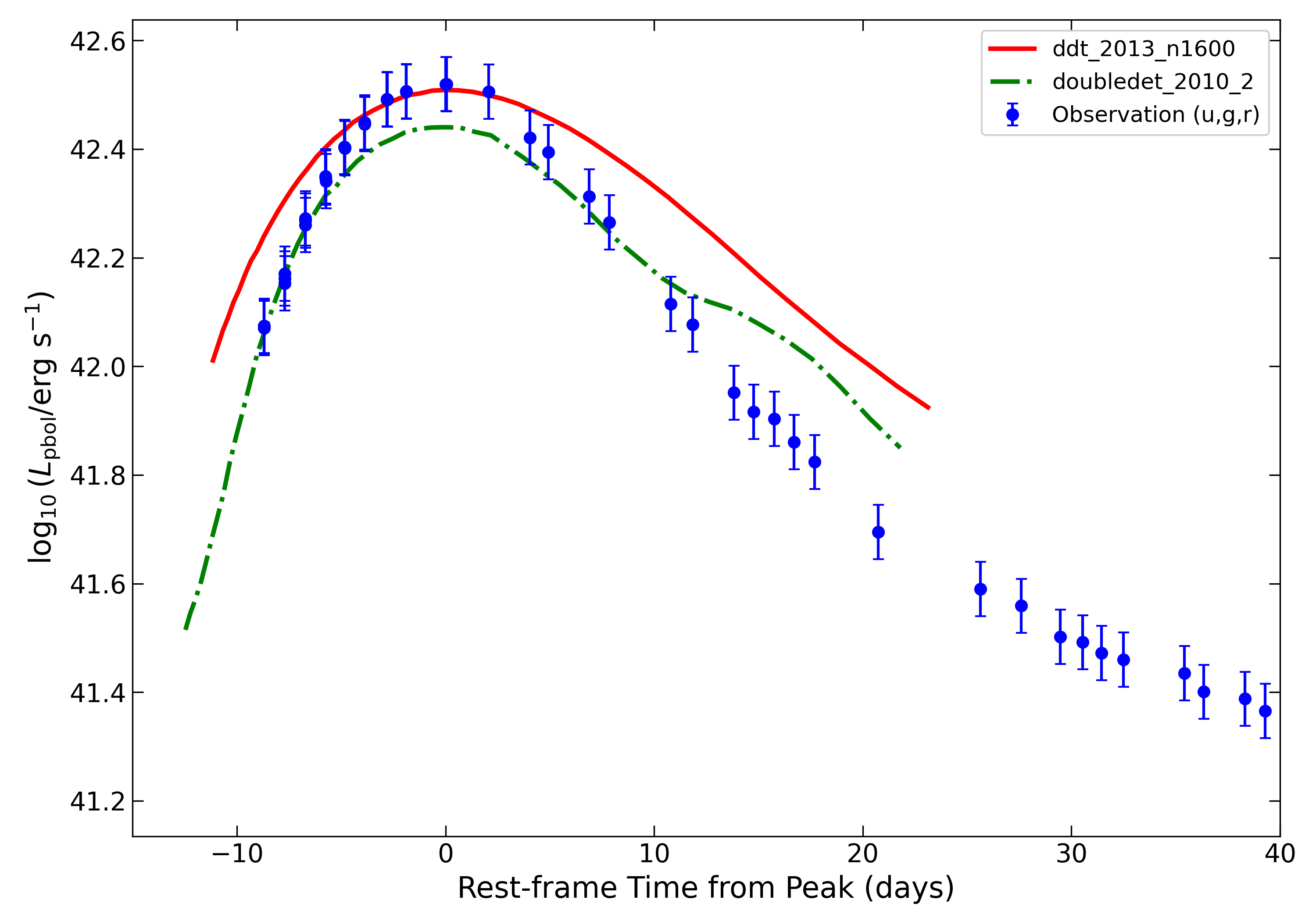}
\caption{Comparison of the pseudo-bolometric light curve of SN 2024aedt (blue dots) with selected explosion models. The red solid line represents the best-fit delayed-detonation model (DDT), and the green dot-dash line represents the double-detonation model (DDet). The pseudo-bolometric luminosity is integrated over the wavelength range covered by the $u$, $g$, and $r$ bands. For the observations, the light curve was derived by integrating the spectral energy distribution approximated from the $u$, $g$, and $r$ photometry.}
\label{fig:model_plc}
\end{figure}

Given the systematic uncertainties associated with the blackbody corrections used to derive the full bolometric luminosity, we performed an alternative comparison using pseudo-bolometric light curves. These were constructed by integrating the flux over the observed wavelength range (covering the $u, g$, and $r$ bands) for both the observed SEDs and the model spectra. In this pseudo-bolometric analysis, we identify \texttt{ddt\_2013\_n1600} and \texttt{doubledet\_2010\_2} as the best-matching models (Figure \ref{fig:model_plc}. This result differs from the best matches obtained via the full bolometric light curve comparison, underscoring the impact of blackbody correction on model selection. Notably, the \texttt{ddt\_2013\_n1600} model provides the best fit to the earliest epoch spectrum. However, although the integrated pseudo-bolometric luminosity shows good agreement, discrepancies remain in the individual bands. When aligned to the peak time of the $r$-band, the synthetic $u$, $g$, and $r$ light curves of these models do not simultaneously reproduce the detailed evolution observed in SN 2024aedt.

As shown in Figure \ref{fig:exp}, early light curve behavior indicates a smooth power-law rise without an early-excess (EEx) feature (although we acknowledge that an inconspicuous EEx occurring in the gap between the last non-detection and the first detection cannot be excluded). In the DDet models, the detonation of an e.g., $0.055 M_\odot$ He shell can produce clear EEx emissions in the light curves \citep{Noebauer2017,Maeda2018}. The absence of EEx in SN 2024aedt may suggest that the He-shell mass is significantly lower or that the SN was viewing away from the detonation direction where the most inconspicuous EEx is expected, which have been proposed for interpreting normal SNe Ia without EEx features \citep{Shen2021, https://doi.org/10.48550/arxiv.2507.15609}. The result is also supported by the relatively normal peak color and the lack of line blanking caused by IGE in the spectra. Alternatively, tension in early photometric evolution could be mitigated by an IME-dominated shell, which produces a peak color bluer than that of an IGE-dominated shell \citep{Magee2021}. Furthermore, \cite{collins2021multi} found that their synthetic photometry from the \texttt{ARTIS-NLTE} model can well reproduce the observations of SN 2011fe without the reddening seen in classic models. For the DDT model, a smooth rise is easy to reproduce \citep{Noebauer2017}, and it has also been used to explain other transitional events (e.g., \citealp{Ashall2018}).

\begin{figure}[ht]
\centering
\includegraphics[width=0.9\columnwidth]{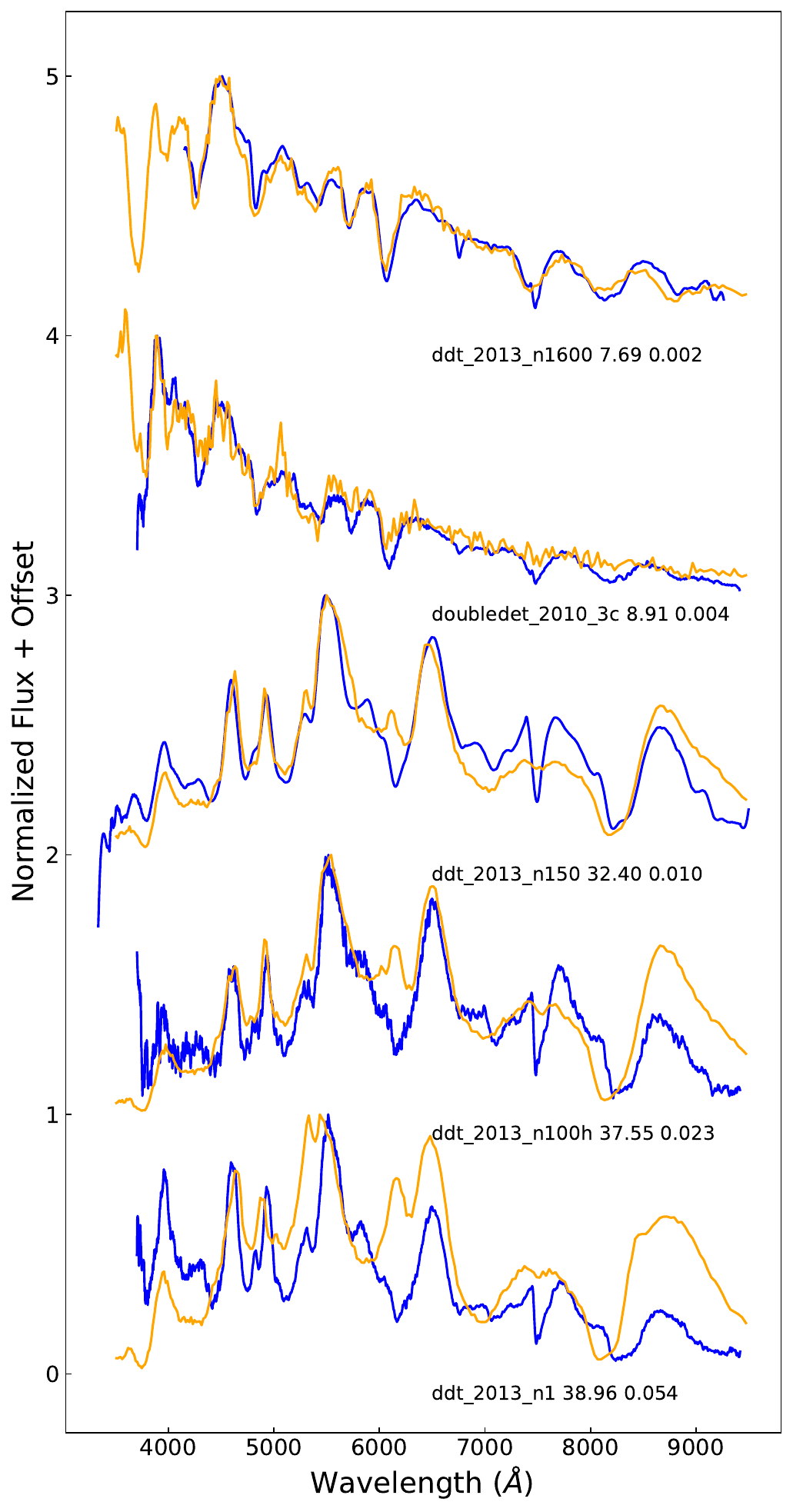}
\caption{Best-match HESMA model spectra for each observed epoch of SN 2024aedt. The observed spectra are shown in blue, and the model spectra are shown in orange. The top two display the two pre-maximum phases, and the bottom three shows the three post-maximum phases. The best-fit model name, phase, and mean squared error (MSE) are indicated below each spectrum.}
\label{fig:model_spec}
\end{figure}

For the spectral comparison, we quantify the similarity between observed spectra and the theoretical models by calculating the mean squared error (MSE) after normalization. The comparison was performed at similar epochs, using the peak of the bolometric light curve as the reference time for the model spectra. For each of our five observed spectra, we consider all model spectra within a $\pm$ 2-day window of the observation epochs and identify the model with the lowest MSE as the best match. The results are presented in Figure \ref{fig:model_spec}.

The comparison shows that the premaximum spectrum of SN 2024aedt is well-matched by the models in terms of line velocities, absorption strengths, and the overall continuum shape. The subsequent two spectra also show a relatively good match, though with larger deviations appearing at the red end compared to the blue end. The quality of the match for the final, latest-phase spectrum is poorer than for the earlier epochs. Statistically, the DDT models provide a better overall fit to our spectral series. In four of the five epochs, the single best-matching model is a DDT model from \cite{2013MNRAS.429.1156S, 2013MNRAS.436..333S}. Furthermore, the majority of the top five best-fitting models in each epoch are also from the DDT scenario. A DDet model from \cite{2010ApJ...719.1067K} provides the best fit in only one epoch. We also considered the 1D DDet model from \cite{Callan2025}, available in HESMA, which features a $0.04\,M_\odot$ He shell and $1.01\,M_\odot$ of ejecta. This model is constructed by radially averaging cells from the 3D model of \cite{Gronow2021} (which had $0.018\,M_\odot$ of He and $0.83\,M_\odot$ of ejecta) within a solid-angle cone centered on the negative z-axis. While this model provides a good match to our second spectrum, it fails to reproduce the earliest one, showing significant blue-end suppression when normalized to redder wavelengths. This discrepancy is likely caused by the large amount of synthesized IGE in the model's outer layers, which is inconsistent with our observations, highlighting the crucial role of early-time spectra and blue-end photometry in constraining explosion scenarios. Consequently, a relatively thick He shell is disfavored for SN 2024aedt. This finding is consistent with the parameter space explored by \cite{Polin2019}, where thin He shells ($< 0.02\,M_\odot$) were shown to reproduce smooth early-time light curves as well as appropriate color and spectroscopic features. However, whether such low-mass shells can successfully initiate a detonation remains under debate \citep{Iwata2024}.

Also, the apparent statistical preference for the DDT scenario must be interpreted with caution. The apparent good fit of the DDT scenario is undermined by two key problems. First, the best-fitting DDT model is not consistent across all epochs, with the required number of ignition kernels varying between epochs. This is also found in \cite{2013MNRAS.436..333S}. They note that the evolution of their synthetic spectra is too quick, an imperfection also demonstrated by their \texttt{SNID} matching results. We further match each of the best-matching DDT models against the entire observed time series, focusing mainly on the pre-maximum epochs. The match for the first epoch always yields relatively good results, but the subsequent epoch shows significant underestimates of the flux at the blue end (the \texttt{ddt\_2013\_n1} model does not have a match with the first epoch, but it shows a good match with the second one). This corresponds to the redder color found in \cite{2013MNRAS.436..333S} due to excessive line blanketing caused by the stable IGEs formed in the deflagration process. The elements float to the surface and dominate the layers with intermediate velocities, suggesting that SN 2024aedt either experiences a lesser extent of deflagration than the model predicts (which creates difficulties for the light curve matching, since models with fewer ignition kernels exhibit higher luminosity), or is caused by other mechanisms without deflagration. Second, the models that provide the best fit to our crucial pre-maximum spectra are noted by \cite{2013MNRAS.436..333S} to be a poor match for subluminous SNe based on their photometric properties. For the DDet scenario, the inherent asymmetry of explosions means that significant viewing-angle effects are expected, their angle-averaged model spectra are not representative of a single observational line-of-sight. However, there are no angle-dependent synthetic observations available for the selected scenarios in HESMA.

Far from being a simple caveat, the asymmetry of DDet scenario provides a natural explanation for the observed diversity in early-time spectral features (such as the blue-end variations in the early-time spectra) across the transitional population. The DDet can potentially unify the transitional class within a single physical framework, and possibly include the normal and 91bg-like populations as well through different distributions of its underlying physical parameters. On the other hand, the viewing-angle effects for most DDT models are less significant (with the exception of models with few ignition kernels, e.g., the N3 model in \citealp{2013MNRAS.436..333S}). Given these considerations, DDet model with thin He shell can offer a compelling and unified explanation for the entire transitional group. The diversity found in the early-time line velocities and blue-end absorption features (Figures \ref{fig:comp_spec_1} $\&$ \ref{fig:comp_spec_3}) contrasts with the greater homogeneity observed in the post-maximum spectra, illustrating the importance and diagnostic power of early-phase observations for constraining progenitor properties and explosion mechanisms. Detailed analysis of early-time observations of a larger sample is required to test this hypothesis.

\section{Conclusion} \label{sec:conclusion}

In this paper, we present a comprehensive analysis of the transitional SN Ia 2024aedt. The photometric evolution shows a clear power-law rise consistent with an expanding fireball model. The peak absolute magnitude $M_B = -18.49 \pm 0.03$ mag, decline rate $\Delta m_{15}(B) = 1.53 \pm 0.36$ mag, and early color evolution place it in the parameter space between normal and 91bg-like SN Ia populations, along with other transitional events. From the modeling of light curves, we derive a synthesized $^{56}\mathrm{Ni}$ mass of $0.414 \pm 0.042\,M_{\odot}$.  Spectroscopically, measurements of the pEW and velocity of the Si II lines in the early spectra show that the pEW evolution of transitional SNe, including SN 2024aedt, is much slower and flatter than that of normal SNe Ia at similar phases, consistent with their evolution in the Branch diagram ``Cool" (CL) region. For the spectral evolution, SN 2024aedt shows an overall similarity to other transitional objects. However, a detailed comparison reveals diversity in the early-time blue features, which becomes more homogeneous at later phases. Based on our comparison with available models, we find that both the DDT and DDet scenarios are capable of reproducing the observed properties of SN 2024aedt to a reasonable extent. Further detailed analysis and a larger sample of observations with improved early-phase coverage are required to distinguish between these models.

With the continuing WFST survey operation, a larger population of transitional SNe, similar to SN 2024aedt, is expected to be discovered. The survey's high-cadence strategy and depth enable the early discovery of valuable transients and prompt follow-up observations. With this greatly enriched sample, it will be possible to precisely constrain the parameter space that connects the normal and subluminous populations. Moreover, statistical studies of early light curves and color curves may reveal diversity within the transitional class, providing crucial constraints on the progenitor channels and explosion mechanisms of SNe Ia.

\begin{acknowledgments}
This work was supported by the National Key R\&D Program of China (Grant No. 2023YFA1608100), the National Natural Science Foundation of China (Grant No. 12393811, 12233008), and the Strategic Priority Research Program of the Chinese Academy of Science (Grant No. XDB0550300). J.J. acknowledges support from the Japan Society for the Promotion of Science (JSPS) KAKENHI grants JP22K14069. K.M. acknowledges support from the JSPS KAKENHI grant JP24KK0070 and 24H01810. K.M. and H.K. acknowledge support from the JSPS bilateral JPJSBP120229923. L.G. acknowledges financial support from AGAUR, CSIC, MCIN and AEI 10.13039/501100011033 under projects PID2023-151307NB-I00, PIE 20215AT016, CEX2020-001058-M, ILINK23001, COOPB2304, and 2021-SGR-01270.

Based on observations made with the Nordic Optical Telescope, owned in collaboration by the University of Turku and Aarhus University, and operated jointly by Aarhus University, the University of Turku and the University of Oslo, representing Denmark, Finland and Norway, the University of Iceland and Stockholm University at the Observatorio del Roque de los Muchachos, La Palma, Spain, of the Instituto de Astrofisica de Canarias. The NOT data were obtained under program ID Pnn-nnn.
This work has made use of data from the Asteroid Terrestrial-impact Last Alert System (ATLAS) project. The Asteroid Terrestrial-impact Last Alert System (ATLAS) project is primarily funded to search for near earth asteroids through NASA grants NN12AR55G, 80NSSC18K0284, and 80NSSC18K1575; byproducts of the NEO search include images and catalogs from the survey area. This work was partially funded by Kepler/K2 grant J1944/80NSSC19K0112 and HST GO-15889, and STFC grants ST/T000198/1 and ST/S006109/1. The ATLAS science products have been made possible through the contributions of the University of Hawaii Institute for Astronomy, the Queen’s University Belfast, the Space Telescope Science Institute, the South African Astronomical Observatory, and The Millennium Institute of Astrophysics (MAS), Chile.

This research made use of Photutils, an Astropy package for
detection and photometry of astronomical sources \citep{larry_bradley_2024_13989456}.

The Pan-STARRS1 Surveys (PS1) and the PS1 public science archive have been made possible through contributions by the Institute for Astronomy, the University of Hawaii, the Pan-STARRS Project Office, the Max-Planck Society and its participating institutes, the Max Planck Institute for Astronomy, Heidelberg and the Max Planck Institute for Extraterrestrial Physics, Garching, The Johns Hopkins University, Durham University, the University of Edinburgh, the Queen's University Belfast, the Harvard-Smithsonian Center for Astrophysics, the Las Cumbres Observatory Global Telescope Network Incorporated, the National Central University of Taiwan, the Space Telescope Science Institute, the National Aeronautics and Space Administration under Grant No. NNX08AR22G issued through the Planetary Science Division of the NASA Science Mission Directorate, the National Science Foundation Grant No. AST-1238877, the University of Maryland, Eotvos Lorand University (ELTE), the Los Alamos National Laboratory, and the Gordon and Betty Moore Foundation.

This work makes use of observations from the Las Cumbres Observatory global telescope network.

Based on observations obtained with the Apache Point Observatory 3.5-meter telescope, which is owned and operated by the Astrophysical Research Consortium.

This work made use of the Heidelberg Supernova Model Archive (HESMA), https://hesma.h-its.org.
\end{acknowledgments}

\facilities{HST(STIS), Swift(XRT and UVOT), AAVSO, CTIO:1.3m, CTIO:1.5m, CXO}

\software{Astropy \citep{astropy:2013, astropy:2018, astropy:2022}, 
          Source Extractor \citep{1996A&AS..117..393B}, SWarp \citep{2002ASPC..281..228B}, astroquery \citep{2019AJ....157...98G}, Photutils \citep{larry_bradley_2024_13989456}, PyKOSMOS \citep{davenport_2021_5120786}, superbol \citep{Nicholl_2018}, SALT2 \citep{refId0}, MOSFiT \citep{2018ApJS..236....6G}, CIGALE \citep{2019A&A...622A.103B}, SNooPy \citep{2011AJ....141...19B}
          }

\bibliography{sample701}{}

@Article{Wang2023,
author={Wang, Tinggui
and Liu, Guilin
and Cai, Zhenyi
and Geng, Jinjun
and Fang, Min
and He, Haoning
and Jiang, Ji-an
and Jiang, Ning
and Kong, Xu
and Li, Bin
and Li, Ye
and Luo, Wentao
and Pan, Zhizheng
and Wu, Xuefeng
and Yang, Ji
and Yu, Jiming
and Zheng, Xianzhong
and Zhu, Qingfeng
and Cai, Yi-Fu
and Chen, Yuanyuan
and Chen, Zhiwei
and Dai, Zigao
and Fan, Lulu
and Fan, Yizhong
and Fang, Wenjuan
and He, Zhicheng
and Hu, Lei
and Hu, Maokai
and Jin, Zhiping
and Jiang, Zhibo
and Li, Guoliang
and Li, Fan
and Li, Xuzhi
and Liang, Runduo
and Lin, Zheyu
and Liu, Qingzhong
and Liu, Wenhao
and Liu, Zhengyan
and Liu, Wei
and Liu, Yao
and Lou, Zheng
and Qu, Han
and Sheng, Zhenfeng
and Shi, Jianchun
and Shu, Yiping
and Su, Zhenbo
and Sun, Tianrui
and Wang, Hongchi
and Wang, Huiyuan
and Wang, Jian
and Wang, Junxian
and Wei, Daming
and Wei, Junjie
and Xue, Yongquan
and Yan, Jingzhi
and Yang, Chao
and Yuan, Ye
and Yuan, Yefei
and Zhang, Hongxin
and Zhang, Miaomiao
and Zhao, Haibin
and Zhao, Wen},
title={Science with the 2.5-meter Wide Field Survey Telescope (WFST)},
journal={Science China Physics, Mechanics {\&} Astronomy},
year={2023},
month={Sep},
day={14},
volume={66},
number={10},
pages={109512},
issn={1869-1927},
doi={10.1007/s11433-023-2197-5},
url={https://doi.org/10.1007/s11433-023-2197-5}
}

@BOOK{1991rc3..book.....D,
       author = {{de Vaucouleurs}, Gerard and {de Vaucouleurs}, Antoinette and {Corwin}, Jr., Herold G. and {Buta}, Ronald J. and {Paturel}, Georges and {Fouque}, Pascal},
        title = "{Third Reference Catalogue of Bright Galaxies}",
         year = 1991,
       adsurl = {https://ui.adsabs.harvard.edu/abs/1991rc3..book.....D}
}

@ARTICLE{2003AJ....126.2268W,
       author = {{Wegner}, G. and {Bernardi}, M. and {Willmer}, C.~N.~A. and {da Costa}, L.~N. and {Alonso}, M.~V. and {Pellegrini}, P.~S. and {Maia}, M.~A.~G. and {Chaves}, O.~L. and {Rit{\'e}}, C.},
        title = "{Redshift-Distance Survey of Early-Type Galaxies: Spectroscopic Data}",
      journal = {\aj},
         year = 2003,
        month = nov,
       volume = {126},
       number = {5},
        pages = {2268-2280},
          doi = {10.1086/378959},
archivePrefix = {arXiv},
       eprint = {astro-ph/0308357},
 primaryClass = {astro-ph},
       adsurl = {https://ui.adsabs.harvard.edu/abs/2003AJ....126.2268W}
}

@article{refId0,
	author = {{Guy, J.} and {Astier, P.} and {Baumont, S.} and {Hardin, D.} and {Pain, R.} and {Regnault, N.} and {Basa, S.} and {Carlberg, R. G.} and {Conley, A.} and {Fabbro, S.} and {Fouchez, D.} and {Hook, I. M.} and {Howell, D. A.} and {Perrett, K.} and {Pritchet, C. J.} and {Rich, J.} and {Sullivan, M.} and {Antilogus, P.} and {Aubourg, E.} and {Bazin, G.} and {Bronder, J.} and {Filiol, M.} and {Palanque-Delabrouille, N.} and {Ripoche, P.} and {Ruhlmann-Kleider, V.}},
	title = {SALT2: using distant supernovae to improve the use   of type Ia 
supernovae as distance indicators ***},
	DOI= "10.1051/0004-6361:20066930",
	url= "https://doi.org/10.1051/0004-6361:20066930",
	journal = {A\&A},
	year = 2007,
	volume = 466,
	number = 1,
	pages = "11-21",
}

@ARTICLE{2023arXiv230516279M,
       author = {{Masci}, Frank J. and {Laher}, Russ R. and {Rusholme}, Benjamin and {Shupe}, David and {Paladini}, Roberta and {Groom}, Steve and {Wold}, Avery and {Miller}, Adam A. and {Drake}, Andrew},
        title = "{A New Forced Photometry Service for the Zwicky Transient Facility}",
      journal = {arXiv e-prints},
         year = 2023,
        month = may,
          eid = {arXiv:2305.16279},
        pages = {arXiv:2305.16279},
          doi = {10.48550/arXiv.2305.16279},
archivePrefix = {arXiv},
       eprint = {2305.16279},
 primaryClass = {astro-ph.IM},
       adsurl = {https://ui.adsabs.harvard.edu/abs/2023arXiv230516279M}
}

@ARTICLE{2018PASP..130f4505T,
       author = {{Tonry}, J.~L. and {Denneau}, L. and {Heinze}, A.~N. and {Stalder}, B. and {Smith}, K.~W. and {Smartt}, S.~J. and {Stubbs}, C.~W. and {Weiland}, H.~J. and {Rest}, A.},
        title = "{ATLAS: A High-cadence All-sky Survey System}",
      journal = {\pasp},
         year = 2018,
        month = jun,
       volume = {130},
       number = {988},
        pages = {064505},
          doi = {10.1088/1538-3873/aabadf},
archivePrefix = {arXiv},
       eprint = {1802.00879},
 primaryClass = {astro-ph.IM},
       adsurl = {https://ui.adsabs.harvard.edu/abs/2018PASP..130f4505T}
}

@ARTICLE{2020PASP..132h5002S,
       author = {{Smith}, K.~W. and {Smartt}, S.~J. and {Young}, D.~R. and {Tonry}, J.~L. and {Denneau}, L. and {Flewelling}, H. and {Heinze}, A.~N. and {Weiland}, H.~J. and {Stalder}, B. and {Rest}, A. and {Stubbs}, C.~W. and {Anderson}, J.~P. and {Chen}, T. -W. and {Clark}, P. and {Do}, A. and {F{\"o}rster}, F. and {Fulton}, M. and {Gillanders}, J. and {McBrien}, O.~R. and {O'Neill}, D. and {Srivastav}, S. and {Wright}, D.~E.},
        title = "{Design and Operation of the ATLAS Transient Science Server}",
      journal = {\pasp},
         year = 2020,
        month = aug,
       volume = {132},
       number = {1014},
          eid = {085002},
        pages = {085002},
          doi = {10.1088/1538-3873/ab936e},
archivePrefix = {arXiv},
       eprint = {2003.09052},
 primaryClass = {astro-ph.IM},
       adsurl = {https://ui.adsabs.harvard.edu/abs/2020PASP..132h5002S}
}

@ARTICLE{2021TNSAN...7....1S,
       author = {{Shingles}, L. and {Smith}, K.~W. and {Young}, D.~R. and {Smartt}, S.~J. and {Tonry}, J. and {Denneau}, L. and {Heinze}, A. and {Weiland}, H. and {Flewelling}, H. and {Stalder}, B. and {Clocchiatti}, A. and {F{\"o}rster}, F. and {Pignata}, G. and {Rest}, A. and {Anderson}, J. and {Stubbs}, C. and {Erasmus}, N.},
        title = "{Release of the ATLAS Forced Photometry server for public use}",
      journal = {Transient Name Server AstroNote},
         year = 2021,
        month = jan,
       volume = {7},
        pages = {1-7},
       adsurl = {https://ui.adsabs.harvard.edu/abs/2021TNSAN...7....1S}
}

@ARTICLE{2019AJ....157...98G,
       author = {{Ginsburg}, Adam and {Sip{\H{o}}cz}, Brigitta M. and {Brasseur}, C.~E. and {Cowperthwaite}, Philip S. and {Craig}, Matthew W. and {Deil}, Christoph and {Guillochon}, James and {Guzman}, Giannina and {Liedtke}, Simon and {Lian Lim}, Pey and {Lockhart}, Kelly E. and {Mommert}, Michael and {Morris}, Brett M. and {Norman}, Henrik and {Parikh}, Madhura and {Persson}, Magnus V. and {Robitaille}, Thomas P. and {Segovia}, Juan-Carlos and {Singer}, Leo P. and {Tollerud}, Erik J. and {de Val-Borro}, Miguel and {Valtchanov}, Ivan and {Woillez}, Julien and {Astroquery Collaboration} and {a subset of astropy Collaboration}},
        title = "{astroquery: An Astronomical Web-querying Package in Python}",
      journal = {\aj},
         year = 2019,
        month = mar,
       volume = {157},
       number = {3},
          eid = {98},
        pages = {98},
          doi = {10.3847/1538-3881/aafc33},
archivePrefix = {arXiv},
       eprint = {1901.04520},
 primaryClass = {astro-ph.IM},
       adsurl = {https://ui.adsabs.harvard.edu/abs/2019AJ....157...98G}
}

@Article{Olling2015,
author={Olling, Rob P.
and Mushotzky, Richard
and Shaya, Edward J.
and Rest, Armin
and Garnavich, Peter M.
and Tucker, Brad E.
and Kasen, Daniel
and Margheim, Steve
and Filippenko, Alexei V.},
title={No signature of ejecta interaction with a stellar companion in three type Ia supernovae},
journal={Nature},
year={2015},
month={May},
day={01},
volume={521},
number={7552},
pages={332-335},
issn={1476-4687},
doi={10.1038/nature14455},
url={https://doi.org/10.1038/nature14455}
}

@article{Guy2007,
  title = {SALT2: using distant supernovae to improve the use of type Ia supernovae as distance indicators},
  volume = {466},
  ISSN = {1432-0746},
  url = {http://dx.doi.org/10.1051/0004-6361:20066930},
  DOI = {10.1051/0004-6361:20066930},
  number = {1},
  journal = {Astronomy \& Astrophysics},
  publisher = {EDP Sciences},
  author = {Guy,  J. and Astier,  P. and Baumont,  S. and Hardin,  D. and Pain,  R. and Regnault,  N. and Basa,  S. and Carlberg,  R. G. and Conley,  A. and Fabbro,  S. and Fouchez,  D. and Hook,  I. M. and Howell,  D. A. and Perrett,  K. and Pritchet,  C. J. and Rich,  J. and Sullivan,  M. and Antilogus,  P. and Aubourg,  E. and Bazin,  G. and Bronder,  J. and Filiol,  M. and Palanque-Delabrouille,  N. and Ripoche,  P. and Ruhlmann-Kleider,  V.},
  year = {2007},
  month = feb,
  pages = {11–21}
}

@article{Hicken_2009,
doi = {10.1088/0004-637X/700/1/331},
url = {https://dx.doi.org/10.1088/0004-637X/700/1/331},
year = {2009},
month = {jul},
publisher = {The American Astronomical Society},
volume = {700},
number = {1},
pages = {331},
author = {Hicken, Malcolm and Challis, Peter and Jha, Saurabh and Kirshner, Robert P. and Matheson, Tom and Modjaz, Maryam and Rest, Armin and Michael Wood-Vasey, W. and Bakos, Gaspar and Barton, Elizabeth J. and Berlind, Perry and Bragg, Ann and Briceño, Cesar and Brown, Warren R. and Caldwell, Nelson and Calkins, Mike and Cho, Richard and Ciupik, Larry and Contreras, Maria and Dendy, Kristi-Concannon and Dosaj, Anil and Durham, Nick and Eriksen, Kris and Esquerdo, Gil and Everett, Mark and Falco, Emilio and Fernandez, Jose and Gaba, Alejandro and Garnavich, Peter and Graves, Genevieve and Green, Paul and Groner, Ted and Hergenrother, Carl and Holman, Matthew J. and Hradecky, Vit and Huchra, John and Hutchison, Bob and Jerius, Diab and Jordan, Andres and Kilgard, Roy and Krauss, Miriam and Luhman, Kevin and Macri, Lucas and Marrone, Daniel and McDowell, Jonathan and McIntosh, Daniel and McNamara, Brian and Megeath, Tom and Mochejska, Barbara and Munoz, Diego and Muzerolle, James and Naranjo, Orlando and Narayan, Gautham and Pahre, Michael and Peters, Wayne and Peterson, Dawn and Rines, Ken and Ripman, Ben and Roussanova, Anna and Schild, Rudolph and Sicilia-Aguilar, Aurora and Sokoloski, Jennifer and Smalley, Kyle and Smith, Andy and Spahr, Tim and Stanek, K. Z. and Barmby, Pauline and Blondin, Stéphane and Stubbs, Christopher W. and Szentgyorgyi, Andrew and Torres, Manuel A. P. and Vaz, Amili and Vikhlinin, Alexey and Wang, Zhong and Westover, Mike and Woods, Deborah and Zhao, Ping},
title = {CfA3: 185 TYPE Ia SUPERNOVA LIGHT CURVES FROM THE CfA},
journal = {The Astrophysical Journal}
}

@article{Rigault2025,
  title = {ZTF SN Ia DR2: Overview},
  volume = {694},
  ISSN = {1432-0746},
  url = {http://dx.doi.org/10.1051/0004-6361/202450388},
  DOI = {10.1051/0004-6361/202450388},
  journal = {Astronomy \& Astrophysics},
  publisher = {EDP Sciences},
  author = {Rigault,  M. and Smith,  M. and Goobar,  A. and Maguire,  K. and Dimitriadis,  G. and Johansson,  J. and Nordin,  J. and Burgaz,  U. and Dhawan,  S. and Sollerman,  J. and Regnault,  N. and Kowalski,  M. and Nugent,  P. and Andreoni,  I. and Amenouche,  M. and Aubert,  M. and Barjou-Delayre,  C. and Bautista,  J. and Bellm,  E. and Betoule,  M. and Bloom,  J. S. and Carreres,  B. and Chen,  T. X. and Copin,  Y. and Deckers,  M. and de Jaeger,  T. and Feinstein,  F. and Fouchez,  D. and Fremling,  C. and Galbany,  L. and Ginolin,  M. and Graham,  M. and Groom,  S. L. and Harvey,  L. and Kasliwal,  M. M. and Kenworthy,  W. D. and Kim,  Y.-L. and Kuhn,  D. and Kulkarni,  S. R. and Lacroix,  L. and Laher,  R. R. and Masci,  F. J. and M\"{u}ller-Bravo,  T. E. and Miller,  A. and Osman,  M. and Perley,  D. and Popovic,  B. and Purdum,  J. and Qin,  Y.-J. and Racine,  B. and Reusch,  S. and Riddle,  R. and Rosnet,  P. and Rosselli,  D. and Ruppin,  F. and Senzel,  R. and Rusholme,  B. and Schweyer,  T. and Terwel,  J. H. and Townsend,  A. and Tzanidakis,  A. and Wold,  A. and Yan,  L.},
  year = {2025},
  month = feb,
  pages = {A1}
}

@article{Phillips1993,
  title = {The absolute magnitudes of Type IA supernovae},
  volume = {413},
  ISSN = {1538-4357},
  url = {http://dx.doi.org/10.1086/186970},
  DOI = {10.1086/186970},
  journal = {The Astrophysical Journal},
  publisher = {American Astronomical Society},
  author = {Phillips,  M. M.},
  year = {1993},
  month = aug,
  pages = {L105}
}

@inbook{Taubenberger2017,
  title = {The Extremes of Thermonuclear Supernovae},
  ISBN = {9783319218465},
  url = {http://dx.doi.org/10.1007/978-3-319-21846-5_37},
  DOI = {10.1007/978-3-319-21846-5_37},
  booktitle = {Handbook of Supernovae},
  publisher = {Springer International Publishing},
  author = {Taubenberger,  Stefan},
  year = {2017},
  pages = {317–373}
}

@article{Phillips1987,
  title = {The type 1a supernova 1986G in NGC 5128 - Optical photometry and spectra},
  volume = {99},
  ISSN = {1538-3873},
  url = {http://dx.doi.org/10.1086/132020},
  DOI = {10.1086/132020},
  journal = {Publications of the Astronomical Society of the Pacific},
  publisher = {IOP Publishing},
  author = {Phillips,  M. M. and Phillips,  A. C. and Heathcote,  S. R. and Blanco,  V. M. and Geisler,  D. and Hamilton,  D. and Suntzeff,  N. B. and Jablonski,  F. J. and Steiner,  J. E. and Cowley,  A. P. and Schmidtke,  P. and Wyckoff,  S. and Hutchings,  J. B. and Tonry,  J. and Strauss,  M. A. and Thorstensen,  J. R. and Honey,  W. and Maza,  J. and Ruiz,  M. T. and Landolt,  A. U. and Uomoto,  A. and Rich,  R. M. and Grindlay,  J. E. and Cohn,  H. and Smith,  H. A. and Lutz,  J. H. and Lavery,  R. J. and Saha,  A.},
  year = {1987},
  month = jul,
  pages = {592}
}

@article{Schlafly2011,
  title = {MEASURING REDDENING WITH SLOAN DIGITAL SKY SURVEY STELLAR SPECTRA AND RECALIBRATING SFD},
  volume = {737},
  ISSN = {1538-4357},
  url = {http://dx.doi.org/10.1088/0004-637X/737/2/103},
  DOI = {10.1088/0004-637x/737/2/103},
  number = {2},
  journal = {The Astrophysical Journal},
  publisher = {American Astronomical Society},
  author = {Schlafly,  Edward F. and Finkbeiner,  Douglas P.},
  year = {2011},
  month = aug,
  pages = {103}
}

@ARTICLE{2007ApJ...663..320F,
       author = {{Fitzpatrick}, E.~L. and {Massa}, D.},
        title = "{An Analysis of the Shapes of Interstellar Extinction Curves. V. The IR-through-UV Curve Morphology}",
      journal = {\apj},
         year = 2007,
        month = jul,
       volume = {663},
       number = {1},
        pages = {320-341},
          doi = {10.1086/518158},
archivePrefix = {arXiv},
       eprint = {0705.0154},
 primaryClass = {astro-ph},
       adsurl = {https://ui.adsabs.harvard.edu/abs/2007ApJ...663..320F}
}

@INPROCEEDINGS{2010ASSP...14..211D,
       author = {{Djupvik}, Anlaug Amanda and {Andersen}, Johannes},
        title = "{The Nordic Optical Telescope}",
    booktitle = {Highlights of Spanish Astrophysics V},
         year = 2010,
       editor = {{Diego}, Jose M. and {Goicoechea}, Luis J. and {Gonz{\'a}lez-Serrano}, J. Ignacio and {Gorgas}, Javier},
       series = {Astrophysics and Space Science Proceedings},
       volume = {14},
        month = jan,
        pages = {211},
          doi = {10.1007/978-3-642-11250-8_21},
archivePrefix = {arXiv},
       eprint = {0901.4015},
 primaryClass = {astro-ph.IM},
       adsurl = {https://ui.adsabs.harvard.edu/abs/2010ASSP...14..211D}
}

@article{Nicholl_2018,
doi = {10.3847/2515-5172/aaf799},
url = {https://dx.doi.org/10.3847/2515-5172/aaf799},
year = {2018},
month = {dec},
publisher = {The American Astronomical Society},
volume = {2},
number = {4},
pages = {230},
author = {Nicholl, Matt},
title = {SuperBol: A User-friendly Python Routine for Bolometric Light Curves},
journal = {Research Notes of the AAS}
}

@ARTICLE{2018ApJS..236....6G,
       author = {{Guillochon}, James and {Nicholl}, Matt and {Villar}, V. Ashley and {Mockler}, Brenna and {Narayan}, Gautham and {Mandel}, Kaisey S. and {Berger}, Edo and {Williams}, Peter K.~G.},
        title = "{MOSFiT: Modular Open Source Fitter for Transients}",
      journal = {\apjs},
         year = 2018,
        month = may,
       volume = {236},
       number = {1},
          eid = {6},
        pages = {6},
          doi = {10.3847/1538-4365/aab761},
archivePrefix = {arXiv},
       eprint = {1710.02145},
 primaryClass = {astro-ph.IM},
       adsurl = {https://ui.adsabs.harvard.edu/abs/2018ApJS..236....6G}
}

@ARTICLE{1994ApJS...92..527N,
       author = {{Nadyozhin}, D.~K.},
        title = "{The Properties of NI CO Fe Decay}",
      journal = {\apjs},
         year = 1994,
        month = jun,
       volume = {92},
        pages = {527},
          doi = {10.1086/192008},
       adsurl = {https://ui.adsabs.harvard.edu/abs/1994ApJS...92..527N}
}

@ARTICLE{1973ApJ...186.1007W,
       author = {{Whelan}, John and {Iben}, Jr., Icko},
        title = "{Binaries and Supernovae of Type I}",
      journal = {\apj},
         year = 1973,
        month = dec,
       volume = {186},
        pages = {1007-1014},
          doi = {10.1086/152565},
       adsurl = {https://ui.adsabs.harvard.edu/abs/1973ApJ...186.1007W}
}

@ARTICLE{1982ApJ...253..798N,
       author = {{Nomoto}, K.},
        title = "{Accreting white dwarf models for type I supernovae. I - Presupernova evolution and triggering mechanisms}",
      journal = {\apj},
         year = 1982,
        month = feb,
       volume = {253},
        pages = {798-810},
          doi = {10.1086/159682},
       adsurl = {https://ui.adsabs.harvard.edu/abs/1982ApJ...253..798N}
}

@ARTICLE{1981NInfo..49....3T,
       author = {{Tutukov}, A.~V. and {Yungelson}, L.~R.},
        title = "{Evolutionary Scenario for Close Binary Systems of Low and Moderate Masses}",
      journal = {Nauchnye Informatsii},
         year = 1981,
        month = jan,
       volume = {49},
        pages = {3},
       adsurl = {https://ui.adsabs.harvard.edu/abs/1981NInfo..49....3T}
}

@ARTICLE{1984ApJS...54..335I,
       author = {{Iben}, Jr., I. and {Tutukov}, A.~V.},
        title = "{Supernovae of type I as end products of the evolution of binaries with components of moderate initial mass.}",
      journal = {\apjs},
         year = 1984,
        month = feb,
       volume = {54},
        pages = {335-372},
          doi = {10.1086/190932},
       adsurl = {https://ui.adsabs.harvard.edu/abs/1984ApJS...54..335I}
}

@ARTICLE{1984ApJ...277..355W,
       author = {{Webbink}, R.~F.},
        title = "{Double white dwarfs as progenitors of R Coronae Borealis stars and type I supernovae.}",
      journal = {\apj},
         year = 1984,
        month = feb,
       volume = {277},
        pages = {355-360},
          doi = {10.1086/161701},
       adsurl = {https://ui.adsabs.harvard.edu/abs/1984ApJ...277..355W}
}

@ARTICLE{1991A&A...245..114K,
       author = {{Khokhlov}, A.~M.},
        title = "{Delayed detonation model for type IA supernovae}",
      journal = {\aap},
         year = 1991,
        month = may,
       volume = {245},
       number = {1},
        pages = {114-128},
       adsurl = {https://ui.adsabs.harvard.edu/abs/1991A&A...245..114K}
}

@article{10.1093/mnras/stad1226,
    author = {Harvey, L and Maguire, K and Magee, M R and Bulla, M and Dhawan, S and Schulze, S and Sollerman, J and Deckers, M and Dimitriadis, G and Reusch, S and Smith, M and Terwel, J and Coughlin, M W and Masci, F and Purdum, J and Reedy, A and Robert, E and Wold, A},
    title = {Early-time spectroscopic modelling of the transitional Type Ia Supernova 2021rhu with tardis},
    journal = {Monthly Notices of the Royal Astronomical Society},
    volume = {522},
    number = {3},
    pages = {4444-4467},
    year = {2023},
    month = {05},
    issn = {0035-8711},
    doi = {10.1093/mnras/stad1226},
    url = {https://doi.org/10.1093/mnras/stad1226},
    eprint = {https://academic.oup.com/mnras/article-pdf/522/3/4444/53960783/stad1226.pdf},
}

@ARTICLE{2013MNRAS.433.2240G,
       author = {{Ganeshalingam}, Mohan and {Li}, Weidong and {Filippenko}, Alexei V.},
        title = "{Constraints on dark energy with the LOSS SN Ia sample}",
      journal = {\mnras},
         year = 2013,
        month = aug,
       volume = {433},
       number = {3},
        pages = {2240-2258},
          doi = {10.1093/mnras/stt893},
archivePrefix = {arXiv},
       eprint = {1307.0824},
 primaryClass = {astro-ph.CO},
       adsurl = {https://ui.adsabs.harvard.edu/abs/2013MNRAS.433.2240G}
}

@PHDTHESIS{2002PhDT........10J,
       author = {{Jha}, Saurabh},
        title = "{Exploding stars, near and far}",
       school = {Harvard University, Massachusetts},
         year = 2002,
        month = oct,
       adsurl = {https://ui.adsabs.harvard.edu/abs/2002PhDT........10J}
}

@article{Krisciunas2009,
  title = {THE FAST DECLINING TYPE Ia SUPERNOVA 2003gs,  AND EVIDENCE FOR A SIGNIFICANT DISPERSION IN NEAR-INFRARED ABSOLUTE MAGNITUDES OF FAST DECLINERS AT MAXIMUM LIGHT},
  volume = {138},
  ISSN = {1538-3881},
  url = {http://dx.doi.org/10.1088/0004-6256/138/6/1584},
  DOI = {10.1088/0004-6256/138/6/1584},
  number = {6},
  journal = {The Astronomical Journal},
  publisher = {American Astronomical Society},
  author = {Krisciunas,  Kevin and Marion,  G. H. and Suntzeff,  Nicholas B. and Blanc,  Guillaume and Bufano,  Filomena and Candia,  Pablo and Cartier,  Regis and Elias-Rosa,  Nancy and Espinoza,  Juan and Gonzalez,  David and Gonzalez,  Luis and Gonzalez,  Sergio and Gooding,  Samuel D. and Hamuy,  Mario and Knox,  Ethan A. and Milne,  Peter A. and Morrell,  Nidia and Phillips,  Mark M. and Stritzinger,  Maximilian and Thomas-Osip,  Joanna},
  year = {2009},
  month = oct,
  pages = {1584–1596}
}

@article{Gall2018,
  title = {Two transitional type Ia supernovae located in the Fornax cluster member NGC 1404: SN 2007on and SN 2011iv},
  volume = {611},
  ISSN = {1432-0746},
  url = {http://dx.doi.org/10.1051/0004-6361/201730886},
  DOI = {10.1051/0004-6361/201730886},
  journal = {Astronomy \& Astrophysics},
  publisher = {EDP Sciences},
  author = {Gall,  C. and Stritzinger,  M. D. and Ashall,  C. and Baron,  E. and Burns,  C. R. and Hoeflich,  P. and Hsiao,  E. Y. and Mazzali,  P. A. and Phillips,  M. M. and Filippenko,  A. V. and Anderson,  J. P. and Benetti,  S. and Brown,  P. J. and Campillay,  A. and Challis,  P. and Contreras,  C. and Elias de la Rosa,  N. and Folatelli,  G. and Foley,  R. J. and Fraser,  M. and Holmbo,  S. and Marion,  G. H. and Morrell,  N. and Pan,  Y.-C. and Pignata,  G. and Suntzeff,  N. B. and Taddia,  F. and Torres Robledo,  S. and Valenti,  S.},
  year = {2018},
  month = mar,
  pages = {A58}
}

@article{Hsiao2015,
  title = {Strong near-infrared carbon in the Type Ia supernova iPTF13ebh},
  volume = {578},
  ISSN = {1432-0746},
  url = {http://dx.doi.org/10.1051/0004-6361/201425297},
  DOI = {10.1051/0004-6361/201425297},
  journal = {Astronomy \& Astrophysics},
  publisher = {EDP Sciences},
  author = {Hsiao,  E. Y. and Burns,  C. R. and Contreras,  C. and H\"{o}flich,  P. and Sand,  D. and Marion,  G. H. and Phillips,  M. M. and Stritzinger,  M. and González-Gaitán,  S. and Mason,  R. E. and Folatelli,  G. and Parent,  E. and Gall,  C. and Amanullah,  R. and Anupama,  G. C. and Arcavi,  I. and Banerjee,  D. P. K. and Beletsky,  Y. and Blanc,  G. A. and Bloom,  J. S. and Brown,  P. J. and Campillay,  A. and Cao,  Y. and De Cia,  A. and Diamond,  T. and Freedman,  W. L. and Gonzalez,  C. and Goobar,  A. and Holmbo,  S. and Howell,  D. A. and Johansson,  J. and Kasliwal,  M. M. and Kirshner,  R. P. and Krisciunas,  K. and Kulkarni,  S. R. and Maguire,  K. and Milne,  P. A. and Morrell,  N. and Nugent,  P. E. and Ofek,  E. O. and Osip,  D. and Palunas,  P. and Perley,  D. A. and Persson,  S. E. and Piro,  A. L. and Rabus,  M. and Roth,  M. and Schiefelbein,  J. M. and Srivastav,  S. and Sullivan,  M. and Suntzeff,  N. B. and Surace,  J. and Woźniak,  P. R. and Yaron,  O.},
  year = {2015},
  month = may,
  pages = {A9}
}

@article{Leloudas2009,
  title = {The normal Type Ia SN 2003hv out to very late phases},
  volume = {505},
  ISSN = {1432-0746},
  url = {http://dx.doi.org/10.1051/0004-6361/200912364},
  DOI = {10.1051/0004-6361/200912364},
  number = {1},
  journal = {Astronomy \& Astrophysics},
  publisher = {EDP Sciences},
  author = {Leloudas,  G. and Stritzinger,  M. D. and Sollerman,  J. and Burns,  C. R. and Kozma,  C. and Krisciunas,  K. and Maund,  J. R. and Milne,  P. and Filippenko,  A. V. and Fransson,  C. and Ganeshalingam,  M. and Hamuy,  M. and Li,  W. and Phillips,  M. M. and Schmidt,  B. P. and Skottfelt,  J. and Taubenberger,  S. and Boldt,  L. and Fynbo,  J. P. U. and Gonzalez,  L. and Salvo,  M. and Thomas-Osip,  J.},
  year = {2009},
  month = jul,
  pages = {265–279}
}

@ARTICLE{2017ApJ...846...58H,
       author = {{Hoeflich}, P. and {Hsiao}, E.~Y. and {Ashall}, C. and {Burns}, C.~R. and {Diamond}, T.~R. and {Phillips}, M.~M. and {Sand}, D. and {Stritzinger}, M.~D. and {Suntzeff}, N. and {Contreras}, C. and {Krisciunas}, K. and {Morrell}, N. and {Wang}, L.},
        title = "{Light and Color Curve Properties of Type Ia Supernovae: Theory Versus Observations}",
      journal = {\apj},
         year = 2017,
        month = sep,
       volume = {846},
       number = {1},
          eid = {58},
        pages = {58},
          doi = {10.3847/1538-4357/aa84b2},
archivePrefix = {arXiv},
       eprint = {1707.05350},
 primaryClass = {astro-ph.SR},
       adsurl = {https://ui.adsabs.harvard.edu/abs/2017ApJ...846...58H}
}

@article{Contreras2010,
  title = {THE CARNEGIE SUPERNOVA PROJECT: FIRST PHOTOMETRY DATA RELEASE OF LOW-REDSHIFT TYPE Ia SUPERNOVAE},
  volume = {139},
  ISSN = {1538-3881},
  url = {http://dx.doi.org/10.1088/0004-6256/139/2/519},
  DOI = {10.1088/0004-6256/139/2/519},
  number = {2},
  journal = {The Astronomical Journal},
  publisher = {American Astronomical Society},
  author = {Contreras,  Carlos and Hamuy,  Mario and Phillips,  M. M. and Folatelli,  Gastón and Suntzeff,  Nicholas B. and Persson,  S. E. and Stritzinger,  Maximilian and Boldt,  Luis and González,  Sergio and Krzeminski,  Wojtek and Morrell,  Nidia and Roth,  Miguel and Salgado,  Francisco and Maureira,  María José and Burns,  Christopher R. and Freedman,  W. L. and Madore,  Barry F. and Murphy,  David and Wyatt,  Pamela and Li,  Weidong and Filippenko,  Alexei V.},
  year = {2010},
  month = jan,
  pages = {519–539}
}

@article{Sahu2013,
  title = {Photometric and spectroscopic evolution of supernova SN 2009an: another case of a transitional Type Ia event},
  volume = {430},
  ISSN = {0035-8711},
  url = {http://dx.doi.org/10.1093/mnras/sts609},
  DOI = {10.1093/mnras/sts609},
  number = {2},
  journal = {Monthly Notices of the Royal Astronomical Society},
  publisher = {Oxford University Press (OUP)},
  author = {Sahu,  D. K. and Anupama,  G. C. and Anto,  P.},
  year = {2013},
  month = jan,
  pages = {869–887}
}

@article{Burns2018,
  title = {The Carnegie Supernova Project: Absolute Calibration and the Hubble Constant},
  volume = {869},
  ISSN = {1538-4357},
  url = {http://dx.doi.org/10.3847/1538-4357/aae51c},
  DOI = {10.3847/1538-4357/aae51c},
  number = {1},
  journal = {The Astrophysical Journal},
  publisher = {American Astronomical Society},
  author = {Burns,  Christopher R. and Parent,  Emilie and Phillips,  M. M. and Stritzinger,  Maximilian and Krisciunas,  Kevin and Suntzeff,  Nicholas B. and Hsiao,  Eric Y. and Contreras,  Carlos and Anais,  Jorge and Boldt,  Luis and Busta,  Luis and Campillay,  Abdo and Castellón,  Sergio and Folatelli,  Gastón and Freedman,  Wendy L. and González,  Consuelo and Hamuy,  Mario and Heoflich,  Peter and Krzeminski,  Wojtek and Madore,  Barry F. and Morrell,  Nidia and Persson,  S. E. and Roth,  Miguel and Salgado,  Francisco and Serón,  Jacqueline and Torres,  Simón},
  year = {2018},
  month = dec,
  pages = {56}
}

@article{Wyatt2021,
  title = {Strong Near-infrared Carbon Absorption in the Transitional Type Ia SN 2015bp*},
  volume = {914},
  ISSN = {1538-4357},
  url = {http://dx.doi.org/10.3847/1538-4357/abf7c3},
  DOI = {10.3847/1538-4357/abf7c3},
  number = {1},
  journal = {The Astrophysical Journal},
  publisher = {American Astronomical Society},
  author = {Wyatt,  S. D. and Sand,  D. J. and Hsiao,  E. Y. and Burns,  C. R. and Valenti,  S. and Bostroem,  K. A. and Lundquist,  M. and Galbany,  L. and Lu,  J. and Ashall,  C. and Diamond,  T. R. and Filippenko,  A. V. and Graham,  M. L. and Hoeflich,  P. and Kirshner,  R. P. and Krisciunas,  K. and Marion,  G. H. and Morrell,  N. and Persson,  S. E. and Phillips,  M. M. and Stritzinger,  M. D. and Suntzeff,  N. B. and Taddia,  F.},
  year = {2021},
  month = jun,
  pages = {57}
}

@ARTICLE{2017ApJ...835...64G,
       author = {{Guillochon}, James and {Parrent}, Jerod and {Kelley}, Luke Zoltan and {Margutti}, Raffaella},
        title = "{An Open Catalog for Supernova Data}",
      journal = {\apj},
         year = 2017,
        month = jan,
       volume = {835},
       number = {1},
          eid = {64},
        pages = {64},
          doi = {10.3847/1538-4357/835/1/64},
archivePrefix = {arXiv},
       eprint = {1605.01054},
 primaryClass = {astro-ph.SR},
       adsurl = {https://ui.adsabs.harvard.edu/abs/2017ApJ...835...64G}
}

@article{Graham2017,
  title = {Nebular-phase spectra of nearby Type Ia Supernovae},
  volume = {472},
  ISSN = {1365-2966},
  url = {http://dx.doi.org/10.1093/mnras/stx2224},
  DOI = {10.1093/mnras/stx2224},
  number = {3},
  journal = {Monthly Notices of the Royal Astronomical Society},
  publisher = {Oxford University Press (OUP)},
  author = {Graham,  M. L. and Kumar,  S. and Hosseinzadeh,  G. and Hiramatsu,  D. and Arcavi,  I. and Howell,  D. A. and Valenti,  S. and Sand,  D. J. and Parrent,  J. T. and McCully,  C. and Filippenko,  A. V.},
  year = {2017},
  month = aug,
  pages = {3437–3454}
}

@ARTICLE{2015MNRAS.446.3895F,
       author = {{Firth}, R.~E. and {Sullivan}, M. and {Gal-Yam}, A. and {Howell}, D.~A. and {Maguire}, K. and {Nugent}, P. and {Piro}, A.~L. and {Baltay}, C. and {Feindt}, U. and {Hadjiyksta}, E. and {McKinnon}, R. and {Ofek}, E. and {Rabinowitz}, D. and {Walker}, E.~S.},
        title = "{The rising light curves of Type Ia supernovae}",
      journal = {\mnras},
         year = 2015,
        month = feb,
       volume = {446},
       number = {4},
        pages = {3895-3910},
          doi = {10.1093/mnras/stu2314},
archivePrefix = {arXiv},
       eprint = {1411.1064},
 primaryClass = {astro-ph.HE},
       adsurl = {https://ui.adsabs.harvard.edu/abs/2015MNRAS.446.3895F}
}

@ARTICLE{2017MNRAS.464.4476C,
       author = {{Cartier}, R. and {Sullivan}, M. and {Firth}, R.~E. and {Pignata}, G. and {Mazzali}, P. and {Maguire}, K. and {Childress}, M.~J. and {Arcavi}, I. and {Ashall}, C. and {Bassett}, B. and {Crawford}, S.~M. and {Frohmaier}, C. and {Galbany}, L. and {Gal-Yam}, A. and {Hosseinzadeh}, G. and {Howell}, D.~A. and {Inserra}, C. and {Johansson}, J. and {Kasai}, E.~K. and {McCully}, C. and {Prajs}, S. and {Prentice}, S. and {Schulze}, S. and {Smartt}, S.~J. and {Smith}, K.~W. and {Smith}, M. and {Valenti}, S. and {Young}, D.~R.},
        title = "{Early observations of the nearby Type Ia supernova SN 2015F}",
      journal = {\mnras},
         year = 2017,
        month = feb,
       volume = {464},
       number = {4},
        pages = {4476-4494},
          doi = {10.1093/mnras/stw2678},
archivePrefix = {arXiv},
       eprint = {1609.04465},
 primaryClass = {astro-ph.SR},
       adsurl = {https://ui.adsabs.harvard.edu/abs/2017MNRAS.464.4476C}
}

@ARTICLE{2008ApJ...689..377W,
       author = {{Wood-Vasey}, W. Michael and {Friedman}, Andrew S. and {Bloom}, Joshua S. and {Hicken}, Malcolm and {Modjaz}, Maryam and {Kirshner}, Robert P. and {Starr}, Dan L. and {Blake}, Cullen H. and {Falco}, Emilio E. and {Szentgyorgyi}, Andrew H. and {Challis}, Peter and {Blondin}, St{\'e}phane and {Mandel}, Kaisey S. and {Rest}, Armin},
        title = "{Type Ia Supernovae Are Good Standard Candles in the Near Infrared: Evidence from PAIRITEL}",
      journal = {\apj},
         year = 2008,
        month = dec,
       volume = {689},
       number = {1},
        pages = {377-390},
          doi = {10.1086/592374},
archivePrefix = {arXiv},
       eprint = {0711.2068},
 primaryClass = {astro-ph},
       adsurl = {https://ui.adsabs.harvard.edu/abs/2008ApJ...689..377W}
}

@ARTICLE{2016A&A...588A..84D,
       author = {{Dhawan}, S. and {Leibundgut}, B. and {Spyromilio}, J. and {Blondin}, S.},
        title = "{A reddening-free method to estimate the $^{56}$Ni mass of Type Ia supernovae}",
      journal = {\aap},
         year = 2016,
        month = apr,
       volume = {588},
          eid = {A84},
        pages = {A84},
          doi = {10.1051/0004-6361/201527201},
archivePrefix = {arXiv},
       eprint = {1601.04874},
 primaryClass = {astro-ph.SR},
       adsurl = {https://ui.adsabs.harvard.edu/abs/2016A&A...588A..84D}
}

@article{Kasen2006,
  title = {Secondary Maximum in the Near‐Infrared Light Curves of Type Ia Supernovae},
  volume = {649},
  ISSN = {1538-4357},
  url = {http://dx.doi.org/10.1086/506588},
  DOI = {10.1086/506588},
  number = {2},
  journal = {The Astrophysical Journal},
  publisher = {American Astronomical Society},
  author = {Kasen,  Daniel},
  year = {2006},
  month = oct,
  pages = {939–953}
}

@article{Deckers2025,
  title = {ZTF SN Ia DR2: Secondary maximum in type Ia supernovae},
  volume = {694},
  ISSN = {1432-0746},
  url = {http://dx.doi.org/10.1051/0004-6361/202450379},
  DOI = {10.1051/0004-6361/202450379},
  journal = {Astronomy \& Astrophysics},
  publisher = {EDP Sciences},
  author = {Deckers,  M. and Maguire,  K. and Shingles,  L. and Dimitriadis,  G. and Rigault,  M. and Smith,  M. and Goobar,  A. and Nordin,  J. and Johansson,  J. and Amenouche,  M. and Burgaz,  U. and Dhawan,  S. and Ginolin,  M. and Harvey,  L. and Kenworthy,  W. D. and Kim,  Y.-L. and Laher,  R. R. and Luo,  N. and Kulkarni,  S. R. and Masci,  F. J. and Galbany,  L. and M\"{u}ller-Bravo,  T. E. and Nugent,  P. E. and Pletskova,  N. and Purdum,  J. and Racine,  B. and Sollerman,  J. and Terwel,  J. H.},
  year = {2025},
  month = feb,
  pages = {A12}
}

@ARTICLE{2012PASP..124..668Y,
       author = {{Yaron}, Ofer and {Gal-Yam}, Avishay},
        title = "{WISeREP{\textemdash}An Interactive Supernova Data Repository}",
      journal = {\pasp},
         year = 2012,
        month = jul,
       volume = {124},
       number = {917},
        pages = {668},
          doi = {10.1086/666656},
archivePrefix = {arXiv},
       eprint = {1204.1891},
 primaryClass = {astro-ph.IM},
       adsurl = {https://ui.adsabs.harvard.edu/abs/2012PASP..124..668Y}
}

@article{Savitzky1964,
  title = {Smoothing and Differentiation of Data by Simplified Least Squares Procedures.},
  volume = {36},
  ISSN = {1520-6882},
  url = {http://dx.doi.org/10.1021/ac60214a047},
  DOI = {10.1021/ac60214a047},
  number = {8},
  journal = {Analytical Chemistry},
  publisher = {American Chemical Society (ACS)},
  author = {Savitzky,  Abraham. and Golay,  M. J. E.},
  year = {1964},
  month = jul,
  pages = {1627–1639}
}

@article{Blondin2012,
  title = {THE SPECTROSCOPIC DIVERSITY OF TYPE Ia SUPERNOVAE},
  volume = {143},
  ISSN = {1538-3881},
  url = {http://dx.doi.org/10.1088/0004-6256/143/5/126},
  DOI = {10.1088/0004-6256/143/5/126},
  number = {5},
  journal = {The Astronomical Journal},
  publisher = {American Astronomical Society},
  author = {Blondin,  S. and Matheson,  T. and Kirshner,  R. P. and Mandel,  K. S. and Berlind,  P. and Calkins,  M. and Challis,  P. and Garnavich,  P. M. and Jha,  S. W. and Modjaz,  M. and Riess,  A. G. and Schmidt,  B. P.},
  year = {2012},
  month = apr,
  pages = {126}
}

@ARTICLE{2006PASP..118..560B,
       author = {{Branch}, David and {Dang}, Leeann Chau and {Hall}, Nicholas and {Ketchum}, Wesley and {Melakayil}, Mercy and {Parrent}, Jerod and {Troxel}, M.~A. and {Casebeer}, D. and {Jeffery}, David J. and {Baron}, E.},
        title = "{Comparative Direct Analysis of Type Ia Supernova Spectra. II. Maximum Light}",
      journal = {\pasp},
         year = 2006,
        month = apr,
       volume = {118},
       number = {842},
        pages = {560-571},
          doi = {10.1086/502778},
archivePrefix = {arXiv},
       eprint = {astro-ph/0601048},
 primaryClass = {astro-ph},
       adsurl = {https://ui.adsabs.harvard.edu/abs/2006PASP..118..560B}
}

@article{Burgaz2025,
  title = {ZTF SN Ia DR2: The spectral diversity of Type Ia supernovae in a volume-limited sample},
  volume = {694},
  ISSN = {1432-0746},
  url = {http://dx.doi.org/10.1051/0004-6361/202450386},
  DOI = {10.1051/0004-6361/202450386},
  journal = {Astronomy \& Astrophysics},
  publisher = {EDP Sciences},
  author = {Burgaz,  U. and Maguire,  K. and Dimitriadis,  G. and Harvey,  L. and Senzel,  R. and Sollerman,  J. and Nordin,  J. and Galbany,  L. and Rigault,  M. and Smith,  M. and Goobar,  A. and Johansson,  J. and Rosnet,  P. and Alburai,  A. and Amenouche,  M. and Deckers,  M. and Dhawan,  S. and Ginolin,  M. and Kim,  Y.-L. and Miller,  A. A. and Muller-Bravo,  T. E. and Nugent,  P. E. and Terwel,  J. H. and Dekany,  R. and Drake,  A. and Graham,  M. J. and Groom,  S. L. and Kasliwal,  M. M. and Kulkarni,  S. R. and Nolan,  K. and Nir,  G. and Riddle,  R. L. and Rusholme,  B. and Sharma,  Y.},
  year = {2025},
  month = feb,
  pages = {A9}
}

@article{Pereira2013,
  title = {Spectrophotometric time series of SN 2011fe from the Nearby Supernova Factory},
  volume = {554},
  ISSN = {1432-0746},
  url = {http://dx.doi.org/10.1051/0004-6361/201221008},
  DOI = {10.1051/0004-6361/201221008},
  journal = {Astronomy \& Astrophysics},
  publisher = {EDP Sciences},
  author = {Pereira,  R. and Thomas,  R. C. and Aldering,  G. and Antilogus,  P. and Baltay,  C. and Benitez-Herrera,  S. and Bongard,  S. and Buton,  C. and Canto,  A. and Cellier-Holzem,  F. and Chen,  J. and Childress,  M. and Chotard,  N. and Copin,  Y. and Fakhouri,  H. K. and Fink,  M. and Fouchez,  D. and Gangler,  E. and Guy,  J. and Hillebrandt,  W. and Hsiao,  E. Y. and Kerschhaggl,  M. and Kowalski,  M. and Kromer,  M. and Nordin,  J. and Nugent,  P. and Paech,  K. and Pain,  R. and Pécontal,  E. and Perlmutter,  S. and Rabinowitz,  D. and Rigault,  M. and Runge,  K. and Saunders,  C. and Smadja,  G. and Tao,  C. and Taubenberger,  S. and Tilquin,  A. and Wu,  C.},
  year = {2013},
  month = jun,
  pages = {A27}
}

@article{Zhao2020,
  title = {A study of Si <scp>ii</scp> and S <scp>ii</scp> features in spectra of Type Ia supernovae},
  volume = {503},
  ISSN = {1365-2966},
  url = {http://dx.doi.org/10.1093/mnras/staa3985},
  DOI = {10.1093/mnras/staa3985},
  number = {4},
  journal = {Monthly Notices of the Royal Astronomical Society},
  publisher = {Oxford University Press (OUP)},
  author = {Zhao,  Xulin and Maeda,  Keiichi and Wang,  Xiaofeng and Sai,  Hanna},
  year = {2020},
  month = apr,
  pages = {4667–4680}
}

@article{Silverman2012,
  title = {Berkeley Supernova Ia Program - II. Initial analysis of spectra obtained near maximum brightness: BSNIP II: initial spectral analysis},
  volume = {425},
  ISSN = {0035-8711},
  url = {http://dx.doi.org/10.1111/j.1365-2966.2012.21269.x},
  DOI = {10.1111/j.1365-2966.2012.21269.x},
  number = {3},
  journal = {Monthly Notices of the Royal Astronomical Society},
  publisher = {Oxford University Press (OUP)},
  author = {Silverman,  Jeffrey M. and Kong,  Jason J. and Filippenko,  Alexei V.},
  year = {2012},
  month = aug,
  pages = {1819–1888}
}

@ARTICLE{1992ApJ...384L..15F,
       author = {{Filippenko}, Alexei V. and {Richmond}, Michael W. and {Matheson}, Thomas and {Shields}, Joseph C. and {Burbidge}, E. Margaret and {Cohen}, Ross D. and {Dickinson}, Mark and {Malkan}, Matthew A. and {Nelson}, Brant and {Pietz}, Jochen and {Schlegel}, David and {Schmeer}, Patrick and {Spinrad}, Hyron and {Steidel}, Charles C. and {Tran}, Hien D. and {Wren}, William},
        title = "{The Peculiar Type IA SN 1991T: Detonation of a White Dwarf?}",
      journal = {\apjl},
         year = 1992,
        month = jan,
       volume = {384},
        pages = {L15},
          doi = {10.1086/186252},
       adsurl = {https://ui.adsabs.harvard.edu/abs/1992ApJ...384L..15F}
}

@ARTICLE{1992AJ....104.1543F,
       author = {{Filippenko}, Alexei V. and {Richmond}, Michael W. and {Branch}, David and {Gaskell}, Martin and {Herbst}, William and {Ford}, Charles H. and {Treffers}, Richard R. and {Matheson}, Thomas and {Ho}, Luis C. and {Dey}, Arjun and {Sargent}, Wallace L.~W. and {Small}, Todd A. and {van Breugel}, Wil J.~M.},
        title = "{The Subluminous, Spectroscopically Peculiar Type 1a Supernova 1991bg in the Elliptical Galaxy NGC 4374}",
      journal = {\aj},
         year = 1992,
        month = oct,
       volume = {104},
        pages = {1543},
          doi = {10.1086/116339},
       adsurl = {https://ui.adsabs.harvard.edu/abs/1992AJ....104.1543F}
}

@article{Krisciunas2000,
  title = {Uniformity of (V−Near‐Infrared) Color Evolution of Type Ia Supernovae and Implications for Host Galaxy Extinction Determination},
  volume = {539},
  ISSN = {1538-4357},
  url = {http://dx.doi.org/10.1086/309263},
  DOI = {10.1086/309263},
  number = {2},
  journal = {The Astrophysical Journal},
  publisher = {American Astronomical Society},
  author = {Krisciunas,  Kevin and Hastings,  N. C. and Loomis,  Karen and McMillan,  Russet and Rest,  Armin and Riess,  Adam G. and Stubbs,  Christopher},
  year = {2000},
  month = aug,
  pages = {658–674}
}

@article{Garavini2004,
  title = {Spectroscopic Observations and Analysis of the Peculiar SN 1999aa},
  volume = {128},
  ISSN = {1538-3881},
  url = {http://dx.doi.org/10.1086/421747},
  DOI = {10.1086/421747},
  number = {1},
  journal = {The Astronomical Journal},
  publisher = {American Astronomical Society},
  author = {Garavini,  G. and Folatelli,  G. and Goobar,  A. and Nobili,  S. and Aldering,  G. and Amadon,  A. and Amanullah,  R. and Astier,  P. and Balland,  C. and Blanc,  G. and Burns,  M. S. and Conley,  A. and Dahlén,  T. and Deustua,  S. E. and Ellis,  R. and Fabbro,  S. and Fan,  X. and Frye,  B. and Gates,  E. L. and Gibbons,  R. and Goldhaber,  G. and Goldman,  B. and Groom,  D. E. and Haissinski,  J. and Hardin,  D. and Hook,  I. M. and Howell,  D. A. and Kasen,  D. and Kent,  S. and Kim,  A. G. and Knop,  R. A. and Lee,  B. C. and Lidman,  C. and Mendez,  J. and Miller,  G. J. and Moniez,  M. and Mourão,  A. and Newberg,  H. and Nugent,  P. E. and Pain,  R. and Perdereau,  O. and Perlmutter,  S. and Prasad,  V. and Quimby,  R. and Raux,  J. and Regnault,  N. and Rich,  J. and Richards,  G. T. and Ruiz-Lapuente,  P. and Sainton,  G. and Schaefer,  B. E. and Schahmaneche,  K. and Smith,  E. and Spadafora,  A. L. and Stanishev,  V. and Walton,  N. A. and Wang,  L. and Wood-Vasey,  W. M.},
  year = {2004},
  month = jul,
  pages = {387–404}
}

@article{Jha2006,
  title = {UBVRILight Curves of 44 Type Ia Supernovae},
  volume = {131},
  ISSN = {1538-3881},
  url = {http://dx.doi.org/10.1086/497989},
  DOI = {10.1086/497989},
  number = {1},
  journal = {The Astronomical Journal},
  publisher = {American Astronomical Society},
  author = {Jha,  Saurabh and Kirshner,  Robert P. and Challis,  Peter and Garnavich,  Peter M. and Matheson,  Thomas and Soderberg,  Alicia M. and Graves,  Genevieve J. M. and Hicken,  Malcolm and Alves,  João F. and Arce,  Héctor G. and Balog,  Zoltan and Barmby,  Pauline and Barton,  Elizabeth J. and Berlind,  Perry and Bragg,  Ann E. and Briceño,  César and Brown,  Warren R. and Buckley,  James H. and Caldwell,  Nelson and Calkins,  Michael L. and Carter,  Barbara J. and Concannon,  Kristi Dendy and Donnelly,  R. Hank and Eriksen,  Kristoffer A. and Fabricant,  Daniel G. and Falco,  Emilio E. and Fiore,  Fabrizio and Garcia,  Michael R. and Gómez,  Mercedes and Grogin,  Norman A. and Groner,  Ted and Groot,  Paul J. and Haisch,  Jr.,  Karl E. and Hartmann,  Lee and Hergenrother,  Carl W. and Holman,  Matthew J. and Huchra,  John P. and Jayawardhana,  Ray and Jerius,  Diab and Kannappan,  Sheila J. and Kim,  Dong-Woo and Kleyna,  Jan T. and Kochanek,  Christopher S. and Koranyi,  Daniel M. and Krockenberger,  Martin and Lada,  Charles J. and Luhman,  Kevin L. and Luu,  Jane X. and Macri,  Lucas M. and Mader,  Jeff A. and Mahdavi,  Andisheh and Marengo,  Massimo and Marsden,  Brian G. and McLeod,  Brian A. and McNamara,  Brian R. and Megeath,  S. Thomas and Moraru,  Dan and Mossman,  Amy E. and Muench,  August A. and Muñoz,  Jose A. and Muzerolle,  James and Naranjo,  Orlando and Nelson-Patel,  Kristin and Pahre,  Michael A. and Patten,  Brian M. and Peters,  James and Peters,  Wayne and Raymond,  John C. and Rines,  Kenneth and Schild,  Rudolph E. and Sobczak,  Gregory J. and Spahr,  Timothy B. and Stauffer,  John R. and Stefanik,  Robert P. and Szentgyorgyi,  Andrew H. and Tollestrup,  Eric V. and V\"{a}is\"{a}nen,  Petri and Vikhlinin,  Alexey and Wang,  Zhong and Willner,  S. P. and Wolk,  Scott J. and Zajac,  Joseph M. and Zhao,  Ping and Stanek,  Krzysztof Z.},
  year = {2006},
  month = jan,
  pages = {527–554}
}

@article{Matheson2008,
  title = {OPTICAL SPECTROSCOPY OF TYPE Ia SUPERNOVAE},
  volume = {135},
  ISSN = {1538-3881},
  url = {http://dx.doi.org/10.1088/0004-6256/135/4/1598},
  DOI = {10.1088/0004-6256/135/4/1598},
  number = {4},
  journal = {The Astronomical Journal},
  publisher = {American Astronomical Society},
  author = {Matheson,  T. and Kirshner,  R. P. and Challis,  P. and Jha,  S. and Garnavich,  P. M. and Berlind,  P. and Calkins,  M. L. and Blondin,  S. and Balog,  Z. and Bragg,  A. E. and Caldwell,  N. and Concannon,  K. Dendy and Falco,  E. E. and Graves,  G. J. M. and Huchra,  J. P. and Kuraszkiewicz,  J. and Mader,  J. A. and Mahdavi,  A. and Phelps,  M. and Rines,  K. and Song,  I. and Wilkes,  B. J.},
  year = {2008},
  month = mar,
  pages = {1598–1615}
}

@article{AndrewHowell2006,
  title = {The type Ia supernova SNLS-03D3bb from a super-Chandrasekhar-mass white dwarf star},
  volume = {443},
  ISSN = {1476-4687},
  url = {http://dx.doi.org/10.1038/nature05103},
  DOI = {10.1038/nature05103},
  number = {7109},
  journal = {Nature},
  publisher = {Springer Science and Business Media LLC},
  author = {Andrew Howell,  D. and Sullivan,  Mark and Nugent,  Peter E. and Ellis,  Richard S. and Conley,  Alexander J. and Le Borgne,  Damien and Carlberg,  Raymond G. and Guy,  Julien and Balam,  David and Basa,  Stephane and Fouchez,  Dominique and Hook,  Isobel M. and Hsiao,  Eric Y. and Neill,  James D. and Pain,  Reynald and Perrett,  Kathryn M. and Pritchet,  Christopher J.},
  year = {2006},
  month = sep,
  pages = {308–311}
}

@article{Chakradhari2014,
  title = {Supernova SN 2012dn: a spectroscopic clone of SN 2006gz},
  volume = {443},
  ISSN = {1365-2966},
  url = {http://dx.doi.org/10.1093/mnras/stu1258},
  DOI = {10.1093/mnras/stu1258},
  number = {2},
  journal = {Monthly Notices of the Royal Astronomical Society},
  publisher = {Oxford University Press (OUP)},
  author = {Chakradhari,  N. K. and Sahu,  D. K. and Srivastav,  S. and Anupama,  G. C.},
  year = {2014},
  month = jul,
  pages = {1663–1679}
}

@article{Jha2006_02cx,
  title = {Late-Time Spectroscopy of SN 2002cx: The Prototype of a New Subclass of Type Ia Supernovae},
  volume = {132},
  ISSN = {1538-3881},
  url = {http://dx.doi.org/10.1086/504599},
  DOI = {10.1086/504599},
  number = {1},
  journal = {The Astronomical Journal},
  publisher = {American Astronomical Society},
  author = {Jha,  Saurabh and Branch,  David and Chornock,  Ryan and Foley,  Ryan J. and Li,  Weidong and Swift,  Brandon J. and Casebeer,  Darrin and Filippenko,  Alexei V.},
  year = {2006},
  month = jun,
  pages = {189–196}
}

@article{Hamuy2003,
  title = {An asymptotic-giant-branch star in the progenitor system of a type Ia supernova},
  volume = {424},
  ISSN = {1476-4687},
  url = {http://dx.doi.org/10.1038/nature01854},
  DOI = {10.1038/nature01854},
  number = {6949},
  journal = {Nature},
  publisher = {Springer Science and Business Media LLC},
  author = {Hamuy,  Mario and Phillips,  M. M. and Suntzeff,  Nicholas B. and Maza,  José and González,  L. E. and Roth,  Miguel and Krisciunas,  Kevin and Morrell,  Nidia and Green,  E. M. and Persson,  S. E. and McCarthy,  P. J.},
  year = {2003},
  month = aug,
  pages = {651–654}
}

@article{Silverman2013,
  title = {TYPE Ia SUPERNOVAE STRONGLY INTERACTING WITH THEIR CIRCUMSTELLAR MEDIUM},
  volume = {207},
  ISSN = {1538-4365},
  url = {http://dx.doi.org/10.1088/0067-0049/207/1/3},
  DOI = {10.1088/0067-0049/207/1/3},
  number = {1},
  journal = {The Astrophysical Journal Supplement Series},
  publisher = {American Astronomical Society},
  author = {Silverman,  Jeffrey M. and Nugent,  Peter E. and Gal-Yam,  Avishay and Sullivan,  Mark and Howell,  D. Andrew and Filippenko,  Alexei V. and Arcavi,  Iair and Ben-Ami,  Sagi and Bloom,  Joshua S. and Cenko,  S. Bradley and Cao,  Yi and Chornock,  Ryan and Clubb,  Kelsey I. and Coil,  Alison L. and Foley,  Ryan J. and Graham,  Melissa L. and Griffith,  Christopher V. and Horesh,  Assaf and Kasliwal,  Mansi M. and Kulkarni,  Shrinivas R. and Leonard,  Douglas C. and Li,  Weidong and Matheson,  Thomas and Miller,  Adam A. and Modjaz,  Maryam and Ofek,  Eran O. and Pan,  Yen-Chen and Perley,  Daniel A. and Poznanski,  Dovi and Quimby,  Robert M. and Steele,  Thea N. and Sternberg,  Assaf and Xu,  Dong and Yaron,  Ofer},
  year = {2013},
  month = jun,
  pages = {3}
}

@article{Müller-Bravo2022, 
  author = {Tomás E. Müller-Bravo and Lluís Galbany},
  title = {HostPhot: global and local photometry of galaxies hosting supernovae or other transients},
  doi = {10.21105/joss.04508}, 
  url = {https://doi.org/10.21105/joss.04508}, 
  year = {2022}, 
  publisher = {The Open Journal}, 
  volume = {7}, 
  number = {76}, 
  pages = {4508},  
  journal = {Journal of Open Source Software} 
}

@ARTICLE{2010AJ....140.1868W,
       author = {{Wright}, Edward L. and {Eisenhardt}, Peter R.~M. and {Mainzer}, Amy K. and {Ressler}, Michael E. and {Cutri}, Roc M. and {Jarrett}, Thomas and {Kirkpatrick}, J. Davy and {Padgett}, Deborah and {McMillan}, Robert S. and {Skrutskie}, Michael and {Stanford}, S.~A. and {Cohen}, Martin and {Walker}, Russell G. and {Mather}, John C. and {Leisawitz}, David and {Gautier}, III, Thomas N. and {McLean}, Ian and {Benford}, Dominic and {Lonsdale}, Carol J. and {Blain}, Andrew and {Mendez}, Bryan and {Irace}, William R. and {Duval}, Valerie and {Liu}, Fengchuan and {Royer}, Don and {Heinrichsen}, Ingolf and {Howard}, Joan and {Shannon}, Mark and {Kendall}, Martha and {Walsh}, Amy L. and {Larsen}, Mark and {Cardon}, Joel G. and {Schick}, Scott and {Schwalm}, Mark and {Abid}, Mohamed and {Fabinsky}, Beth and {Naes}, Larry and {Tsai}, Chao-Wei},
        title = "{The Wide-field Infrared Survey Explorer (WISE): Mission Description and Initial On-orbit Performance}",
      journal = {\aj},
         year = 2010,
        month = dec,
       volume = {140},
       number = {6},
        pages = {1868-1881},
          doi = {10.1088/0004-6256/140/6/1868},
archivePrefix = {arXiv},
       eprint = {1008.0031},
 primaryClass = {astro-ph.IM},
       adsurl = {https://ui.adsabs.harvard.edu/abs/2010AJ....140.1868W}
}

@ARTICLE{2006AJ....131.1163S,
       author = {{Skrutskie}, M.~F. and {Cutri}, R.~M. and {Stiening}, R. and {Weinberg}, M.~D. and {Schneider}, S. and {Carpenter}, J.~M. and {Beichman}, C. and {Capps}, R. and {Chester}, T. and {Elias}, J. and {Huchra}, J. and {Liebert}, J. and {Lonsdale}, C. and {Monet}, D.~G. and {Price}, S. and {Seitzer}, P. and {Jarrett}, T. and {Kirkpatrick}, J.~D. and {Gizis}, J.~E. and {Howard}, E. and {Evans}, T. and {Fowler}, J. and {Fullmer}, L. and {Hurt}, R. and {Light}, R. and {Kopan}, E.~L. and {Marsh}, K.~A. and {McCallon}, H.~L. and {Tam}, R. and {Van Dyk}, S. and {Wheelock}, S.},
        title = "{The Two Micron All Sky Survey (2MASS)}",
      journal = {\aj},
         year = 2006,
        month = feb,
       volume = {131},
       number = {2},
        pages = {1163-1183},
          doi = {10.1086/498708},
       adsurl = {https://ui.adsabs.harvard.edu/abs/2006AJ....131.1163S}
}

@article{Martin2005,
  title = {TheGalaxy Evolution Explorer: A Space Ultraviolet Survey Mission},
  volume = {619},
  ISSN = {1538-4357},
  url = {http://dx.doi.org/10.1086/426387},
  DOI = {10.1086/426387},
  number = {1},
  journal = {The Astrophysical Journal},
  publisher = {American Astronomical Society},
  author = {Martin,  D. Christopher and Fanson,  James and Schiminovich,  David and Morrissey,  Patrick and Friedman,  Peter G. and Barlow,  Tom A. and Conrow,  Tim and Grange,  Robert and Jelinsky,  Patrick N. and Milliard,  Bruno and Siegmund,  Oswald H. W. and Bianchi,  Luciana and Byun,  Yong-Ik and Donas,  Jose and Forster,  Karl and Heckman,  Timothy M. and Lee,  Young-Wook and Madore,  Barry F. and Malina,  Roger F. and Neff,  Susan G. and Rich,  R. Michael and Small,  Todd and Surber,  Frank and Szalay,  Alex S. and Welsh,  Barry and Wyder,  Ted K.},
  year = {2005},
  month = jan,
  pages = {L1–L6}
}

@article{Abdurrouf2022,
  title = {The Seventeenth Data Release of the Sloan Digital Sky Surveys: Complete Release of MaNGA,  MaStar,  and APOGEE-2 Data},
  volume = {259},
  ISSN = {1538-4365},
  url = {http://dx.doi.org/10.3847/1538-4365/ac4414},
  DOI = {10.3847/1538-4365/ac4414},
  number = {2},
  journal = {The Astrophysical Journal Supplement Series},
  publisher = {American Astronomical Society},
  author = {Abdurro’uf and Accetta,  Katherine and Aerts,  Conny and Silva Aguirre,  Víctor and Ahumada,  Romina and Ajgaonkar,  Nikhil and Filiz Ak,  N. and Alam,  Shadab and Allende Prieto,  Carlos and Almeida,  Andrés and Anders,  Friedrich and Anderson,  Scott F. and Andrews,  Brett H. and Anguiano,  Borja and Aquino-Ortíz,  Erik and Aragón-Salamanca,  Alfonso and Argudo-Fernández,  Maria and Ata,  Metin and Aubert,  Marie and Avila-Reese,  Vladimir and Badenes,  Carles and Barbá,  Rodolfo H. and Barger,  Kat and Barrera-Ballesteros,  Jorge K. and Beaton,  Rachael L. and Beers,  Timothy C. and Belfiore,  Francesco and Bender,  Chad F. and Bernardi,  Mariangela and Bershady,  Matthew A. and Beutler,  Florian and Bidin,  Christian Moni and Bird,  Jonathan C. and Bizyaev,  Dmitry and Blanc,  Guillermo A. and Blanton,  Michael R. and Boardman,  Nicholas Fraser and Bolton,  Adam S. and Boquien,  Médéric and Borissova,  Jura and Bovy,  Jo and Brandt,  W. N. and Brown,  Jordan and Brownstein,  Joel R. and Brusa,  Marcella and Buchner,  Johannes and Bundy,  Kevin and Burchett,  Joseph N. and Bureau,  Martin and Burgasser,  Adam and Cabang,  Tuesday K. and Campbell,  Stephanie and Cappellari,  Michele and Carlberg,  Joleen K. and Wanderley,  Fábio Carneiro and Carrera,  Ricardo and Cash,  Jennifer and Chen,  Yan-Ping and Chen,  Wei-Huai and Cherinka,  Brian and Chiappini,  Cristina and Choi,  Peter Doohyun and Chojnowski,  S. Drew and Chung,  Haeun and Clerc,  Nicolas and Cohen,  Roger E. and Comerford,  Julia M. and Comparat,  Johan and da Costa,  Luiz and Covey,  Kevin and Crane,  Jeffrey D. and Cruz-Gonzalez,  Irene and Culhane,  Connor and Cunha,  Katia and Dai 戴,  Y. Sophia 昱 and Damke,  Guillermo and Darling,  Jeremy and Davidson Jr.,  James W. and Davies,  Roger and Dawson,  Kyle and De Lee,  Nathan and Diamond-Stanic,  Aleksandar M. and Cano-Díaz,  Mariana and Sánchez,  Helena Domínguez and Donor,  John and Duckworth,  Chris and Dwelly,  Tom and Eisenstein,  Daniel J. and Elsworth,  Yvonne P. and Emsellem,  Eric and Eracleous,  Mike and Escoffier,  Stephanie and Fan,  Xiaohui and Farr,  Emily and Feng,  Shuai and Fernández-Trincado,  José G. and Feuillet,  Diane and Filipp,  Andreas and Fillingham,  Sean P and Frinchaboy,  Peter M. and Fromenteau,  Sebastien and Galbany,  Lluís and García,  Rafael A. and García-Hernández,  D. A. and Ge,  Junqiang and Geisler,  Doug and Gelfand,  Joseph and Géron,  Tobias and Gibson,  Benjamin J. and Goddy,  Julian and Godoy-Rivera,  Diego and Grabowski,  Kathleen and Green,  Paul J. and Greener,  Michael and Grier,  Catherine J. and Griffith,  Emily and Guo,  Hong and Guy,  Julien and Hadjara,  Massinissa and Harding,  Paul and Hasselquist,  Sten and Hayes,  Christian R. and Hearty,  Fred and Hernández,  Jesús and Hill,  Lewis and Hogg,  David W. and Holtzman,  Jon A. and Horta,  Danny and Hsieh,  Bau-Ching and Hsu,  Chin-Hao and Hsu,  Yun-Hsin and Huber,  Daniel and Huertas-Company,  Marc and Hutchinson,  Brian and Hwang,  Ho Seong and Ibarra-Medel,  Héctor J. and Chitham,  Jacob Ider and Ilha,  Gabriele S. and Imig,  Julie and Jaekle,  Will and Jayasinghe,  Tharindu and Ji,  Xihan and Johnson,  Jennifer A. and Jones,  Amy and J\"{o}nsson,  Henrik and Katkov,  Ivan and Khalatyan,  Dr. Arman and Kinemuchi,  Karen and Kisku,  Shobhit and Knapen,  Johan H. and Kneib,  Jean-Paul and Kollmeier,  Juna A. and Kong,  Miranda and Kounkel,  Marina and Kreckel,  Kathryn and Krishnarao,  Dhanesh and Lacerna,  Ivan and Lane,  Richard R. and Langgin,  Rachel and Lavender,  Ramon and Law,  David R. and Lazarz,  Daniel and Leung,  Henry W. and Leung,  Ho-Hin and Lewis,  Hannah M. and Li,  Cheng and Li,  Ran and Lian,  Jianhui and Liang,  Fu-Heng and Lin 林,  Lihwai 俐 暉 and Lin,  Yen-Ting and Lin,  Sicheng and Lintott,  Chris and Long,  Dan and Longa-Peña,  Penélope and López-Cobá,  Carlos and Lu,  Shengdong and Lundgren,  Britt F. and Luo,  Yuanze and Mackereth,  J. Ted and de la Macorra,  Axel and Mahadevan,  Suvrath and Majewski,  Steven R. and Manchado,  Arturo and Mandeville,  Travis and Maraston,  Claudia and Margalef-Bentabol,  Berta and Masseron,  Thomas and Masters,  Karen L. and Mathur,  Savita and McDermid,  Richard M. and Mckay,  Myles and Merloni,  Andrea and Merrifield,  Michael and Meszaros,  Szabolcs and Miglio,  Andrea and Di Mille,  Francesco and Minniti,  Dante and Minsley,  Rebecca and Monachesi,  Antonela and Moon,  Jeongin and Mosser,  Benoit and Mulchaey,  John and Muna,  Demitri and Muñoz,  Ricardo R. and Myers,  Adam D. and Myers,  Natalie and Nadathur,  Seshadri and Nair,  Preethi and Nandra,  Kirpal and Neumann,  Justus and Newman,  Jeffrey A. and Nidever,  David L. and Nikakhtar,  Farnik and Nitschelm,  Christian and O’Connell,  Julia E. and Garma-Oehmichen,  Luis and Luan Souza de Oliveira,  Gabriel and Olney,  Richard and Oravetz,  Daniel and Ortigoza-Urdaneta,  Mario and Osorio,  Yeisson and Otter,  Justin and Pace,  Zachary J. and Padilla,  Nelson and Pan,  Kaike and Pan,  Hsi-An and Parikh,  Taniya and Parker,  James and Peirani,  Sebastien and Peña Ramírez,  Karla and Penny,  Samantha and Percival,  Will J. and Perez-Fournon,  Ismael and Pinsonneault,  Marc and Poidevin,  Frédérick and Poovelil,  Vijith Jacob and Price-Whelan,  Adrian M. and Bárbara de Andrade Queiroz,  Anna and Raddick,  M. Jordan and Ray,  Amy and Rembold,  Sandro Barboza and Riddle,  Nicole and Riffel,  Rogemar A. and Riffel,  Rogério and Rix,  Hans-Walter and Robin,  Annie C. and Rodríguez-Puebla,  Aldo and Roman-Lopes,  Alexandre and Román-Zúñiga,  Carlos and Rose,  Benjamin and Ross,  Ashley J. and Rossi,  Graziano and Rubin,  Kate H. R. and Salvato,  Mara and Sánchez,  Sebástian F. and Sánchez-Gallego,  José R. and Sanderson,  Robyn and Santana Rojas,  Felipe Antonio and Sarceno,  Edgar and Sarmiento,  Regina and Sayres,  Conor and Sazonova,  Elizaveta and Schaefer,  Adam L. and Schiavon,  Ricardo and Schlegel,  David J and Schneider,  Donald P. and Schultheis,  Mathias and Schwope,  Axel and Serenelli,  Aldo and Serna,  Javier and Shao,  Zhengyi and Shapiro,  Griffin and Sharma,  Anubhav and Shen,  Yue and Shetrone,  Matthew and Shu,  Yiping and Simon,  Joshua D. and Skrutskie,  M. F. and Smethurst,  Rebecca and Smith,  Verne and Sobeck,  Jennifer and Spoo,  Taylor and Sprague,  Dani and Stark,  David V. and Stassun,  Keivan G. and Steinmetz,  Matthias and Stello,  Dennis and Stone-Martinez,  Alexander and Storchi-Bergmann,  Thaisa and Stringfellow,  Guy S. and Stutz,  Amelia and Su,  Yung-Chau and Taghizadeh-Popp,  Manuchehr and Talbot,  Michael S. and Tayar,  Jamie and Telles,  Eduardo and Teske,  Johanna and Thakar,  Ani and Theissen,  Christopher and Tkachenko,  Andrew and Thomas,  Daniel and Tojeiro,  Rita and Hernandez Toledo,  Hector and Troup,  Nicholas W. and Trump,  Jonathan R. and Trussler,  James and Turner,  Jacqueline and Tuttle,  Sarah and Unda-Sanzana,  Eduardo and Vázquez-Mata,  José Antonio and Valentini,  Marica and Valenzuela,  Octavio and Vargas-González,  Jaime and Vargas-Magaña,  Mariana and Alfaro,  Pablo Vera and Villanova,  Sandro and Vincenzo,  Fiorenzo and Wake,  David and Warfield,  Jack T. and Washington,  Jessica Diane and Weaver,  Benjamin Alan and Weijmans,  Anne-Marie and Weinberg,  David H. and Weiss,  Achim and Westfall,  Kyle B. and Wild,  Vivienne and Wilde,  Matthew C. and Wilson,  John C. and Wilson,  Robert F. and Wilson,  Mikayla and Wolf,  Julien and Wood-Vasey,  W. M. and Yan 严,  Renbin 人斌 and Zamora,  Olga and Zasowski,  Gail and Zhang,  Kai and Zhao,  Cheng and Zheng,  Zheng and Zheng,  Zheng and Zhu,  Kai},
  year = {2022},
  month = mar,
  pages = {35}
}

@ARTICLE{2019A&A...622A.103B,
       author = {{Boquien}, M. and {Burgarella}, D. and {Roehlly}, Y. and {Buat}, V. and {Ciesla}, L. and {Corre}, D. and {Inoue}, A.~K. and {Salas}, H.},
        title = "{CIGALE: a python Code Investigating GALaxy Emission}",
      journal = {\aap},
         year = 2019,
        month = feb,
       volume = {622},
          eid = {A103},
        pages = {A103},
          doi = {10.1051/0004-6361/201834156},
archivePrefix = {arXiv},
       eprint = {1811.03094},
 primaryClass = {astro-ph.GA},
       adsurl = {https://ui.adsabs.harvard.edu/abs/2019A&A...622A.103B}
}

@article{Bruzual2003,
  title = {Stellar population synthesis at the resolution of 2003},
  volume = {344},
  ISSN = {1365-2966},
  url = {http://dx.doi.org/10.1046/j.1365-8711.2003.06897.x},
  DOI = {10.1046/j.1365-8711.2003.06897.x},
  number = {4},
  journal = {Monthly Notices of the Royal Astronomical Society},
  publisher = {Oxford University Press (OUP)},
  author = {Bruzual,  G. and Charlot,  S.},
  year = {2003},
  month = oct,
  pages = {1000–1028}
}

@article{Charlot2000,
  title = {A Simple Model for the Absorption of Starlight by Dust in Galaxies},
  volume = {539},
  ISSN = {1538-4357},
  url = {http://dx.doi.org/10.1086/309250},
  DOI = {10.1086/309250},
  number = {2},
  journal = {The Astrophysical Journal},
  publisher = {American Astronomical Society},
  author = {Charlot,  Stephane and Fall,  S. Michael},
  year = {2000},
  month = aug,
  pages = {718–731}
}

@ARTICLE{2014ApJ...780..172D,
       author = {{Draine}, B.~T. and {Aniano}, G. and {Krause}, Oliver and {Groves}, Brent and {Sandstrom}, Karin and {Braun}, Robert and {Leroy}, Adam and {Klaas}, Ulrich and {Linz}, Hendrik and {Rix}, Hans-Walter and {Schinnerer}, Eva and {Schmiedeke}, Anika and {Walter}, Fabian},
        title = "{Andromeda's Dust}",
      journal = {\apj},
         year = 2014,
        month = jan,
       volume = {780},
       number = {2},
          eid = {172},
        pages = {172},
          doi = {10.1088/0004-637X/780/2/172},
archivePrefix = {arXiv},
       eprint = {1306.2304},
 primaryClass = {astro-ph.CO},
       adsurl = {https://ui.adsabs.harvard.edu/abs/2014ApJ...780..172D}
}

@ARTICLE{2012MNRAS.420.2756S,
       author = {{Stalevski}, Marko and {Fritz}, Jacopo and {Baes}, Maarten and {Nakos}, Theodoros and {Popovi{\'c}}, Luka {\v{C}}.},
        title = "{3D radiative transfer modelling of the dusty tori around active galactic nuclei as a clumpy two-phase medium}",
      journal = {\mnras},
         year = 2012,
        month = mar,
       volume = {420},
       number = {4},
        pages = {2756-2772},
          doi = {10.1111/j.1365-2966.2011.19775.x},
archivePrefix = {arXiv},
       eprint = {1109.1286},
 primaryClass = {astro-ph.CO},
       adsurl = {https://ui.adsabs.harvard.edu/abs/2012MNRAS.420.2756S}
}

@article{Stalevski2016,
  title = {The dust covering factor in active galactic nuclei},
  volume = {458},
  ISSN = {1365-2966},
  url = {http://dx.doi.org/10.1093/mnras/stw444},
  DOI = {10.1093/mnras/stw444},
  number = {3},
  journal = {Monthly Notices of the Royal Astronomical Society},
  publisher = {Oxford University Press (OUP)},
  author = {Stalevski,  Marko and Ricci,  Claudio and Ueda,  Yoshihiro and Lira,  Paulina and Fritz,  Jacopo and Baes,  Maarten},
  year = {2016},
  month = mar,
  pages = {2288–2302}
}

@ARTICLE{2024TNSTR4922....1T,
       author = {{Tonry}, J. and {Denneau}, L. and {Weiland}, H. and {Siverd}, R. and {Erasmus}, N. and {Koorts}, W. and {Jordan}, A. and {Suc}, V. and {Smartt}, S.~J. and {Smith}, K.~W. and {Young}, D.~R. and {Nicholl}, M. and {Fulton}, M. and {McCollum}, M. and {Moore}, T. and {Weston}, J. and {Sheng}, X. and {Angus}, C.~R. and {Wilson}, A. and {Aamer}, A. and {Magill}, D. and {Ramsden}, P. and {Shingles}, L. and {Srivastav}, S. and {Gillanders}, J.~H. and {Stevance}, H. and {Cooper}, A.~J. and {Stoppa}, F. and {Rhodes}, L. and {Rest}, A. and {Chen}, T.~W. and {Stubbs}, C. and {Sommer}, J.~S. and {Schmidt}, B.~P.},
        title = "{ATLAS Transient Discovery Report for 2024-12-16}",
      journal = {Transient Name Server Discovery Report},
         year = 2024,
        month = dec,
       volume = {2024-4922},
        pages = {1},
       adsurl = {https://ui.adsabs.harvard.edu/abs/2024TNSTR4922....1T}
}

@ARTICLE{1999AJ....118.2675R,
       author = {{Riess}, Adam G. and {Filippenko}, Alexei V. and {Li}, Weidong and {Treffers}, Richard R. and {Schmidt}, Brian P. and {Qiu}, Yulei and {Hu}, Jingyao and {Armstrong}, Mark and {Faranda}, Chuck and {Thouvenot}, Eric and {Buil}, Christian},
        title = "{The Rise Time of Nearby Type IA Supernovae}",
      journal = {\aj},
         year = 1999,
        month = dec,
       volume = {118},
       number = {6},
        pages = {2675-2688},
          doi = {10.1086/301143},
archivePrefix = {arXiv},
       eprint = {astro-ph/9907037},
 primaryClass = {astro-ph},
       adsurl = {https://ui.adsabs.harvard.edu/abs/1999AJ....118.2675R}
}

@article{Jiang2017,
  title = {A hybrid type Ia supernova with an early flash triggered by helium-shell detonation},
  volume = {550},
  ISSN = {1476-4687},
  url = {http://dx.doi.org/10.1038/nature23908},
  DOI = {10.1038/nature23908},
  number = {7674},
  journal = {Nature},
  publisher = {Springer Science and Business Media LLC},
  author = {Jiang,  Ji-an and Doi,  Mamoru and Maeda,  Keiichi and Shigeyama,  Toshikazu and Nomoto,  Ken’ichi and Yasuda,  Naoki and Jha,  Saurabh W. and Tanaka,  Masaomi and Morokuma,  Tomoki and Tominaga,  Nozomu and Ivezić,  Željko and Ruiz-Lapuente,  Pilar and Stritzinger,  Maximilian D. and Mazzali,  Paolo A. and Ashall,  Christopher and Mould,  Jeremy and Baade,  Dietrich and Suzuki,  Nao and Connolly,  Andrew J. and Patat,  Ferdinando and Wang,  Lifan and Yoachim,  Peter and Jones,  David and Furusawa,  Hisanori and Miyazaki,  Satoshi},
  year = {2017},
  month = oct,
  pages = {80–83}
}

@misc{https://doi.org/10.48550/arxiv.1706.09879,
  doi = {10.48550/ARXIV.1706.09879},
  url = {https://arxiv.org/abs/1706.09879},
  author = {Kromer,  Markus and Ohlmann,  Sebastian T. and Roepke,  Friedrich K.},
  title = {Simulating the observed diversity of Type Ia supernovae - Introducing a model data base},
  publisher = {arXiv},
  year = {2017},
  copyright = {arXiv.org perpetual,  non-exclusive license}
}

@ARTICLE{2013MNRAS.429.1156S,
       author = {{Seitenzahl}, Ivo R. and {Ciaraldi-Schoolmann}, Franco and {R{\"o}pke}, Friedrich K. and {Fink}, Michael and {Hillebrandt}, Wolfgang and {Kromer}, Markus and {Pakmor}, R{\"u}diger and {Ruiter}, Ashley J. and {Sim}, Stuart A. and {Taubenberger}, Stefan},
        title = "{Three-dimensional delayed-detonation models with nucleosynthesis for Type Ia supernovae}",
      journal = {\mnras},
         year = 2013,
        month = feb,
       volume = {429},
       number = {2},
        pages = {1156-1172},
          doi = {10.1093/mnras/sts402},
archivePrefix = {arXiv},
       eprint = {1211.3015},
 primaryClass = {astro-ph.SR},
       adsurl = {https://ui.adsabs.harvard.edu/abs/2013MNRAS.429.1156S}
}

@ARTICLE{2013MNRAS.436..333S,
       author = {{Sim}, S.~A. and {Seitenzahl}, I.~R. and {Kromer}, M. and {Ciaraldi-Schoolmann}, F. and {R{\"o}pke}, F.~K. and {Fink}, M. and {Hillebrandt}, W. and {Pakmor}, R. and {Ruiter}, A.~J. and {Taubenberger}, S.},
        title = "{Synthetic light curves and spectra for three-dimensional delayed-detonation models of Type Ia supernovae}",
      journal = {\mnras},
         year = 2013,
        month = nov,
       volume = {436},
       number = {1},
        pages = {333-347},
          doi = {10.1093/mnras/stt1574},
archivePrefix = {arXiv},
       eprint = {1308.4833},
 primaryClass = {astro-ph.HE},
       adsurl = {https://ui.adsabs.harvard.edu/abs/2013MNRAS.436..333S}
}

@ARTICLE{2021A&A...649A.155G,
       author = {{Gronow}, Sabrina and {Collins}, Christine E. and {Sim}, Stuart A. and {R{\"o}pke}, Friedrich K.},
        title = "{Double detonations of sub-M$_{Ch}$ CO white dwarfs: variations in Type Ia supernovae due to different core and He shell masses}",
      journal = {\aap},
         year = 2021,
        month = may,
       volume = {649},
          eid = {A155},
        pages = {A155},
          doi = {10.1051/0004-6361/202039954},
archivePrefix = {arXiv},
       eprint = {2102.06719},
 primaryClass = {astro-ph.SR},
       adsurl = {https://ui.adsabs.harvard.edu/abs/2021A&A...649A.155G}
}

@ARTICLE{2010ApJ...719.1067K,
       author = {{Kromer}, M. and {Sim}, S.~A. and {Fink}, M. and {R{\"o}pke}, F.~K. and {Seitenzahl}, I.~R. and {Hillebrandt}, W.},
        title = "{Double-detonation Sub-Chandrasekhar Supernovae: Synthetic Observables for Minimum Helium Shell Mass Models}",
      journal = {\apj},
         year = 2010,
        month = aug,
       volume = {719},
       number = {2},
        pages = {1067-1082},
          doi = {10.1088/0004-637X/719/2/1067},
archivePrefix = {arXiv},
       eprint = {1006.4489},
 primaryClass = {astro-ph.HE},
       adsurl = {https://ui.adsabs.harvard.edu/abs/2010ApJ...719.1067K}
}

@INPROCEEDINGS{2010SPIE.7740E..15A,
       author = {{Axelrod}, T. and {Kantor}, J. and {Lupton}, R.~H. and {Pierfederici}, F.},
        title = "{An open source application framework for astronomical imaging pipelines}",
    booktitle = {Software and Cyberinfrastructure for Astronomy},
         year = 2010,
       editor = {{Radziwill}, Nicole M. and {Bridger}, Alan},
       series = {Society of Photo-Optical Instrumentation Engineers (SPIE) Conference Series},
       volume = {7740},
        month = jul,
          eid = {774015},
        pages = {774015},
          doi = {10.1117/12.857297},
       adsurl = {https://ui.adsabs.harvard.edu/abs/2010SPIE.7740E..15A}
}

@ARTICLE{2018PASJ...70S...5B,
       author = {{Bosch}, James and {Armstrong}, Robert and {Bickerton}, Steven and {Furusawa}, Hisanori and {Ikeda}, Hiroyuki and {Koike}, Michitaro and {Lupton}, Robert and {Mineo}, Sogo and {Price}, Paul and {Takata}, Tadafumi and {Tanaka}, Masayuki and {Yasuda}, Naoki and {AlSayyad}, Yusra and {Becker}, Andrew C. and {Coulton}, William and {Coupon}, Jean and {Garmilla}, Jose and {Huang}, Song and {Krughoff}, K. Simon and {Lang}, Dustin and {Leauthaud}, Alexie and {Lim}, Kian-Tat and {Lust}, Nate B. and {MacArthur}, Lauren A. and {Mandelbaum}, Rachel and {Miyatake}, Hironao and {Miyazaki}, Satoshi and {Murata}, Ryoma and {More}, Surhud and {Okura}, Yuki and {Owen}, Russell and {Swinbank}, John D. and {Strauss}, Michael A. and {Yamada}, Yoshihiko and {Yamanoi}, Hitomi},
        title = "{The Hyper Suprime-Cam software pipeline}",
      journal = {\pasj},
         year = 2018,
        month = jan,
       volume = {70},
          eid = {S5},
        pages = {S5},
          doi = {10.1093/pasj/psx080},
archivePrefix = {arXiv},
       eprint = {1705.06766},
 primaryClass = {astro-ph.IM},
       adsurl = {https://ui.adsabs.harvard.edu/abs/2018PASJ...70S...5B}
}

@ARTICLE{2019ApJ...873..111I,
       author = {{Ivezi{\'c}}, {\v{Z}}eljko and {Kahn}, Steven M. and {Tyson}, J. Anthony and {Abel}, Bob and {Acosta}, Emily and {Allsman}, Robyn and {Alonso}, David and {AlSayyad}, Yusra and {Anderson}, Scott F. and {Andrew}, John and {Angel}, James Roger P. and {Angeli}, George Z. and {Ansari}, Reza and {Antilogus}, Pierre and {Araujo}, Constanza and {Armstrong}, Robert and {Arndt}, Kirk T. and {Astier}, Pierre and {Aubourg}, {\'E}ric and {Auza}, Nicole and {Axelrod}, Tim S. and {Bard}, Deborah J. and {Barr}, Jeff D. and {Barrau}, Aurelian and {Bartlett}, James G. and {Bauer}, Amanda E. and {Bauman}, Brian J. and {Baumont}, Sylvain and {Bechtol}, Ellen and {Bechtol}, Keith and {Becker}, Andrew C. and {Becla}, Jacek and {Beldica}, Cristina and {Bellavia}, Steve and {Bianco}, Federica B. and {Biswas}, Rahul and {Blanc}, Guillaume and {Blazek}, Jonathan and {Blandford}, Roger D. and {Bloom}, Josh S. and {Bogart}, Joanne and {Bond}, Tim W. and {Booth}, Michael T. and {Borgland}, Anders W. and {Borne}, Kirk and {Bosch}, James F. and {Boutigny}, Dominique and {Brackett}, Craig A. and {Bradshaw}, Andrew and {Brandt}, William Nielsen and {Brown}, Michael E. and {Bullock}, James S. and {Burchat}, Patricia and {Burke}, David L. and {Cagnoli}, Gianpietro and {Calabrese}, Daniel and {Callahan}, Shawn and {Callen}, Alice L. and {Carlin}, Jeffrey L. and {Carlson}, Erin L. and {Chandrasekharan}, Srinivasan and {Charles-Emerson}, Glenaver and {Chesley}, Steve and {Cheu}, Elliott C. and {Chiang}, Hsin-Fang and {Chiang}, James and {Chirino}, Carol and {Chow}, Derek and {Ciardi}, David R. and {Claver}, Charles F. and {Cohen-Tanugi}, Johann and {Cockrum}, Joseph J. and {Coles}, Rebecca and {Connolly}, Andrew J. and {Cook}, Kem H. and {Cooray}, Asantha and {Covey}, Kevin R. and {Cribbs}, Chris and {Cui}, Wei and {Cutri}, Roc and {Daly}, Philip N. and {Daniel}, Scott F. and {Daruich}, Felipe and {Daubard}, Guillaume and {Daues}, Greg and {Dawson}, William and {Delgado}, Francisco and {Dellapenna}, Alfred and {de Peyster}, Robert and {de Val-Borro}, Miguel and {Digel}, Seth W. and {Doherty}, Peter and {Dubois}, Richard and {Dubois-Felsmann}, Gregory P. and {Durech}, Josef and {Economou}, Frossie and {Eifler}, Tim and {Eracleous}, Michael and {Emmons}, Benjamin L. and {Fausti Neto}, Angelo and {Ferguson}, Henry and {Figueroa}, Enrique and {Fisher-Levine}, Merlin and {Focke}, Warren and {Foss}, Michael D. and {Frank}, James and {Freemon}, Michael D. and {Gangler}, Emmanuel and {Gawiser}, Eric and {Geary}, John C. and {Gee}, Perry and {Geha}, Marla and {Gessner}, Charles J.~B. and {Gibson}, Robert R. and {Gilmore}, D. Kirk and {Glanzman}, Thomas and {Glick}, William and {Goldina}, Tatiana and {Goldstein}, Daniel A. and {Goodenow}, Iain and {Graham}, Melissa L. and {Gressler}, William J. and {Gris}, Philippe and {Guy}, Leanne P. and {Guyonnet}, Augustin and {Haller}, Gunther and {Harris}, Ron and {Hascall}, Patrick A. and {Haupt}, Justine and {Hernandez}, Fabio and {Herrmann}, Sven and {Hileman}, Edward and {Hoblitt}, Joshua and {Hodgson}, John A. and {Hogan}, Craig and {Howard}, James D. and {Huang}, Dajun and {Huffer}, Michael E. and {Ingraham}, Patrick and {Innes}, Walter R. and {Jacoby}, Suzanne H. and {Jain}, Bhuvnesh and {Jammes}, Fabrice and {Jee}, M. James and {Jenness}, Tim and {Jernigan}, Garrett and {Jevremovi{\'c}}, Darko and {Johns}, Kenneth and {Johnson}, Anthony S. and {Johnson}, Margaret W.~G. and {Jones}, R. Lynne and {Juramy-Gilles}, Claire and {Juri{\'c}}, Mario and {Kalirai}, Jason S. and {Kallivayalil}, Nitya J. and {Kalmbach}, Bryce and {Kantor}, Jeffrey P. and {Karst}, Pierre and {Kasliwal}, Mansi M. and {Kelly}, Heather and {Kessler}, Richard and {Kinnison}, Veronica and {Kirkby}, David and {Knox}, Lloyd and {Kotov}, Ivan V. and {Krabbendam}, Victor L. and {Krughoff}, K. Simon and {Kub{\'a}nek}, Petr and {Kuczewski}, John and {Kulkarni}, Shri and {Ku}, John and {Kurita}, Nadine R. and {Lage}, Craig S. and {Lambert}, Ron and {Lange}, Travis and {Langton}, J. Brian and {Le Guillou}, Laurent and {Levine}, Deborah and {Liang}, Ming and {Lim}, Kian-Tat and {Lintott}, Chris J. and {Long}, Kevin E. and {Lopez}, Margaux and {Lotz}, Paul J. and {Lupton}, Robert H. and {Lust}, Nate B. and {MacArthur}, Lauren A. and {Mahabal}, Ashish and {Mandelbaum}, Rachel and {Markiewicz}, Thomas W. and {Marsh}, Darren S. and {Marshall}, Philip J. and {Marshall}, Stuart and {May}, Morgan and {McKercher}, Robert and {McQueen}, Michelle and {Meyers}, Joshua and {Migliore}, Myriam and {Miller}, Michelle and {Mills}, David J.},
        title = "{LSST: From Science Drivers to Reference Design and Anticipated Data Products}",
      journal = {\apj},
         year = 2019,
        month = mar,
       volume = {873},
       number = {2},
          eid = {111},
        pages = {111},
          doi = {10.3847/1538-4357/ab042c},
archivePrefix = {arXiv},
       eprint = {0805.2366},
 primaryClass = {astro-ph},
       adsurl = {https://ui.adsabs.harvard.edu/abs/2019ApJ...873..111I}
}
\bibliographystyle{aasjournalv7}

\appendix

\section{Fringe correction procedure} \label{sec:fringe}

The LCO images, particularly in the $z$-band and to a lesser extent in the $i$-band, were affected by significant fringing patterns. To correct for this, a fringe-correction template was constructed. The method assumes a linear relationship for the fringe pattern intensity at any given pixel across different images. We used archival images from the \texttt{LCO Science Archive} that matched the observatory, telescope, instrument, and filter of our science images but targeted different sky locations. Frames containing bright or extended sources near the center were excluded.

The correction process involved the following steps. First, each image was normalized by its exposure time, divided into a grid of $20\times20$ pixel blocks, and the mean value of each block was calculated. A single image was then chosen as a reference frame. For every other image, a scatter plot was generated by plotting the mean block values of the reference frame on the x-axis against the corresponding values from the other image on the y-axis. These plots typically exhibit a multi-branched structure: divergent branches correspond to blocks containing astronomical sources, while a tight, linear branch near the origin represents blocks containing only background flux. This linear branch reflects the true scaling relationship of the fringe pattern between the two images. A linear regression was performed exclusively on this branch to determine the scaling factor and offset needed to match the backgrounds.

After all template images were scaled to the reference frame, they were median-combined to create a master fringe template. This process removes most astronomical sources. Our science target was deliberately positioned at an offset from the image center, ensuring that our photometry was unaffected by potential residual artifacts at the center of the template. While this method proved effective for most of the data, a subset of images could not be corrected satisfactorily. These included frames with significant residual patterns post-subtraction or those from which a reliable fringe pattern could not be extracted. To ensure the quality of the final light curve, these problematic images were excluded from the analysis.

\section{Corner plot of MOSFiT} \label{sec:corner}

\begin{figure}[h!]
\centering
\includegraphics[width=0.85\textwidth]{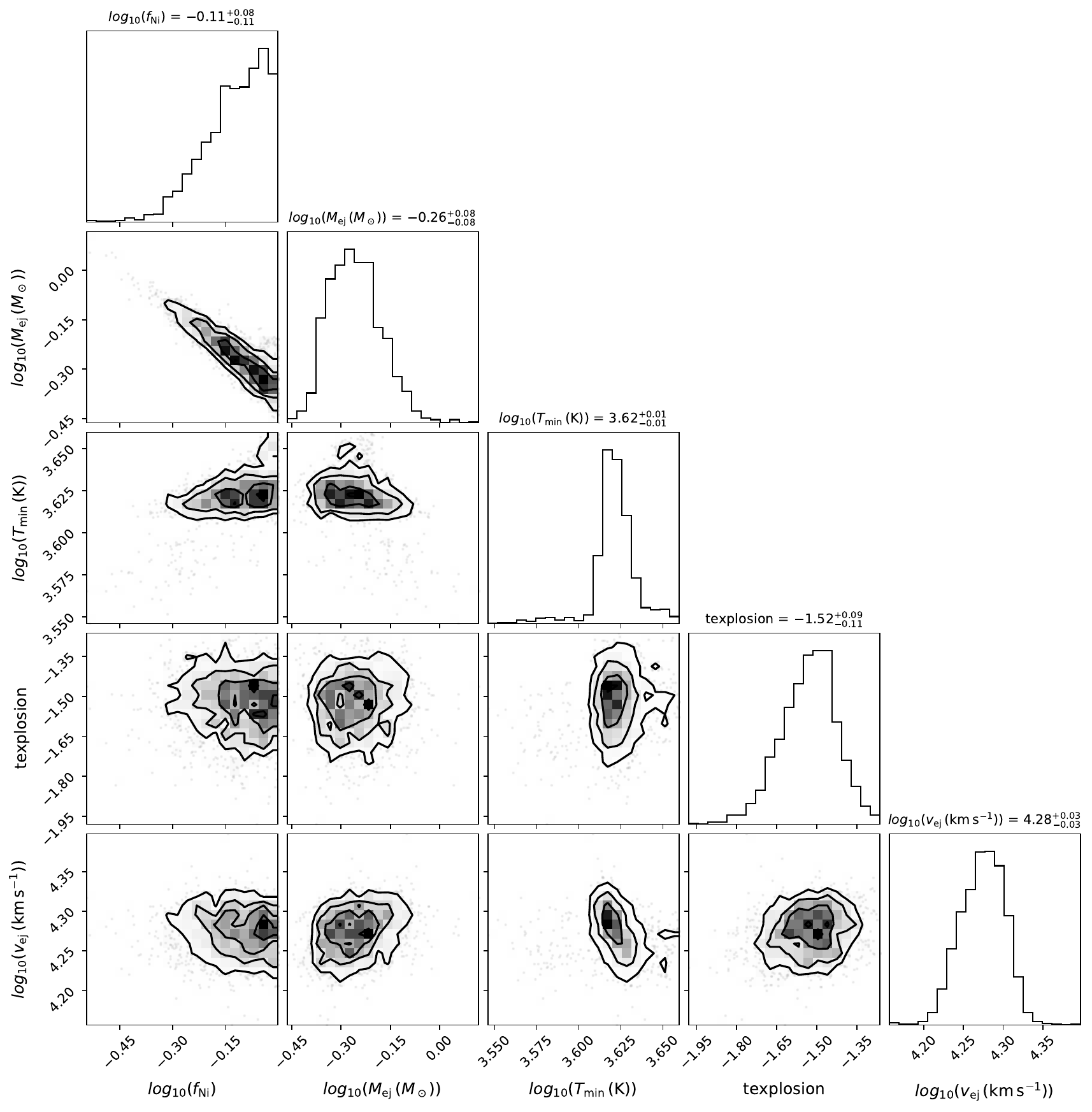}
\caption{Corner plot showing the posterior probability distributions for the parameters of the \texttt{MOSFiT} fit.}
\label{fig:param}
\end{figure}

\section{Photometry of SN 2024aedt}

\begin{table}[h!]
\centering
\centering
\begin{tabular}{cccccc}
\hline
Source & Type & MJD & Filter & Apparent Magnitude & Phase \\

\hline
WFST & DET & 60659.526 & g & $18.961 \pm 0.022$ & -12.946 \\
WFST & DET & 60659.559 & g & $18.871 \pm 0.025$ & -12.914 \\
WFST & DET & 60660.525 & g & $18.348 \pm 0.013$ & -11.965 \\
WFST & DET & 60661.516 & g & $17.894 \pm 0.039$ & -10.992 \\
WFST & DET & 60662.526 & g & $17.547 \pm 0.004$ & -10.001 \\
WFST & DET & 60662.559 & g & $17.515 \pm 0.005$ & -9.969 \\
WFST & DET & 60663.531 & u & $18.042 \pm 0.021$ & -9.013 \\
WFST & DET & 60663.579 & g & $17.187 \pm 0.003$ & -8.967 \\
WFST & DET & 60663.655 & r & $17.191 \pm 0.005$ & -8.892 \\
WFST & DET & 60663.655 & r & $17.215 \pm 0.005$ & -8.891 \\
\dots & \dots & \dots & \dots & \dots & \dots \\
\hline
\end{tabular}
\label{tab:aedt_photometry}
\caption{Detections and 5$\sigma$ upper limits of SN 2024aedt, corrected for Milky Way (MW) extinction. The first few lines are shown here for guidance regarding its form and content; the full version is available in machine-readable format.}
\end{table}

\end{document}